\documentclass[12pt,a4paper,twoside]{article}
\pdfoutput=1
\usepackage[english]{babel}
\usepackage{hyperref} 
\usepackage[numbers,sort&compress]{natbib}
\usepackage{amsmath, amsthm, amssymb, upref, mathrsfs}
\usepackage{amsfonts} 
\usepackage{graphicx}
\usepackage[left=2cm,right=2cm,top=2.5cm,bottom=2.5cm]{geometry}
\usepackage[sc,small]{caption}
\setcaptionwidth{14cm}
\newcommand{\be}{\begin{eqnarray}}
\newcommand{\ee}{\end{eqnarray}}

\newcommand{\mat}[1]{\begin{pmatrix}#1\end{pmatrix}} 

\newcommand{\wt}[1]{\widetilde{#1}}

\renewcommand{\cal}[1]{\mathcal{#1}}
\newcommand{\p}{\partial}
\newcommand{\ds}[1]{\setbox0=\hbox{$#1$}#1\hskip-\wd0\hbox to\wd0{\hss\sl/\/\hss}}

\newcommand{\hc}{\mbox{ h.c.}}

\newcommand{\lag}{\mathscr{L}}
\newcommand{\hu}{H_u}
\newcommand{\hd}{H_d}
\newcommand{\thu}{\widetilde{H}_u}
\newcommand{\thd}{\widetilde{H}_d}
\newcommand{\ie}{i.e.\ }

 \newcommand{\vertex}[2]
{\begin{center}
\begin{tabular}{cc}
\begin{tabular}{c} 
\includegraphics{#1}
\end{tabular}
&
\begin{tabular}{c}
$#2$
\end{tabular}
\end{tabular}
\end{center}}

\begin{document}
\hfill   0906.0583
\vspace{1cm}

\begin{center}
{\bf\LARGE
Neutralino Dark Matter in 
\\ BMSSM Effective Theory  \\  }

\vspace{2.5cm}

{\large
{\bf Marcus Berg$^{+}$, Joakim Edsj\"o$^{+}$,}
{\bf Paolo Gondolo$^{\dag}$, }
{\bf Erik Lundstr\"om$^{+}$ and Stefan Sj\"ors$^{+}$}
\vspace{1cm}

{\it
$^{+}$ Oskar Klein Center for Cosmoparticle Physics \\   
and \\
Cosmology, Particle Astrophysics and String Theory (CoPS) \\
Department of Physics, 
Stockholm University, Albanova\\ SE-106 91 Stockholm, Sweden\\[4mm]
$^{\dag}$ 
Department of Physics and Astronomy, University of Utah,\\ [2mm]
115 South 1400 East, 
Salt Lake City, UT 84112, USA
}}

\end{center}
\vspace{8mm}

% \tableofcontents
%\abstract
\begin{center}
{\bf Abstract}\\
\end{center}
We study thermal  neutralino dark matter in an effective field
theory extension of the MSSM, called ``Beyond the MSSM" (BMSSM) in 
Dine, Seiberg and Thomas (2007). In this class
of effective field theories, the field 
content of the MSSM is unchanged, but the little hierarchy problem is alleviated by allowing small
corrections to the Higgs/higgsino part of the Lagrangian. We perform parameter scans and compute the dark matter relic density. The light Higgsino LSP scenario is modified
the most; we find new regions of parameter space compared to the standard MSSM. This involves
interesting interplay between the WMAP dark matter bounds
and the LEP chargino bound. We also find some changes for gaugino LSPs,
partly due to annihilation through a Higgs resonance, and partly
due to coannihilation with light stops in models that are ruled in
by the new effective terms.

\clearpage

\tableofcontents
\section{Introduction}
In the next few years, 
there will be a significant amount of new data
in dark matter physics
and TeV scale particle physics. 
If  low-energy supersymmetry is relevant to the real world,
and in particular if it provides dark matter with the right properties, 
it is important to keep asking the question as to what extent the Minimal Supersymmetric
Standard Model (MSSM) is the best framework for analyzing the new data. 

The MSSM is ``minimal", in 
that it is the minimal supersymmetric completion of the standard model,
with the only supersymmetry breaking coming
from operators of dimension three or lower. 
From the low-energy point of view, this is 
a very special implementation of low-energy supersymmetry.
For example, the MSSM quartic Higgs terms 
are completely determined in terms
of known electroweak gauge couplings by supersymmetry,
and so are small (of the order $\lambda \sim 0.07$). 
 This is
quite unlike the nonsupersymmetric
standard model, where the quartic Higgs coupling $\lambda$ is fairly unrestricted,
and need not be small.
This is one reason why the nonsupersymmetric
standard model 
has no problem with the present LEP lower bound on the lightest Higgs mass, whereas
it is generally considered to be
somewhat of a problem for the MSSM (more on this in the next section). 

Part of the reason for imposing this simplicity at low energy in the MSSM 
 is standard coupling unification, where the ``energy desert" (the assumption that essentially no  new physics enters between the TeV scale
and the GUT scale) means that all new operators due to new physics are suppressed
by the enormous scale $M_{\rm GUT}$. Another reason to impose
simplicity is purely practical --- there
are already over 100 free parameters in the MSSM, so why add even more uncertainty?

Indeed, if the MSSM holds up to experiment, there is no need to add more uncertainty. 
However, what if it does not? For example, the recent flurry of model building around the PAMELA results \cite{Adriani:2008zr}
is partially motivated by the fact that the standard neutralino does not seem to be able to reproduce the PAMELA results. We do not attempt to model any aspects of any particular new experiment in this paper, 
but the model building motivated by the new PAMELA data provides an example of how 
we may need to go beyond the  ``maximally minimal'' 
framework for neutralino dark matter when attempting to ascribe any new phenomena
to dark matter physics. 

But with guidance from experiment still scarce, how can we possibly generalize the MSSM, with its 124 parameters, in a meaningful way? 
The typical approach in particle phenomenology is to consider non-minimal extensions of the MSSM,
where new fields and new parameters are added to the theory
to solve specific problems, with some success and some degree of arbitrariness. 
Since so little is known about this non-minimal physics, it seems meaningful
to attempt to be systematic about this generalization. 
A first step could be not to add any new particles at all (``minimally non-minimal"), 
but only allow indirect effects by any new heavy particles on the
parameters in the Lagrangian, i.e.\ think of the MSSM as an effective field theory 
with a ``scale of new physics" $M \ll M_{\rm GUT}$, which can be as low as phenomenologically allowed.\footnote{In the past one could have tried to argue
against such new thresholds based on intuition from string models,
but with the advent of stabilized string models with
many interesting thresholds far below $M_{\rm GUT}$ (e.g.\ \cite{Balasubramanian:2005zx,Conlon:2005ki}), this seems overly restrictive. Incidentally, such string models often sport non-neutralino dark matter, as in 
\cite{Conlon:2007gk, Dudas:2009uq},
or  general non-thermal dark matter, as in
\cite{Acharya:2008bk}.}
  Even within
this limited framework,
there is already a bewildering variety of interactions and couplings one could add to 
or modify in the MSSM Lagrangian. 

A very rough way to classify such new parameters is by whether turning
them on tends to make experimental constraints harder or easier to satisfy. 
For example, most parameters that already exist in the MSSM-124, such as off-diagonal squark masses, immediately bring the model beyond experimental bounds on flavor and CP violation when turned on to sizeable values. Therefore, one useful way to make the MSSM slightly non-minimal would be 
if we could identify some class of new parameters that  
rule in a given model, instead of ruling it out.
Such parameters were identified by Dine, Seiberg,
and Thomas \cite{Dine:2007xi}.\footnote{For earlier work in specific models
that identified similar operators, see
\cite{Brignole:2003cm,Casas:2003jx} and \cite{Strumia:1999jm}.}
They enumerated some operators at dimension four, five and six
 that go beyond the MSSM (hence the acronym BMSSM). These operators
 can be added with small coefficients to capture
possible non-minimal physics at the scale $M$, which we take to be $M\gtrsim$ 5 TeV or somewhat higher.
In particular, we impose that the scale $M$
is sufficiently high --- or the underlying microscopic model sufficiently restricted ---
that we can neglect operators  suppressed as $1/M^2$ or more. 
By the logic above, we further restrict the operators at order $1/M$ 
by not turning on operators in the squark and slepton sectors
that would worsen problems with flavor-changing neutral currents. 
This leaves only two operators in the Higgs sector, so two new parameters $\epsilon_1$
and $\epsilon_2$. 
(It would be interesting 
to go beyond our analysis by performing a detailed study of flavor physics,
CP violation 
or $1/M^2$ operators, but we leave that to future work. See also
section \ref{sec:outlook}.)

Roughly speaking, if the MSSM was already ``natural'',
then adding some effective operators as small perturbations would not
affect it  much. It is a sign that the MSSM is very special
that this small addition can 
change the way data would be interpreted
in important ways. For example, tree-level deviations
from the supersymmetric quartic Higgs coupling $\lambda$ could
rule in models that were previously
ruled out by LEP bounds.  These models may
have new phenomenological properties, e.g.\ for dark matter.
Other aspects of the new models
will be discussed in detail in this paper. 

The kind of dimension-four operator that shifts the quartic Higgs coupling
has been analyzed before (e.g. \cite{Martin:1999hc}),
but not together with the ``companion'' dimension-five 
nonrenormalizable Higgs-Higgs-higgsino-higgsino
interactions 
we need to include here. These
operators are ``companions'' because they are both
effective dimension five, i.e.\ the new quartic Higgs terms have
coefficients that scale as $1/M$. (There is of course another mass scale
in the numerator to make the quartic Higgs coupling dimensionless, but this is 
a smaller scale intrinsic to the low-energy theory, such as the
supersymmetric Higgs mass parameter $\mu$.) The new quartic terms include
a term that breaks supersymmetry and is technically hard. 
The new nonrenormalizable Higgs-Higgs-higgsino-higgsino
interaction terms also produce contributions to chargino and neutralino
mass matrices and mixings.

Completing the program of \cite{Dine:2007xi},
Antoniadis et al \cite{Antoniadis:2008ur} 
classified higher-dimensional operators
including the effects of general field redefinitions
and loop corrections. We will not include any
of their operators beyond those of 
\cite{Dine:2007xi}; this will be explained further below. 

Shortly before this work was concluded, Cheung et al considered
effective operators in the light Higgsino scenario \cite{Cheung:2009qk}.
We will comment on the relation between their work and our work below.

\section{Review of the MSSM Lagrangian}
\label{MSSMrev}
This section is a very brief review of the MSSM Lagrangian,
in order to set notation.
Our conventions are summarized in appendix \ref{appconv}.

\subsection{The MSSM superpotential}
The MSSM superpotential is given by:
\be  \label{WMSSM}
	W_\text{MSSM} = -\bar u \mathbf y_u Q H_u  + \bar d \mathbf y_d Q H_d + \bar e \mathbf y_e L H_d - \mu H_u H_d.
\ee
where $Q$ are the left-handed quark superfields, 
$u$ and $d$ are right-handed quark superfields, $H_u$ and $H_d$ are the Higgs superfields,
$L$ are left-handed lepton superfields, $e$ are right-handed lepton superfields, and ${\mathbf y_u}, {\mathbf y_d}, {\mathbf y_e}$ are Yukawa coupling matrices. We have 
suppressed family indices and $SU(2)$ contractions. (We 
 use the conventions of Haber and Kane \cite{Haber:1984rc}, that
are summarized in appendix \ref{appconv}.)

The soft supersymmetry breaking Lagrangian
is (except for the soft terms in the Higgs potential, that we write 
in (\ref{Vhiggs}) below):
\be  \label{VsoftMSSM}
V_{\rm soft}
&=& (-\tilde{\bar{e}} {\mathbf A_e}{\mathbf y_e}\tilde{L}H_d 
- \tilde{\bar{d}} {\mathbf A_d}{\mathbf y_d}\tilde{Q}H_d 
 +\tilde{\bar{u}} {\mathbf A_u}{\mathbf y_u}\tilde{Q}H_u 
+ \mbox{h.c.})  \nonumber
\\ [1mm]
&+&
\tilde{Q}^{\dagger} {\mathbf m}^2_Q \tilde{Q} 
+\tilde{L}^{\dagger} {\mathbf m}^2_L \tilde{L} 
+\tilde{u}^{\dagger} {\mathbf m}^2_u \tilde{u} 
+\tilde{d}^{\dagger} {\mathbf m}^2_d \tilde{d} 
+\tilde{e}^{\dagger} {\mathbf m}^2_e \tilde{e} 
\\
&+&{1 \over 2}\left(
M_1 \tilde{B} \tilde{B}
+M_2 (\tilde{W}^3\tilde{W}^3+2 \tilde{W}^+\tilde{W}^-)
+M_3 \tilde{g}\tilde{g} + \mbox{h.c.} \right)   \nonumber 
\ee
where the ${\mathbf A}$ are general soft trilinear terms,
the ${\mathbf m}$ are general soft sfermion masses,
$M_1,M_2,M_3$ are gaugino masses,
and the tilded fields are superpartners of Standard Model fields.
In the {\tt DarkSUSY} introduction paper \cite{Gondolo:2004sc},
the $SU(2)$ contraction is the opposite of Haber and Kane, so
we have converted  eq.\ (\ref{VsoftMSSM}) to our conventions (again, see appendix \ref{appconv}
for some more details). 

\subsection{The MSSM-7}
\label{The MSSM-7}
For later phenomenological purposes
we will truncate the 124 MSSM parameters (the ``MSSM-124") to
seven parameters, the MSSM-7. These seven parameters
are given 
at the weak scale. They are: a single gaugino mass scale $M_2$,
a single universal scalar mass $m_0$ for squarks and sleptons
(replacing the general ${\mathbf m}$ matrices in (\ref{VsoftMSSM})), the two trilinear soft terms $A_t$ and $A_b$ (replacing the general ${\mathbf A}$ matrices in (\ref{VsoftMSSM})
by ${\mathbf A}_e=0,{\mathbf A}_d={\rm diag}(0,0,A_b), {\mathbf A}_u={\rm diag}(0,0,A_t)$),
the ratio of Higgs vacuum expectation values $\tan \beta$, the supersymmetric Higgs mass $\mu$
and the CP-odd Higgs scalar mass $m_{A^0}$. The last three parameters will  be discussed
in more detail in the next section. 

To reduce the three gaugino masses $M_1$, $M_2$ and $M_3$ to a single scale $M_2$, we impose the gaugino mass unification condition 
\be  \label{Mgut}
M_3 = {\alpha_s \over \alpha} \sin^2 \theta_{\rm W} M_2 = 
{3 \over 5} {\alpha_s \over \alpha} \cos^2 \theta_{\rm W} M_1
\ee
%giving 
%$M_1\sim M_2/2$ and $M_3 \sim c M_2$
so we can choose
$M_2$ as the free parameter. From the effective field theory 
perspective  we adopt in this paper, we 
have no real reason to expect standard gauge unification to hold, but since we want
to compare to existing models, we will simply focus on this line through $(M_1,M_2,M_3)$ space for simplicity.

\subsection{The MSSM Higgs potential}
\label{sec:mssmhiggs}
The tree-level
MSSM Higgs potential is given by
\be  \label{Vhiggs}
V_{\rm MSSM}&=& (|\mu|^2 + m_{H_u}^2) H_u^{\dagger} H_u  + 
(|\mu|^2 + m_{H_d}^2) H_d^{\dagger} H_d
-\left( b H_u H_d + \hc \right) \nonumber \\
&& +{g^2 \over 8}\left[ (H_u^{\dagger}H_u + H_d^{\dagger}H_d)^2 - 4 (H_u H_d)^{\dagger}(H_u H_d)
\right] + {g'^2 \over 8} (H_u^{\dagger} H_u - H_d^{\dagger}H_d)^2 \; ,
\ee
and after writing out $SU(2)$ contractions (using the conventions in appendix \ref{appconv})
and rearranging terms:
\begin{align}  \label{Vhiggs2}
	V_\text{MSSM} & = (|\mu|^2+m^2_{H_u})(|H^0_u|^2+|H^+_u|^2) + (|\mu|^2+m^2_{H_d})(|H^0_d|^2+|H^-_d|^2) \nonumber \\
	& + \left( b(H_u^+H_d^- - H_u^0 H_d^0) + \hc \right) \nonumber \\
	& + \frac{1}{8}(g^2 + g'^2) \left(|H^0_u|^2 + |H^+_u|^2 - |H^0_d|^2 - |H^-_d|^2 \right)^2 + \frac{1}{2} g^2 \left| H_u^+ H_d^{0*} + H_u^0 H_d^{-*} \right|^2 \; ,
\end{align}
 which reproduces the expression in \cite{Martin:1997ns}.
It will be relevant in the following that the MSSM quartic Higgs terms are exactly supersymmetric
(i.e.\ completely determined by the couplings $g$ and $g'$ of the ordinary 
electroweak theory, and they are in turn determined by experiment). The reason that the
quartic Higgs terms are exactly supersymmetric is simply by construction:
the MSSM only includes {\it soft} supersymmetry breaking operators, which by definition
are of dimension three or lower. In particular, the quartic Higgs coefficients (let
us schematically call them $\lambda$) are quite small,
and any additional small contributions from effective operators may have important effects. 
For example, in  eq.\ (\ref{Vhiggs}) we have at the Z mass
\be  \label{lambda}
\lambda \; \sim\;  { g^2+g'^2 \over 8} \approx 0.07 \; .
\ee

The parameters of the Higgs potential are restricted such that neutral components of the Higgs doublets $H_u^0$ and $H_d^0$ receive real and positive vaccum expectation values
\be
 v_u = \langle H_u^0 \rangle, \quad v_d = \langle H_d^0 \rangle.
\ee
At tree level, the vacuum expectation values are related  to the $Z^0$ and $W^\pm$ boson masses
by
\be \label{v}
 v^2 = v_u^2 + v_d^2 = \frac{2 m_Z^2}{g^2+g'^2} = \frac{2 m^2_W}{g^2} \approx (174 \text{ GeV})^2
\ee
thus only the ratio between $v_u$ and $v_d$ is unspecified. This ratio is denoted
\be
 \tan\beta = \frac{v_u}{v_d}.
\ee
In terms of the Lagrangian parameters \{$m_{H_u}, m_{H_d}, \mu, b$\}
and the coupling constants $g,g'$, the vacuum expectation values are determined by
\be \label{betaparam}
 \sin{2\beta} = \frac{2b}{m^2_{H_u} + m^2_{H_d} + 2 |\mu|^2}
\ee
\be  \label{vparam}
 v^2 = \frac{2}{g^2+g'^2} \left( \frac{|m^2_{H_d} - m^2_{H_u}|}{\sqrt{1 - \sin^2{2\beta}}} - m^2_{H_u} - m^2_{H_d} - 2 |\mu|^2  \right) .
\ee
The expectation value $v$ is fixed to the experimental value eq.\ (\ref{v}),
so the point of equations (\ref{betaparam}),  (\ref{vparam}) is simply that the original 4 parameters
\{$m_{H_u}, m_{H_d}, \mu, b$\} in the potential in eq.\ (\ref{Vhiggs2})
can be traded for 3 parameters \{$\tan \beta, \mu, b$\} using one constraint (\ref{v}).
We will also trade $b$ for $m_{A^0}$ below. 

The Higgs fields are expanded around their vacuum expectation values
\be
 \begin{pmatrix}
  H_u^0 \\ H_d^0
 \end{pmatrix}
 =
 \begin{pmatrix}
  v_u \\ v_d
 \end{pmatrix}
  +\frac{1}{\sqrt{2}} R_\alpha
 \begin{pmatrix}
  h^0 \\ H^0
 \end{pmatrix}
 +\frac{i}{\sqrt{2}} R_{\beta^0}
 \begin{pmatrix}
  G^0 \\ A^0
 \end{pmatrix}
\ee
and
\be
 \begin{pmatrix}
  H_u^+ \\ H_d^{-*}
 \end{pmatrix}
 = R_{\beta^\pm}
 \begin{pmatrix}
  G^+ \\ H^+
 \end{pmatrix}.
\ee
We diagonalize the quadratic part of the Higgs potential
by the rotation matrices
\be \label{rotmat}
 R_\alpha =
\begin{pmatrix}
 \cos\alpha & \sin\alpha \\
 -\sin\alpha & \cos\alpha
 \end{pmatrix}, \quad
 R_{\beta^0}=
 \begin{pmatrix}
  \sin\beta^0 & \cos\beta^0 \\
  -\cos\beta^0 & \sin\beta^0
 \end{pmatrix}, \quad
 R_{\beta^\pm}=
 \begin{pmatrix}
  \sin\beta^\pm & \cos\beta^\pm \\
  -\cos\beta^\pm & \sin\beta^\pm
\end{pmatrix} \; . \nonumber
\ee
This gives us the following tree-level mass spectrum
\be
 m_{G^0}^2 = m_{G^\pm}^0 = 0,
\ee
\be  \label{A0param}
 m_{A^0}^2 = m^2_{H_u} + m^2_{H_d} + 2 |\mu|^2,
\ee
\be  \label{treehiggs}
 m^2_{h^0,H^0} = \frac{1}{2} \left(m_{A^0}^2 + m_Z^2 \mp \sqrt{(m_{A^0}^2-m_Z^2)^2 + 4m_{A^0}^2m_Z^2 \sin^2{(2\beta)}} \right),
\ee
where clearly the lighter Higgs $h^0$  has the upper sign, and
the heavier Higgs $H^0$ has the lower sign.
We have now traded the original 4 parameters
\{$m_{H_u}, m_{H_d}, b,\mu$\} in the potential in eq.\ (\ref{Vhiggs2})
for the 3 parameters  \{$\tan \beta,\mu, m_{A^0}$\} using the constraint (\ref{v})
and the ``trading equations" (\ref{betaparam}), (\ref{vparam}), (\ref{A0param}).
The charged Higgs fields have masses that are simply
%\be
% m^2_{h^0,H^0} = \frac{1}{2} \left(m_{A^0}^2 + m_Z^2 \mp \sqrt{\Delta} \right), \quad \Delta = (m_{A^0}^2-m_Z^2)^2 + 4m_{A^0}^2m_Z^2 \sin^2{(2\beta)}
%\ee
\be
 m_{H^\pm}^2 = m_{A^0}^2 + m_W^2,
\ee
while for the angles in the rotation matrices (\ref{rotmat}) we find $\beta^0 = \beta^\pm = \beta$, and 
the angle $\alpha$ is determined by
\be
 \sin(2\alpha) = - \frac{m_{A^0}^2 + m_Z^2}{m_{H^0}^2 - m_{h^0}^2} \sin{(2\beta)}, \quad
 \cos(2\alpha) = - \frac{m_{A^0}^2 - m_Z^2}{m_{H^0}^2 - m_{h^0}^2} \cos{(2\beta)}.
\ee
Now that we have the Higgs sector masses and mixings,
we can read off from eq.\ (\ref{treehiggs}) the familiar 
statement that the tree-level lightest Higgs mass satisfies $m_{h^0} \leq m_Z$. 
This is the Higgs version of the ``little hierarchy problem",
that fairly large loop corrections are needed to bring the Higgs mass above the
LEP bound $m_{h^0}\gtrsim$ 114 GeV.  (See section (\ref{computational}) for some more
details on how we implement this experimental bound, including loop corrections.)
The little hierarchy problem 
has been studied for a long time 
(see e.g.\ \cite{Haber:1990aw}) and
was the motivation for
 \cite{Brignole:2003cm,Casas:2003jx,Dine:2007xi} to consider the corrections 
 we are about to include.
 
Why is the tree-level Higgs $h^0$ mass so low in the MSSM? Let us consider the toy potential 
\be \label{toypotential}
V_{\rm toy} = -\mu^2 \phi^2 + \lambda \phi^4
\ee
which has a minimum at $\phi^2_{\rm min} \equiv v^2 = \mu^2/(2\lambda)$, and the mass squared at the minimum is $m^2= V''(v) = 
4\mu^2$, that is 
\be
m &=& 2 \sqrt{2 \lambda} v   \nonumber  \\
& \lesssim & 2\sqrt{2\cdot 0.07}\cdot 174 \text{ GeV}  \approx  120  \text{ GeV} \; ,   \label{mtoy}
 \ee
using the rough value of the quartic coupling $\lambda$ from (\ref{lambda}).
As emphasized above, the MSSM quartic Higgs coupling $\lambda$ is small due to the exact supersymmetry
of MSSM dimension four operators. It is this smallness  in eq.\ (\ref{lambda}) that we want to relax a little,
so that the mass in  eq.\ (\ref{mtoy}) can be somewhat larger. 

\subsection{MSSM neutralinos and charginos}
\label{mssmchi}

The MSSM spectrum also includes neutralinos
and charginos. We will not repeat the standard discussion here,  but
 only give results in the modified theory
 in sections \ref{sec:neumass} and \ref{sec:chargino}, except for two things:
 first, let us already note that we impose that the neutralino $\wt \chi^0_1$ 
is  the  Lightest Supersymmetric Particle (LSP),
  so when we write $m_{\rm LSP}$ it always means $m_{\wt \chi^0_1}$. 
  The mass eigenstate is found from
  rotating gauge eigenstates $\wt\psi^0 = (\wt B^0, \wt W^0, \wt H_d^0, \wt H_u^0)^T$ as
\be \label{LSProtate}
\wt \chi^0_1 = N_{11} \wt B^0 + N_{12} \wt W^0 + N_{13} \wt H_d^0 + N_{14} \wt H_u^0 \; .
\ee
Second, we define the {\it gaugino fraction} of the LSP in terms
of the matrix elements $N_{1j}$  as
\be  \label{Zg}
Z_{\rm g} = |N_{11}|^2 +  |N_{12}|^2 \; .
\ee
As is clear from (\ref{LSProtate}), this simply says how much
of the LSP is gaugino (i.e.\ $\wt B^0$ or $\wt W^0$). 

When presenting numerical results, we typically prefer to use
the physical parameters $(Z_{\rm g},m_{\rm LSP})$
rather than model parameters like $(\mu,M_2)$. 
To provide an idea of the relation between the gaugino fraction and the model parameters, 
it is useful to note that (see e.g.\ \cite{Gunion:1987yh})
\be
1-Z_g \; \sim\;  \sin^2\theta_W \left ({m_Z \over \mu}\right)^2
  \quad \mbox{ when } m_Z \ll M_1\ll \mu
\ee
and
\be
Z_g \; \sim \;  \frac{1 \pm \sin2\beta}{2} \left(
\frac{\sin^2\theta_W}{M_1^2}+\frac{\cos^2\theta_W}{M_2^2} \right) m_Z^2
 \quad \mbox{ when } m_Z \ll \mu \ll M_1  \; . 
\ee
Here the sign is minus for the neutralino we call $N3$ in eq.\ (\ref{neumass3}) below,
and plus for $N4$.  We impose the gaugino mass unification condition (\ref{Mgut}),
so we can always express the $M_1$ in these equations in terms of $M_2$. 
Thus 
 if we set 
 \be
m_Z \ll M_2\ll \mu = \mbox{1000 GeV }
 \ee 
 we find $1-Z_{\rm g} \sim 10^{-2}$ (mostly gaugino), and if we set
\be
 m_Z \ll \mu \ll M_2 = \mbox{1000 GeV }
 \ee 
we obtain  $Z_{\rm g} \sim 10^{-2}$ (mostly higgsino).
This is somewhat representative of certain parameter sets
we will use later, e.g.\ table \ref{bench}. 

This concludes our short review of the MSSM. 
For more details, we refer to review articles such as \cite{Martin:1997ns}. 

\section{Effective theory}
\label{sec:eff}
We now consider allowing for higher-dimensional operators
in the Lagrangian of the MSSM, with small free parameters as coefficients.
For more on how those parameters can be related
to parameters of specific underlying theories, see appendix \ref{eft}.
To be precise,  we will incorporate operators of {\it effective} dimension $> 4$
(counting the power of $1/M$ only),
but {\it scaling} (naive)  dimension $\geq 4$ (counting the total mass dimension of the coefficient,
including $M$ but also $\mu$ and the scale of supersymmetry breaking $m_{\rm SUSY}$).

To be useful, the effective theory should not add too many new parameters.
The number of effective operators one could a priori consider adding to the
MSSM at scaling dimension four and five is in the several hundreds. We will restrict consideration
to 
\begin{itemize}
\item operators that preserve baryon and lepton number, which  must 
be a good approximation. 
\item 
dropping the fairly large number of operators that
can be added in the squark sector \cite{Antoniadis:2008es}.\footnote{These include certain
operators that have recently been relevant
in a different dark matter context  in \cite{Mardon:2009gw},
based on general gauge mediation as discussed in 
\cite{Csaki:2008sr,Komargodski:2008ax}.}
This may 
or may not be a good approximation, but is only a simplifying assumption.
To seriously study physics of effective operators in the squark sector, we should consider more general squark couplings
and masses than the MSSM-7 we use in the parameter scan 
below, which has universal masses and couplings at the electroweak scale. 
\item
dropping CP-violating higher-dimensional operators.
To go beyond this, we should again go to a more general starting point than the MSSM-7 we use in the parameter scan 
below, which already has almost all the potential CP violation in the MSSM 
set to zero in order to easily pass experimental constraints. 
\end{itemize}

With this combination of approximations and simplifications,
it was pointed out by \cite{Dine:2007xi} that
there are only two dimension-five operators that
can be added in the superpotential in the Higgs sector, and a reasonably small number of dimension-six operators
in the K\"ahler potential.
In this paper, we consider the mass scale of new physics $M$ 
to be sufficiently large that the dimension-six ($1/M^2$) operators do not contribute appreciably,
e.g.\ $M \sim 5-10$ TeV.

Consider the BMSSM superpotential
\be  \label{newW}
W = W_\text{MSSM} + W_5
\ee
where $W_{\rm MSSM}$ is 
given in eq.\ (\ref{WMSSM}), and
$W_5$ is  the following operator at effective dimension five:\footnote{This is not
the same $\lambda$ as in the previous section; this $\lambda$
only occurs in this section and in the Appendix.}
\be \label{Weff}
W_5 =  \frac{\lambda}{M} (\hu\hd)^2 \; .
\ee
Here $\lambda$ is a dimensionless number and $M$ is a large mass parameter,
though $M \ll M_{\rm GUT}$. 
Supersymmetry breaking operators can be parametrized by introducing a coupling
in (\ref{Weff}) that depends on a spurion field
$\cal Z$ as 
\be  \label{susybr}
 \lambda \to  \lambda(1+ \cal Z)\; , \qquad \cal Z = m_{\rm SUSY}\theta^2
\ee
where $m_{\rm SUSY}$ is the scale of supersymmetry breaking.

The new terms in the Lagrangian are now calculated (see appendix \ref{appconv})
from the superpotential $W$ in eq.\ (\ref{newW}). From the supersymmetry-preserving part
(\ref{Weff}) we find the following contribution to the Higgs scalar potential, for any scalar field $\phi_i$:
\be
V_F &=& \sum_i \left| {\partial W \over \partial \phi_i } \right|^2
=  \sum_i \left| {\partial W_{\rm MSSM}  \over \partial \phi_i } +  {\partial W_5 \over \partial \phi_i}
 \right|^2  \nonumber \\
 &=& \sum_i \Bigg[
\left| {\partial W_{\rm MSSM} \over \partial \phi_i } \right|^2+
 \underbrace{  {\partial W_5 \over \partial\phi_i}  \left({\partial W_{\rm MSSM}\over \partial \phi_i}\right)^{\!*} 
 + \mbox{h.c.}}_{\textstyle \delta V_1} +\;  {\mathcal O}( 1 /M^{2}) \Bigg]  \label{crossterm}
\ee
The cross term gives the $1/M$ operator we are interested in. 
Explicitly, from the supersymmetry preserving operator
in (\ref{Weff}) we obtain the following contribution to the Higgs 
scalar potential
\begin{align}  \label{V1}
 \delta V_1 = 2 \epsilon_1 (\hu\hd)(\hu^\dagger \hu + \hd^\dagger \hd) + \hc 
\end{align}
where we introduced the dimensionless number
\be\label{epsilon1}
 \epsilon_1 = - \frac{\lambda}{M} \mu^*.
\ee
Notice that $\epsilon_1$ is dimensionless and involves
both $\mu$ and $M$, since it comes from the cross term in (\ref{crossterm}).
Thus, (\ref{V1}) has effective dimension five
and (obviously) scaling dimension four. 

Similarly, 
from the supersymmetry breaking operator in (\ref{susybr}) we get the following 
``soft" contribution to the Higgs scalar potential
%\be
%V_{\rm soft} = V_{\rm soft, MSSM} + \delta V_2
%\ee
%
\begin{align}  \label{V2}
 \delta V_2 = \epsilon_2 (\hu\hd)^2 + \hc  \quad .
\end{align}
Here we  introduced the dimensionless number
\be  \label{epsilon2}
 \epsilon_2 = -\frac{\lambda}{M} m_{\rm SUSY}.
\ee
Although we called the contribution $\delta V_2$ ``soft", it is clearly not soft in the sense
of being dimension three or less. Martin \cite{Martin:1999hc} calls it technically hard.
The point of the classification into
hard and soft in the MSSM is that hard terms could spoil the relatively
nice UV behavior of the MSSM. 
However, since $\delta V_2$ is only an effective theory contribution, that will be matched
to an underlying theory at the scale $M$  (see appendix \ref{eft}),
it will not affect the UV behavior in any important way.

It is already clear at this point that when we give expectation values to  the Higgs fields, the operators 
$\delta V_1$ and $\delta V_2$ 
will affect the minimization of the Higgs potential and could potentially affect the mass of the lightest Higgs field.

Finally, using (\ref{fermionint}) in appendix \ref{appconv}, the supersymmetry-preserving operator
(\ref{Weff}) also produces the following Higgs-Higgs-higgsino-higgsino interactions\footnote{Here 
we disagree with \cite[eq.\ 18]{Dine:2007xi} on the overall sign,
since our $\epsilon_1$ is defined by eq.\ (\ref{epsilon1}). See also appendix \ref{appconv}. }
\begin{align}  \label{lag3}
 \delta\lag_{3} = + \frac{\epsilon_1}{\mu^*} \left( 2 (\thu \thd) (\hu \hd) + 2 (\hu \thd) (\thu \hd)
 + (\thu \hd)^2 + (\hu \thd)^2 \right) + \hc
\end{align}
These terms introduce new interactions, and when the Higgs fields acquire vacuum expectation values,
also new mass terms for neutralinos and charginos.

In general, there will also be dimension-five cross terms 
of (\ref{Weff}) with Yukawa couplings in the MSSM superpotential (\ref{WMSSM}), 
producing for example
\be
| \ldots + \tilde{Q}\tilde{q} + {\lambda \over M}H^3 |^2
\ee
where $H$ is some trilinear combination of Higgs fields. 
However, these only give contributions to the additional dimension-five operators in the 
squark sector, that we have already set to zero. 
Further, the higher-dimension terms do not contribute directly to Yukawa couplings.
They can contribute to processes with 3-particle final states, but we neglect those processes.

\section{Effective theory: final result}
In the previous section we constructed our  BMSSM theory.
To be precise, we constructed an ``MSSM${}_5$'' theory, which is the subset of 
BMSSM to order $1/M$. In addition,
we do not consider the most general MSSM${}_5$ theory, but imposed further restrictions on the $1/M$ operators, as detailed in the beginning of the previous
section. 
With this understanding, we will consider
the MSSM Lagrangian $ \lag$ plus the terms (\ref{V1}), (\ref{V2}), (\ref{lag3}):
\be  \label{expllag}
 \delta\lag &=&
 -\delta V_1 - \delta V_2  + \delta \lag_3   \nonumber \\ [2mm]
&=& - \left( 2\epsilon_1 (\hu\hd)(\hu^\dagger \hu + \hd^\dagger \hd) + \hc \right)
 - \left(\epsilon_2 (\hu\hd)^2 + \hc \right)  \\
 &&\hspace{-1cm} + \left[ \frac{\epsilon_1}{\mu^*} \left( 2 (\thu \thd) (\hu \hd) + 2 (\hu \thd) (\thu \hd)
 + (\thu \hd)^2 + (\hu \thd)^2 \right) + \hc \right]  .  \nonumber
\ee
As before, our conventions are spelled out in appendix \ref{appconv}.
The terms in the second line change the Higgs potential,
and those in the third line add new Higgs-Higgs-higgsino-higgsino couplings
at scaling dimension five, and affect the neutralino and chargino masses, mixings
and couplings to Higgs bosons.
Thus, we have two new free model parameters: 
the small dimensionless numbers $\epsilon_1$ and $\epsilon_2$. 
Our restriction to CP-conserving new interactions (see the previous section)
lets us restrict $\epsilon_{1}$ and $\epsilon_{2}$ to be real,
but they can be positive or negative. 
See also section \ref{sec:outlook} for some comments
on more general BMSSM models. 

As long as we stay at energies sufficiently below the energy scale $M$,
it is a standard exercise to calculate
cross sections including also the nonrenormalizable interactions
in the second line of (\ref{expllag}). For 
a brief review of how this is done, see appendix \ref{eft}. 
The Feynman rules due to the new operators  are presented in appendix \ref{feynmanrules}.

We now proceed to see how the corrections
affect the Higgs potential, the neutralino
masses and mixings, and the chargino masses and mixings.
These different aspects of the BMSSM corrections
can all affect the relic density of dark matter in different regions of parameter space,
as is investigated in section \ref{sec:results}.

\subsection{BMSSM Higgs potential}
\label{sec:bmssmhiggs}
Now we turn on  $\epsilon_{1}$ and $\epsilon_{2}$ 
from (\ref{expllag}) to small but nonzero values. 
The minimum expressed in terms of the vacuum expectation values $v_u,v_d$
will change, so the ratio $\tan{\beta}$ will change. 
However, since the vacuum expectation value $ v = \sqrt{v_u^2 + v_d^2 } \approx 174 \text{ GeV}$ is 
fixed to the experimental value, we keep $v$ fixed when we introduce the $\epsilon_{1,2}$ corrections.

The ``trading equations" (\ref{betaparam}), (\ref{vparam}) and (\ref{A0param}) are replaced 
by 
\be
\frac{2 b}{\sin2\beta} &=& m_{H_u}^2+m_{H_d}^2+2|\mu|^2  +
4 v^2 \epsilon_1 \left(\frac{1}{\sin2\beta}+\sin 2\beta\right) + 2 v^2 \epsilon_2\\
v^2 &=& \label{vshifted}
\frac{2}{g^2+g^{\prime 2}+8 \epsilon_1\sin2\beta} 
\left( \frac{|m^2_{H_d} - m^2_{H_u}|}{\sqrt{1 - \sin^2{2\beta}}} - m^2_{H_u} - m^2_{H_d} - 2 |\mu|^2  \right) \\ [1mm]
m_{A^0}^2 &=& m_{H_u}^2+m_{H_d}^2+2|\mu|^2 + 4 v^2 \epsilon_1 \sin2\beta-2 v^2
\epsilon_2  \label{A0paramshifted} \; . 
\ee
In the effective theory, we use these new equations to eliminate $\{m_{H_u},
m_{H_d},b\}$ in favor of $\{\tan\beta, v, m_{A^0}\}$. As a consequence, the
expressions for the charged Higgs mass and the CP-even Higgs mass matrix
depend explicitly on $\epsilon_1$ and $\epsilon_2$.
(We of course still have  $m_{G^0}^2 = m_{G^\pm}^2 = 0$ for the Goldstone bosons.) For the CP-even Higgs bosons we find
the mass matrix
%In terms of the Lagrangian parameters we find
%\be
% \sin{(2\beta)} = \frac{2b}{m^2_{H_u} + m^2_{H_d} + 2 |\mu|^2} + \delta(\sin{2\beta})
%\ee
%where
%\be
%\delta{(\sin{(2\beta)})} = \frac{2v^2}{m^2_{H_u} + m^2_{H_d} + 2 |\mu|^2} \left[ 2\epsilon_1 \left(1 + \left(  \frac{2b}{m^2_{H_u} + m^2_{H_d} + 2 |\mu|^2} \right)^2 \right) -\epsilon_2  \frac{2b}{m^2_{H_u} + m^2_{H_d} + 2 |\mu|^2} \right]
%\nonumber 
%\ee
%and neglecting quadratic order in $\epsilon$ this reduces to
%\be
%\delta{(\sin{(2\beta)})} = \frac{2v^2}{m_{A^0}^2} \left(2\epsilon_1 (1 + \sin^2{(2\beta)}) -\epsilon_2 \sin{(2\beta)}\right)\;
%.
%\ee
\be
{\mathcal M} = \begin{pmatrix}
m_{A^0}^2 c_{\beta}^2 + m_Z^2 s_{\beta}^2 + 4 v^2 \epsilon_1 s_{2\beta} + 4 v^2
\epsilon_2 c_{\beta}^2 &-(m_{A^0}^2+m_Z^2) s_{\beta} c_{\beta} + 4 v^2 \epsilon_1
 \\
-(m_{A^0}^2+m_Z^2) s_{\beta} c_{\beta} + 4 v^2 \epsilon_1 & m_{A^0}^2 s_{\beta}^2 + m_Z^2 c_{\beta}^2 + 4 v^2 \epsilon_1 s_{2\beta} + 4 v^2
\epsilon_2 s_{\beta}^2
\end{pmatrix}
\label{evenM}
\ee
 in terms of $s_\beta = \sin\beta$, $c_\beta=\cos\beta$, and in the basis $(H_{Ru}^0,H_{Rd}^0)$
 for the real parts of the fields.
This yields,
to first order in $\epsilon_1$ and $\epsilon_2$:
\be  \label{higgsshift}
m_{h^0,H^0}^2 =  \frac{1}{2} \left(m_{A^0}^2 + m_Z^2 \mp \sqrt{(m_{A^0}^2-m_Z^2)^2 + 4m_{A^0}^2m_Z^2 \sin^2{(2\beta)}} \right) + \delta m_{h^0,H^0}^2
\ee
where
\be  \label{higgscorr}
 \delta m_{h^0}^2 &=& 2 v^2
 \Bigg(
 \epsilon_{2}
 +2\epsilon_{1}\sin{(2\beta)}  \\
&& \hspace{3cm}+\frac{2 \epsilon_{1} (m_{A^0}^2+m_Z^2) \sin{(2\beta)} - \epsilon_{2} (m_{A^0}^2-m_Z^2) \cos^2{(2\beta)} }{\sqrt{(m_{A^0}^2-m_Z^2)^2 + 4m_{A^0}^2m_Z^2 \sin^2{(2\beta)}}}
 \Bigg)   \nonumber  \\
 \delta m_{H^0}^2 &=& 2 v^2    \label{higgscorr2}
 \Bigg(
 \epsilon_{2}
 +2\epsilon_{1}\sin{(2\beta)}   \\
 && \hspace{3cm}
 -\frac{2 \epsilon_{1} (m_{A^0}^2+m_Z^2) \sin{(2\beta)} - \epsilon_{2} (m_{A^0}^2-m_Z^2) \cos^2{(2\beta)} }{\sqrt{(m_{A^0}^2-m_Z^2)^2 + 4m_{A^0}^2m_Z^2 \sin^2{(2\beta)}}}
 \Bigg) \, .  \nonumber 
\ee
The charged Higgs fields are now given masses
\be  \label{chargedcorr}
m_{H^{\pm}}^2=m_{A^0}^2+m_W^2+2v^2\epsilon_2 \; .
\ee
For an illustration of eq.\ (\ref{higgscorr}), see fig.\ \ref{fig:higgs} below. 

Since we want to compare BMSSM models with the corresponding
MSSM models for $\epsilon_1=\epsilon_2=0$, we 
have to make a choice how to treat the parameters $\tan \beta$
and $m_{A^0}$. We will treat them the same way as we treated $v$, i.e.\ 
we assign them the same value 
as we would have in the corresponding $\epsilon_1=\epsilon_2=0$ MSSM model. 
One 
reason this is useful is the following. We know that 
the partial width of
$A^0 \rightarrow q \bar{q}$, which is in principle measurable,
is given directly in terms of $\tan \beta$. Thus, keeping it fixed  will prove convenient when we study the $A^0$ resonance
region (see e.g.\ fig.\ \ref{fig:slice1} below), because the width will not change 
appreciably when we turn on $\epsilon_{1,2}$.
In other words, the BMSSM model is  physically very similar to the corresponding MSSM model
in this particular respect,
which would not have been the case had we let $\tan \beta$ vary.
Similarly, the pole mass of $m_{A^0}$ is a physical parameter
that we would like to keep fixed. 
The shift in (\ref{A0paramshifted}) can then be viewed as a finite renormalization,
that we  absorb in the definition of the pole mass, which therefore stays the same.
The value for $m_{A^0}$ we input in (\ref{higgsshift})
is therefore the same before as after turning on $\epsilon_{1,2}$,
and as a check the $A^0$ resonance does not move (see fig.\ \ref{fig:slice1} below).

For the rotation angles we find $\beta^0 = \beta^\pm = \beta$ even with $\epsilon_{1,2}\neq 0$.
That is, the functional form of $\beta^0$ and $\beta^\pm$ 
as functions of $\beta$ does not change, although $\beta$ changes (which we reabsorb
in the free parameter $\tan \beta$, as noted above). On the other hand, the angle $\alpha$ 
receives a correction, which 
to first order in $\epsilon_{1,2}$
can  be expressed as 
\be  \label{alphashift}
 \delta{(\sin{(2\alpha)})} = \frac{4v^2 \cos^2(2\beta) \left( 2\epsilon_1(m_{A^0}^2-m_Z^2)^2 + \epsilon_2 (m_{A^0}^4-m_Z^4) \sin{2\beta} \right)}{\left((m_{A^0}^2-m_Z^2)^2 + 4m_{A^0}^2m_Z^2 \sin^2{(2\beta)}\right)^{3/2}}
\ee
or, which is sometimes more convenient,
\be  \label{alphashift2}
 \delta{(\cos{(2\alpha)})} = \frac{4v^2 \sin(2\beta)  \cos^2(2\beta)\left( 2\epsilon_1(m_{A^0}^2-m_Z^2)^2 + \epsilon_2 (m_{A^0}^4-m_Z^4) \sin{2\beta} \right)}{\left((m_{A^0}^2-m_Z^2)^2 + 4m_{A^0}^2m_Z^2 \sin^2{(2\beta)}\right)^{3/2}} \; .
\ee

\subsection{Effects on the Higgs mass}
\label{sec:effhiggs}
\begin{figure}[h]
\begin{center}
\includegraphics[width=7.8cm]{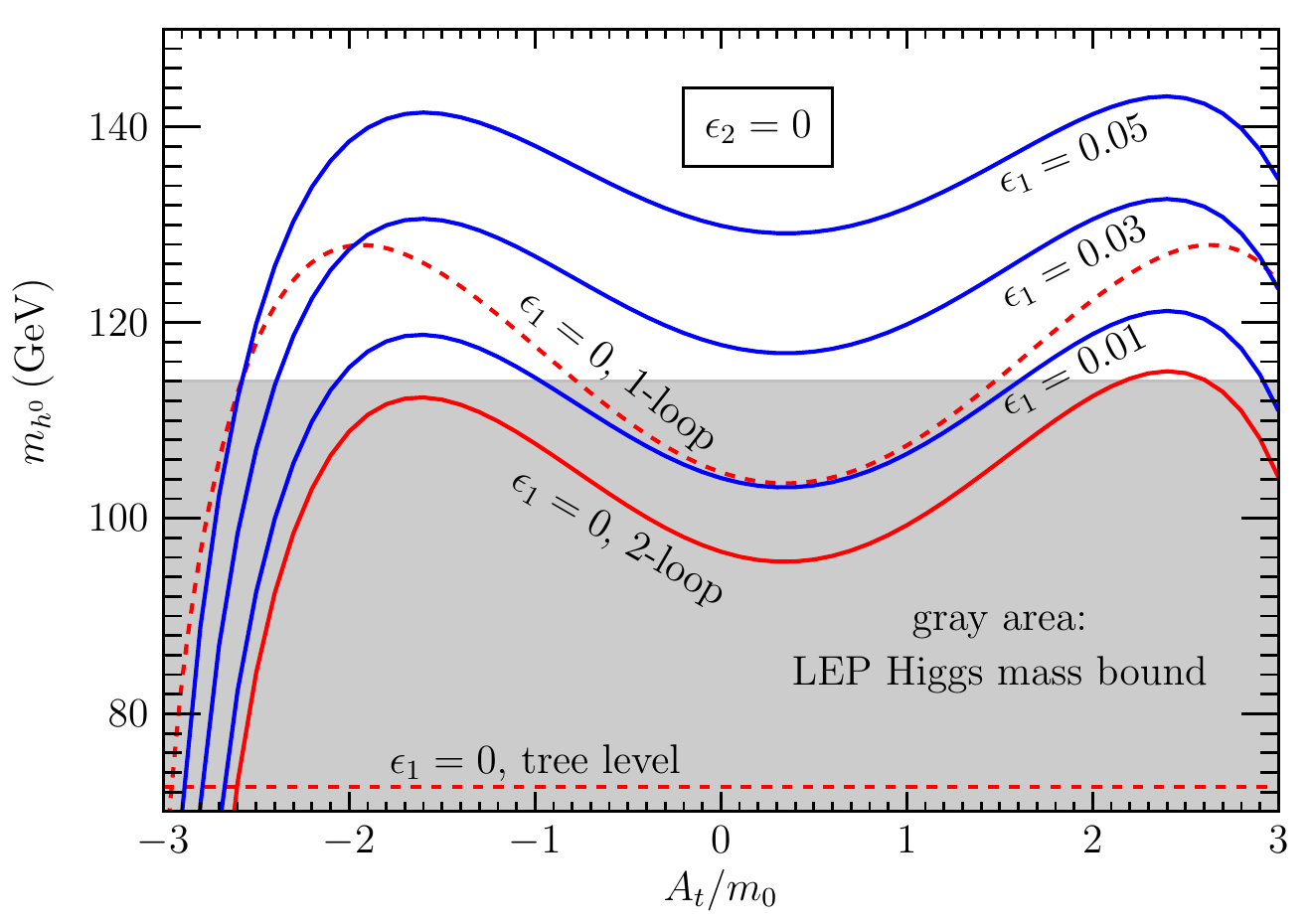}
\includegraphics[width=7.8cm]{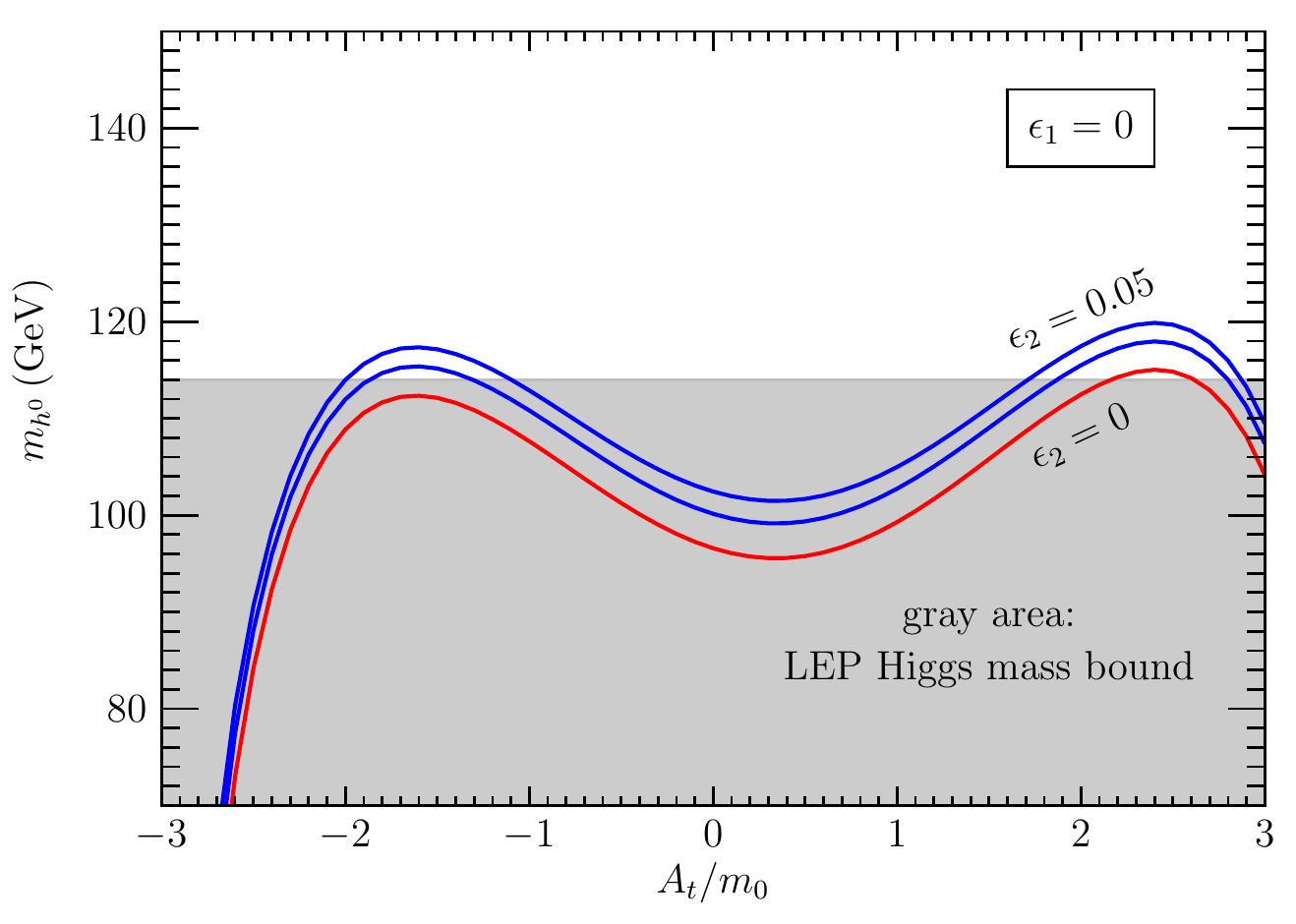}
\end{center}
\vspace{-8mm}
\caption{Higgs mass. The solid curves all include two-loop corrections.
Notice the MSSM lightest Higgs mass increases with 1-loop corrections,
but decreases somewhat when 2-loop corrections are taken into account. 
For this figure we used $M_2 = 500$ GeV,
$\mu=700$ GeV, $\tan \beta=3$, $m_{A^0}=500$ GeV, $m_0=700$ GeV and
$A_b=A_t$. The gray area is the rough estimate $m_{h^0} \leq $ 114 GeV for
the region ruled out by the LEP Higgs bound,
though we remind the reader that we do not use this bound literally
(see section \ref{computational}). 
}
\label{fig:higgs}
\end{figure}

One important question is the size of $\delta m_{h^0}$
since as previously mentioned, this affects the  Higgs little hierarchy problem.
In figure \ref{fig:higgs} we show an example of how $m_{h^0}$ varies with 
$\epsilon_1$ and $\epsilon_2$, and also how the Higgs mass in the ordinary MSSM  depends on loop order. The calculation is performed using {\tt FeynHiggs}
\cite{Heinemeyer:1998yj,Heinemeyer:1998np,Degrassi:2002fi,Frank:2006yh}
(see section \ref{computational} for further details).

To see why a small quartic coupling $\lambda$ makes the mass
shift $\delta m_{h^0}$ large even for fairly small $\epsilon_{1,2}$, let us consider the toy potential $V_{\rm toy}$ of
 (\ref{toypotential}) again.
 If we shift this to
\be
\hspace{3cm} V_{\rm toy, \, shift} = -\mu^2 \phi^2 + \lambda' \phi^4 \; ,  \qquad  \lambda ' = \lambda + \epsilon
\ee
 we  find  $v^2 = \mu^2/(2\lambda')$ (compare this to eq.\ (\ref{vshifted})) and $m^2 = V''(v) = 
4\mu^2$. Since we want to keep $v$ fixed by experimental input,
the toy mass $m^2 = V''(v)$ will shift:
\be
\delta m^2 =  8 \epsilon v^2 \; . 
\ee
Hence the relative change in $m$ is
\be  \label{toyexp}
{\delta m \over m} = { \epsilon \over 2 \lambda}
 + {\mathcal O}\!\left(\!\left({\epsilon \over \lambda}\right)^{\!2}\right) \; . 
\ee
Thus, 
for the estimate $\lambda \sim 0.07$ from (\ref{lambda}),
the relative shift in the toy mass $m$ is of order 50\% 
for $\epsilon \sim 0.07$. (For the actual
BMSSM Higgs mass shifts above, the
corresponding value is slightly higher than 0.07.)\footnote{Pedantically speaking, the actual expansion in (\ref{toyexp}) of course breaks down when $\epsilon \sim \lambda$.
We just want to illustrate that there are large effects when $\epsilon \sim \lambda$.}

In \cite{Antoniadis:2008ur}, it was argued that to avoid ``perturbations"
as big as 50\%, we should in principle impose $\epsilon \ll 0.07$.
However, how strictly this should be imposed depends on the point of view. We argued
that the Higgs mass is unnaturally small due to the smallness of $\lambda$
in the MSSM,
so the Higgs mass could receive relatively
large shifts as we ``naturalize" the unnaturally small coupling $\lambda$
by small additions in $\epsilon_{1,2}$. In practice,
we will allow for shifts up to order 20\%-30\% as
displayed in figure \ref{fig:higgs}. This
restricts $\epsilon_{1}$ to $\lesssim 0.05$. 

\subsection{Bounds on $\epsilon_1$ and $\epsilon_2$}
\label{bounds}

As mentioned above, one could argue that $\epsilon_{1,2}$ too large will cause
unreasonably large mass shifts in $m_{h^0}$ and hence be inconsistent. We will restrict
attention to $m_{h^0}$ mass shifts of at most  20\%-30\%, which gives
\be \label{eps1bound}
|\epsilon_1| \lesssim 0.05 . 
\ee
The $\epsilon_2$ correction does affect the Higgs mass less than $\epsilon_1$, 
so in principle one could consider turning on a larger value than (\ref{eps1bound})
for $\epsilon_2$ as far as the Higgs mass shift is concerned.
However, as it turns out $\epsilon_2$ does not impact our calculations very much, so
 we will simply turn on $\epsilon_2$
conservatively
as in (\ref{eps1bound}). 

Next in the list of restrictions on $\epsilon_{1,2}$, we have the following estimate \cite{Blum:2009na}
on precision electroweak bounds on the $S$ and $T$ variables \cite{Amsler:2008zzb}
expressed as a bound on the scale of new physics $M$:
\be  \label{M_ST}
M \stackrel{?}{>} 8 \mbox{ TeV} \; .
\ee
If we 
would apply this bound to estimate allowed values of $\epsilon_1$, then for 
a perturbative microscopic theory (see appendix \ref{eft})
 the region of low
$\mu$ would only support very small values
$\epsilon_1\lesssim 0.01$ (see eq.\ (\ref{epsilon1})). 
However, the tension with the bounds 
on the precision electroweak observables $S$ and $T$ comes from effective
dimension six (i.e.\ $1/M^2$) terms in the K\"ahler potential
that provide new interactions between Higgs fields and gauge fields. 
These new interactions come with coefficients called $\xi_i$ in \cite{Dine:2007xi} 
that are independent
of $\epsilon_{1,2}$ in the effective theory, and we have set $\xi_i=0$,
as stated in section \ref{sec:eff}. Since our software tools do 
perform checks on precision electroweak observables
(see section \ref{computational}), we will assume 
that (\ref{M_ST}) does not need to be strictly applied to our BMSSM models. 
See also section \ref{sec:outlook}.

When modifying the Higgs potential, one should check vacuum stability.
A priori one might think that small corrections can never affect stability. However,
the usual derivation of 
the condition that the potential be bounded from below is in a $D$-flat direction, 
and the MSSM quartic Higgs term
is zero in this direction (it is a $D$-term), so in principle $\epsilon_{1,2}$ could affect boundedness
from below.  
Nevertheless, for $\epsilon_{1,2}$ sufficiently small and $\epsilon_1$ nonzero, the potential is still bounded from below
since (\ref{V1}) is supersymmetric ($V_F$ 
in eq.\ (\ref{crossterm}) is an absolute value squared). The
issue then becomes whether to see this, we need to keep the effective dimension six operator in (\ref{crossterm}).
For the purposes of this paper, we will not investigate this further.\footnote{Note that 
 if we prefer, we
can always ensure that the quartic term is positive by
turning on a sufficiently large positive $\epsilon_2$, although
we did not do so explicitly. We believe our results will remain mostly unchanged if we did.}

For models with large Higgs mass shifts, i.e.\  large $\epsilon_{1,2}$
beyond (\ref{epsbounds}) below, there is then 
the related issue of whether the $\epsilon_{1,2}$
corrections could deform the potential so severely that a new deeper global 
minimum develops, and if so  how long it will take it to
tunnel to this new true vacuum.
Vaccuum stability under such conditions was checked recently by Blum, Delaunay and Hochberg (BDH)
 in \cite{Blum:2009na}, who found a criterion to exclude such transitions in the BMSSM
 even for large Higgs mass shifts.
Since for our purposes we do not particularly
focus on obtaining large Higgs mass shifts, we will not impose the BDH criterion,
though again it would be interesting to study this issue in more detail. 

To summarize, we will require the following reasonably conservative bounds on 
the BMSSM model parameters $\epsilon_1$
and $\epsilon_2$:
\be \label{epsbounds}
-0.05 \leq  \epsilon_1  \leq 0.05 \; , \qquad -0.05 \leq \epsilon_2  \leq 0.05 \; . 
\ee

\subsection{BMSSM neutralino masses and mixings}
\label{sec:neumass}
In the basis $\wt\psi^0 = (\wt B, \wt W, \wt H_d^0, \wt H_u^0)^T$, the neutralino mass part of the Lagrangian is given by
\begin{align}
 \lag \ni - \frac{1}{2} (\wt\psi^0)^T \cal M_{\wt \chi^0} \wt\psi^0 +  \hc
\end{align}
where
%\be
% \cal M_{\wt \chi^0} =
% \begin{pmatrix}
%  M_1 & 0 & -\frac{g'v_d}{\sqrt{2}} & \frac{g'v_u}{\sqrt{2}} \\
%  0 & M_2 & \frac{gv_d}{\sqrt{2}} & -\frac{gv_u}{\sqrt{2}} \\
%  -\frac{g'v_d}{\sqrt{2}} & \frac{gv_d}{\sqrt{2}} & 0 & -\mu \\
%  \frac{g'v_u}{\sqrt{2}} & -\frac{gv_u}{\sqrt{2}} & -\mu & 0
% \end{pmatrix}
%\ee
%or
 in terms of $s_\beta = \sin\beta$, $c_\beta=\cos\beta$ and the usual Standard Model parameters $m_Z$, $s_{W}=\sin{\theta_W}$ and $c_{W}=\cos{\theta_W}$ we have
\be
 \cal M_{\wt \chi^0} =  \widehat{\cal M}_{\wt \chi^0} +  \delta \cal M_{\wt\chi^0}
\ee
where at tree-level
\be   \label{neumass_tree}
\widehat{ \cal M}_{\wt \chi^0} =
 \begin{pmatrix}
  M_1 & 0 & -m_Z s_{W} c_{\beta} & m_Z s_{W} s_{\beta} \\
  0 & M_2 & m_Z c_{W} c_{\beta} & -m_Z c_{W} s_{\beta} \\
  -m_Z s_{W} c_{\beta} & m_Z c_{W} c_{\beta} & 0 & -\mu \\
  m_Z s_{W} s_{\beta} & -m_Z c_{W} s_{\beta} & -\mu & 0
 \end{pmatrix} \; 
\ee
(i.e.\ the same functional form as in the MSSM)
and the new couplings introduced above introduce the following corrections to the neutralino mass matrix
\be  \label{neucorr}
 \delta \cal M_{\wt\chi^0} = - \frac{2\epsilon_1}{\mu^*}
 \begin{pmatrix}
  0 & 0 & 0 & 0 \\
  0 & 0 & 0 & 0 \\
  0 & 0 & v_u^2 & 2 v_u v_d \\
  0 & 0 & 2 v_u v_d & v_d^2
 \end{pmatrix}
 =
 - \frac{2\epsilon_1}{\mu^*} v^2
 \begin{pmatrix}
  0 & 0 & 0 & 0 \\
  0 & 0 & 0 & 0 \\
  0 & 0 & \sin^2{\beta} & \sin{2\beta} \\
  0 & 0 & \sin{2\beta} & \cos^2{\beta}
 \end{pmatrix}.
\ee
The mass matrix is diagonalized using a unitary matrix $N$, which is such that
\be
 N^* \cal M_{\wt \chi^0} N^\dagger = \text{diag}(m_{\wt \chi_1^0}, m_{\wt \chi_2^0}, m_{\wt \chi_3^0}, m_{\wt \chi_4^0})
\ee
is diagonal and ordered according to $m_{\wt \chi_1^0} \leq m_{\wt \chi_2^0} \leq m_{\wt \chi_3^0} \leq m_{\wt \chi_4^0}$. The mass eigenstates are given by
\be
	\wt \chi^0_i = N_{ij} \wt \psi^0_j, \quad \wt  \psi^0_i=N^\dagger_{ij} \wt \chi^0_j = N^*_{ji} \wt \chi^0_j.
\ee
In the limit where the electroweak symmetry breaking terms can be considered to be small, i.e.\ $\mu$ or $M_1, M_2 \gg m_Z$, and for real parameters,
 the eigenvalues of the neutralino mass matrix are given by
 (see e.g.\ \cite{Gunion:1987yh})
 \begin{align}
 m_{N1} & = M_1 + \frac{M_Z^2 \sin^2{\theta_W}(M_1+\mu \sin{2\beta})}{M_1^2-\mu^2} + \ldots \label{neumass1} \\
 m_{N2} & = M_2 + \frac{M_Z^2 \cos^2{\theta_W}(M_2+\mu \sin{2\beta})}{M_2^2-\mu^2} + \ldots \label{neumass2} \\
 m_{N3} & = |\mu| 
 - \frac{\epsilon_1}{|\mu|} v^2 (1 - 2\sin{2\beta}) 
 - \text{sign}(\mu) \frac{m_Z^2(1 + \sin{2\beta})(M_1 \cos^2{\theta_W} + M_2\sin^2{\theta_W} - \mu)}{2(M_1 - \mu)(M_2 - \mu)} + \ldots   \label{neumass3} \\
 m_{N4} & = |\mu|
 + \frac{\epsilon_1}{|\mu|} v^2 (1 + 2\sin{2\beta})
 + \text{sign}(\mu) \frac{m_Z^2(1 - \sin{2\beta})(M_1 \cos^2{\theta_W} + M_2\sin^2{\theta_W} + \mu)}{2(M_1 + \mu)(M_2 + \mu)} + \ldots   \label{neumass4}
\end{align}
where we have introduced the non-mass-ordered 
neutralinos $N1$ through $N4$. Note that even though
the ordering can change due to the $\epsilon_1$ corrections,
this  is not a problem for the consistency of the theory; the mass corrections due
to $\epsilon_1$ themselves remain reasonably small
for reasonably large $\mu$, as we show in
an example in figure \ref{fig:massdiff} below.

We include the most significant MSSM loop corrections 
to the neutralino mass matrix \cite{Drees:1996pk,Pierce:1993gj,Pierce:1994ew,Lahanas:1993ib}, as implemented
in {\tt DarkSUSY} (see section \ref{computational}). Although these loop corrections 
are small (on the order of a few GeV), the corrections 
in the $(3,3)$ and $(4,4)$ elements of the neutralino 
mass matrix can be important, since those elements are zero in the MSSM
at tree-level. The BMSSM corrections  also contribute to
those elements of the neutralino mass matrix (see 
eq.\ (\ref{neucorr})), so  it is conceptually important 
to include loop corrections in the BMSSM as well, since otherwise the effect of $\epsilon_{1,2}$ would 
appear greater than it really is. 
The loop corrections are themselves
in principle affected by the BMSSM operators,
but this effect is higher order and is neglected here.

\subsection{BMSSM chargino masses and mixings}
\label{sec:chargino}
\begin{figure}[h]
\begin{center}
\includegraphics[width=9cm]{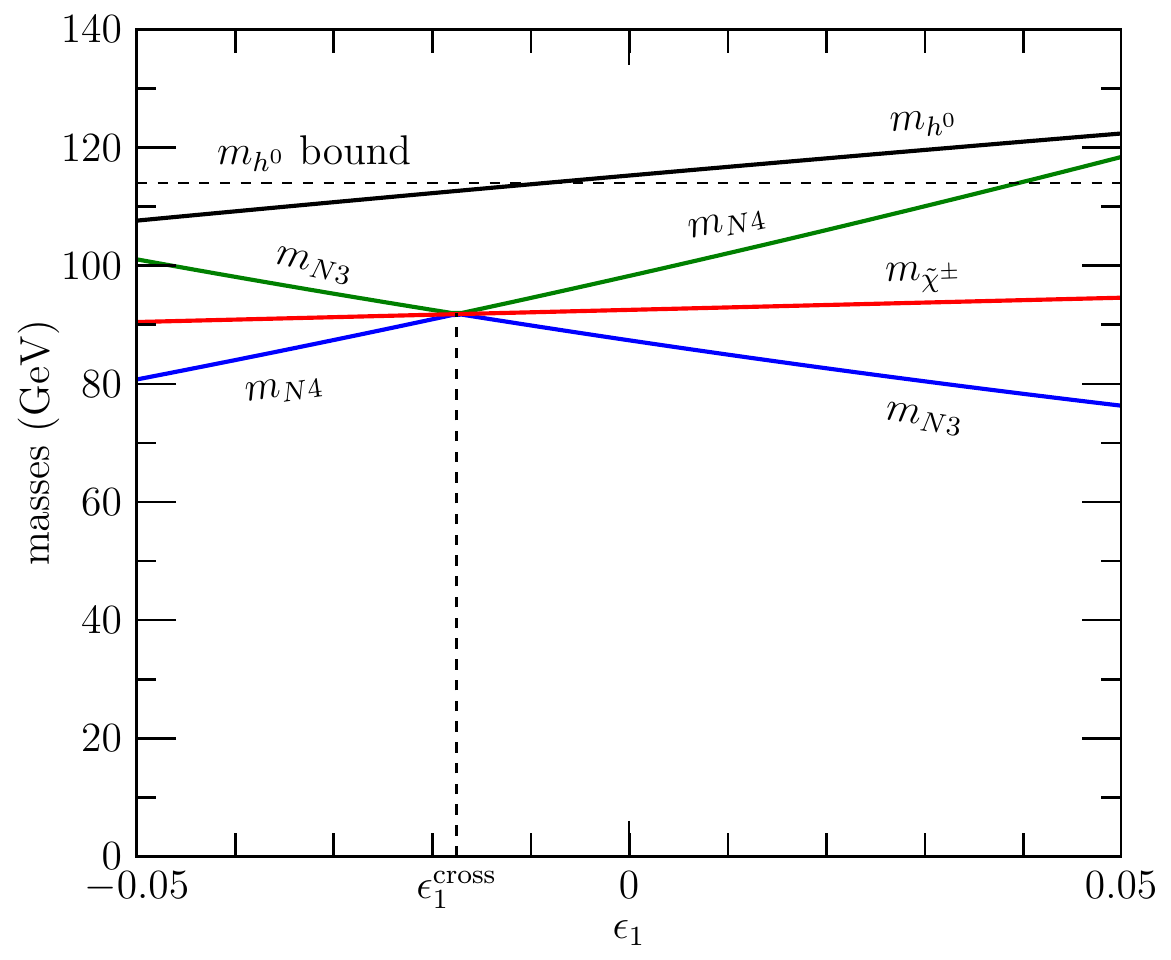}
\caption{Masses and mass crossing for the StHelena${}^{(+)}$ model
(see tables \ref{bench} and \ref{benchspec} below) for $\epsilon_2=0$ and varying $\epsilon_1$. 
The rough estimate $m_{h^0} \geq $ 114 GeV of the LEP Higgs bound  is shown,
though we remind the reader that we do not use this bound literally
(see section \ref{computational}). 
The point $\epsilon_1=\epsilon_1^{\rm cross}$ marks
the crossing point, which is given in eq.\ (\ref{epscross}).  With the rough estimate
of the LEP Higgs bound, the point $\epsilon_1^{\rm cross}$ is in the excluded region
in this example. 
For further discussion, see the main text.}
\label{fig:massdiff}
\end{center}
\end{figure}

In the basis $\wt \psi_\pm= (\wt\psi^+,\wt\psi^-)$ where $\wt\psi^+ = (\wt W^+, \wt H_u^+)^T, \wt\psi^- = (\wt W^-, \wt H_d^-)^T$, the chargino mass terms in the Lagrangian are given by
\begin{align}
 \lag & \ni -\frac{1}{2} (\wt\psi_\pm)^T \cal M_{\wt\chi^+} \wt\psi_{\pm} + \hc \nonumber \\
 & = -\frac{1}{2} (\wt\psi_+)^T X\wt \psi_- -\frac{1}{2} (\wt\psi_-)^T X^T \wt\psi_+ + \hc
\end{align}
where the mass matrix $\cal M_{\wt\chi^+}$ takes the following block-diagonal form
\be
 \cal M_{\wt\chi^+} =
 \mat{0 & X^T \\ X & 0}
\ee
with
\be
X = \widehat{X} + \delta X
\ee
where at tree level
\be  \label{MSSMX}
\widehat{ X} = \begin{pmatrix} M_2 & gv_u \\ gv_d & \mu \end{pmatrix} = \begin{pmatrix} M_2 & \sqrt{2} M_W \sin{\beta} \\ \sqrt{2} M_W \cos{\beta} & \mu \end{pmatrix}
\ee
(i.e.\ the same functional form as in the MSSM)
and the dimension-five operators introduced 
in eq.\ (\ref{expllag}) give the following correction to the chargino mass matrix:
\be  \label{chacorr}
 \delta X = \frac{\epsilon_1}{\mu^*}v^2\sin{2\beta} \mat{0 & 0 \\ 0 & 1}.
\ee
The matrices $X$ and $X^T$ are diagonalized using two unitary matrices $U$ and $V$ such that
\be
 U^* X V^\dagger = \text{diag} (m_{\wt\chi_1},m_{\wt\chi_2}), \quad V^* X^T U^\dagger = \text{diag} (m_{\wt\chi_1},m_{\wt\chi_2})
\ee
with the chargino masses on the diagonal, and where the mass eigenstates are given by
\begin{align}
 \wt \chi^+_i & = V_{ij} \wt\psi^+_j, \quad \wt\psi^+_i = V^\dagger_{ij} \wt \chi^+_j = V^*_{ji} \wt \chi^+_j \\
 \wt \chi^-_i & = U_{ij} \psi^-_j, \quad \wt\psi^-_i = U^\dagger_{ij}\wt  \chi^-_j = U^*_{ji} \wt \chi^-_j.
\end{align}
For the standard MSSM case with $\epsilon_{1,2}=0$ one finds \cite{Martin:1997ns}
\begin{align}  \label{masscha0}
 m^2_{\wt\chi^+_{1,2}} & = \frac{1}{2} \Big( |M_2|^2 + |\mu|^2 + 2 m_W^2 \nonumber \\
 & \mp \sqrt{( |M_2|^2 + |\mu|^2 + 2 m_W^2)^2 -4 |\mu M_2 - m_W^2 \sin{2\beta}|^2} \Big)
\end{align}
For future reference, we define $\tilde{\chi}^{\pm}$ (without the index $1,2$) to be the 
 {\it lightest} chargino. 
Now, we see from eqs.\ (\ref{MSSMX}) and (\ref{chacorr}) that the corrected chargino masses are
simply obtained by replacing $\mu$  in (\ref{masscha0}) with 
\be 
\mu' = \mu + \frac{\epsilon_1}{\mu^*}v^2\sin{2\beta} \; .
\ee
(Note that we do not use this $\mu'$ anywhere else.)
In the case of small real $\mu$ we have
\be
 m_{\wt\chi^{\pm}_2} = |\mu| \left(1 + \frac{\epsilon_1}{\mu^2} v^2 \sin{2\beta} + \ldots \right)
\ee
 In particular, we see that the chargino-neutralino mass splitting
 receives an $\epsilon_1$-dependent correction. We plot
 an example in fig.\ \ref{fig:massdiff}. 
  In the higgsino region, we can write a simple expression for the
  neutralino mass crossing $\epsilon_1^{\rm cross}$ (which incidentally does not
  need to be at the same place as the neutralino-chargino mass crossing,
  despite appearances for our benchmark model in fig.\ \ref{fig:massdiff}):
   \be  \label{epscross}
\epsilon_{1}^\text{cross} &=& \text{sign}(\mu) \frac{m_Z^2 |\mu|  }{2v^2(M_1^2 - \mu^2)(M_2^2 - \mu^2)}\times  \\
&&\hspace{-1cm} \Big( -\mu \sin{2\beta} (s_W^2 M_2^2 + c_W^2 M_1^2 - \mu^2)+\mu^2 (s_W^2 M_1 + c_W^2 M_2) - M_1 M_2 (s_W^2 M_2 + c_W^2 M_1) \Big) \; . 
 \nonumber 
 \ee
Then the neutralino-chargino mass splitting is a simple linear function of $\epsilon_1$:
\be \label{massplit}
 \Delta m_{\pm} =  \left\{
 \begin{array}{ll}
 m_{\chi_{\pm}}-m_{N3} & \mbox{if } \epsilon_1\geq \epsilon_1^{\rm cross} \\
 m_{\chi_{\pm}}-m_{N4} & \mbox{if } \epsilon_1<\epsilon_1^{\rm cross}  \; . 
 \end{array} \right.
 \ee
 where $m_{N3}$ and $m_{N4}$
 are the expressions in (\ref{neumass3}) and (\ref{neumass4}).\footnote{In 
 \cite{Cheung:2009qk},
 it appears that 
  when $\epsilon_1$ increases,
 the Higgs mass increases while the chargino mass {\it decreases},
the opposite of figure \ref{fig:massdiff}. This is due
 to the use of two incompatible conventions for $\epsilon_1$ in their expressions.
 Since this probably affects the results for the relic density in that paper,
 we find it difficult to compare their results to ours. 
 See also section \ref{sthelena}. \label{footnote:cheung}}

\section{Computational tools} 
\label{computational}

For every BMSSM model parameter set 
 we consider,
we calculate the resulting par\-ticle spectrum and relic density using {\tt DarkSUSY} \cite{Gondolo:2004sc}, a publicly-available comprehensive numerical package for neutralino dark matter calculations.

Starting from the {\tt DarkSUSY-5.0.4} version, we have extended the code to also include the BMSSM corrections due to the effective terms (\ref{expllag}). The neutralino and chargino mixing matrices of eqs.\ (\ref{neucorr}) and (\ref{chacorr}), as well as the three-point vertex corrections  and the new four-point vertices of appendix \ref{feynmanrules}  have all been fully implemented into the code.\footnote{We intend to make this extension of {\tt DarkSUSY} public in the future,
though perhaps not in the {\it near} future.}

Within the {\tt DarkSUSY} package the Higgs boson spectrum is determined at two-loop level using the external {\tt FeynHiggs-2.6.4} program \cite{Heinemeyer:1998yj,Heinemeyer:1998np,Degrassi:2002fi,Frank:2006yh}. We have implemented the Higgs boson mass corrections of eqs.\  (\ref{higgscorr}), (\ref{higgscorr2}) and (\ref{chargedcorr}), 
the rotation angles (\ref{alphashift}), (\ref{alphashift2}),
and the neutralino and chargino mixing matrix corrections in eqs.\ (\ref{neucorr}) and (\ref{chacorr}) in  {\tt FeynHiggs-2.6.4}. 
We do not include additional radiative corrections due to the 
three-point vertex corrections or
the new four-point coup\-lings, but
we expect our results to be largely insensitive to this deficiency. 
In addition to the masses, we also extract the total and partial widths of the Higgs bosons from  {\tt FeynHiggs}.

It is also important in this context to check the LEP accelerator bounds
on the Higgs boson masses
 carefully.
In the BMSSM
(or any other theory with a
Higgs sector different from the Standard Model Higgs
boson --- including the MSSM) the 
Standard Model lower Higgs mass limit of   114.4 GeV 
\cite{Amsler:2008zzb}
strictly speaking does not apply; in principle, 
the bound should be determined  model by model  from experimental data.
One could
worry that {\it if} the bound would effectively be weakened when going from the MSSM to the BMSSM,
that would take away some of the motivation for including the corrections
(\ref{higgscorr}) in the first place. 

 For this reason we call {\tt HiggsBounds-1.0.3} \cite{Bechtle:2008jh} from {\tt DarkSUSY}. The  {\tt HiggsBounds} code takes actual Higgs boson production cross sections and partial widths as input, computes rates into the important LEP search channels, and determines whether or not the model is excluded by comparing the rates with existing LEP data.  From our parameter
  scans we note that the simple LEP bound $m_{h^0}\gtrsim 114$ GeV in general is respected to good accuracy within the BMSSM (including the MSSM), and the simple interpretation of the BMSSM corrections as providing new models by raising $m_{h^0}$ works well.\footnote{It is  useful to keep in mind that for special parameter sets, the LEP bound 114.4 GeV can be lowered significantly \cite{Amsler:2008zzb} --- this is a manifestation of the model-dependence of the bound we emphasized in the paragraph above.}
 
Finally, we note that {\tt DarkSUSY} imposes many additional accelerator constraints
on the models,
including bounds on sparticle masses, the rate of b $\rightarrow$ s$\gamma$
and the electroweak observable $\rho$.

\section{Relic density: strategy}
\label{strategy}

We explore the features of the BMSSM by scanning extensively  over the full
nine-dimen\-sional parameter space, \ie over the parameters $M_2$, $\mu$,
$\tan\beta$, $m_{A^0}$, $m_0$, $A_t$, $A_b$, $\epsilon_1$ and $\epsilon_2$.
(For a reminder of what these parameters are, see section \ref{MSSMrev}.)
The scans are
carried out with the help of {\tt DarkSUSY}  \cite{Gondolo:2004sc}, in which we have implemented the
BMSSM as described in section \ref{computational} above.

For each model we calculate quantities like mass spectra and relic density, including
coannihilations (see e.g.\ \cite{Edsjo:1997bg}). We check them against various accelerator constraints, including bounds on
sparticle masses, Higgs boson masses, the rate of b $\rightarrow$ s$\gamma$
and the electroweak observable $\rho$.

Of particular importance for the BMSSM are limits on the lightest Higgs boson and chargino.
To determine the former we use {\tt HiggsBounds} \cite{Bechtle:2008jh} (see section \ref{computational}), while for the chargino
bound we adopt
\begin{eqnarray} \label{chbound}
m_{\tilde{\chi}^{\pm}} > 94 \textnormal{ GeV},
\end{eqnarray}
which is the current standard lower mass limit \cite{Amsler:2008zzb}. 
(Recall from section \ref{sec:chargino} that we define $\tilde{\chi}^{\pm}$
without any $1,2$ subscript as the lightest chargino.)
In close analogy to the discussion of the Higgs mass bound in section \ref{computational} above, the chargino bound (\ref{chbound}) is somewhat model dependent, and should not be
thought of as applying with high precision for an {\it arbitrary} model. Depending on 
particular properties of a given model, the limit
may move several GeV in either direction \cite{Amsler:2008zzb}, and it is not simple to
implement an accurate bound even on a model by model basis.\footnote{By analogy
with the {\tt HiggsBounds} package, it would
be most useful if someone wrote a   {\tt CharginoBounds} package!}. For illustration purposes, we simply use a sharp 94
GeV chargino mass bound as in eq.\ (\ref{chbound}), with the understanding
that this is not very precise. 

Since the BMSSM is only valid below the scale of new physics,
which we take to be $M \sim 5-10 $ TeV (see section \ref{sec:eff}), we also impose a maximum cut on sparticle masses 
by requiring $|\mu|, |M_2|, |m_0| < 2 $ TeV. This cut could be 
raised if we raise $M$, but then our new parameters
$\epsilon_1$, $\epsilon_2$ will typically be rather small (see eqs.\ (\ref{epsilon1}),
(\ref{epsilon2})).

For the Lightest Supersymmetric Particle (LSP)  to be a viable dark matter candidate we require it to be a neutralino, and that it provides a relic density in agreement with the value measured by WMAP  \cite{Komatsu:2008hk}. To be precise,
we use the constraint
\begin{eqnarray}
\label{eq:oh2}
\Omega_{\chi} h^2 = 0.1099 \pm 2\cdot 0.0062, 
\end{eqnarray}
which is the $2\sigma$ result when
combining $\Lambda$CDM, Sunyaev-Zeldovich and lensing datasets \cite{WMAP}. Here $\Omega_{\chi}$ is the dark matter relic density as a fraction of the critical density, and
the dimensionless parameter $h$ is the Hubble constant in units of 100 km Mpc$^{-1}$s$^{-1}$.

In addition to our broad parameter scans, we also highlight regions of particular interest for the
BMSSM by investigating details of a number of focused scans and certain
benchmark points.

We will show most of our plots in terms of the physical parameters $m_{\rm LSP}$
and  the gaugino fraction $Z_{\rm g}$.
See the discussion in section \ref{mssmchi} for some
details
on the relation between $Z_{\rm g}$ and
the model parameters $M_2$ and $\mu$. 

\section{Relic density: results}
\label{sec:results}

\subsection{General parameter space scan}
\label{sec:generalscan}
We have scanned the MSSM parameter space between generous bounds allowing mass parameters up to several TeV, $A_t$ and $A_b$ between $-3m_0$ and $3m_0$ and $\tan \beta$ between 1 and 60. Our sample of MSSM models is larger and we have about $110\,000$ models that pass all constraints (including the WMAP constraint on $\Omega_{\chi} h^2$
of eq.\ (\ref{eq:oh2})). For the BMSSM, we have about $11\,000$ models 
that pass all constraints.
To be clear, the fact that we have ten times more MSSM models than BMSSM models does not mean that it was in any sense harder to find BMSSM models
than MSSM models (of course for $\epsilon_{1,2}$ infinitesimally small, the models tend to be indistinguishable),  it merely reflects the fact that
at the time of writing, we have performed more MSSM scans. 
\label{Africa}
\begin{figure}[h]
\begin{center}
\includegraphics[width=12cm]{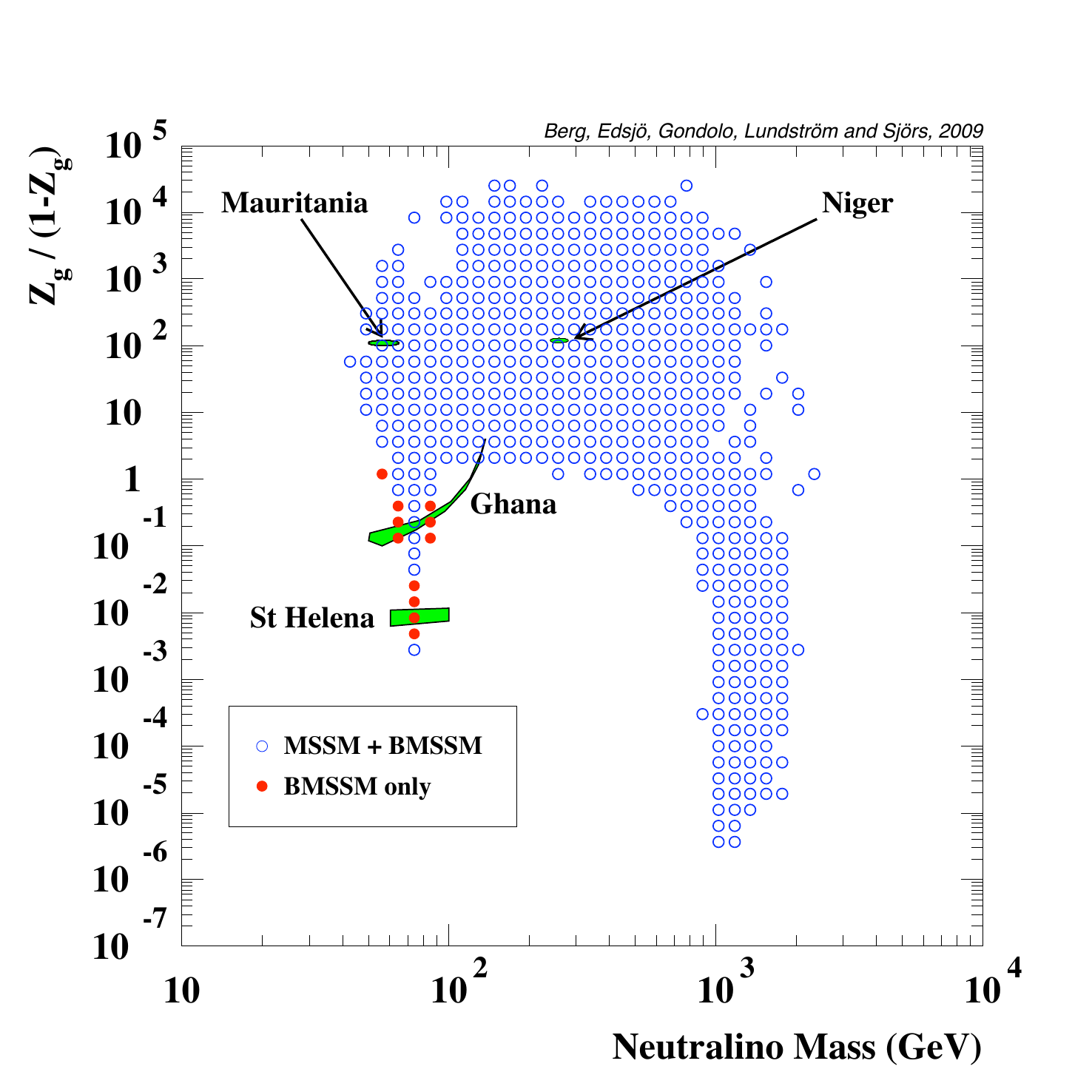}
\caption{MSSM and BMSSM models
that pass all accelerator constraints and furnish
the WMAP relic density. Shown in the figure are 
the locations of our benchmark scans of sections
\ref{sthelena} (St Helena), \ref{sec:saharat} (Niger), \ref{saharah} (Mauritania)
and \ref{nigeria} (Ghana). }
\label{fig:africa}
\end{center}
\end{figure}

In figure \ref{fig:africa} we show the results from this  scan projected onto the $(m_{\rm LSP},$
$Z_{\rm g}/(1-Z_{\rm g}))$ plane, where $m_{\rm LSP}$ is the mass of the lightest neutralino (our dark matter candidate)
and $Z_{\rm g}$ is the gaugino fraction of the LSP,
as defined in eq.\ (\ref{Zg}). 
\begin{figure}[h]
\begin{center}
\includegraphics[width=10cm]{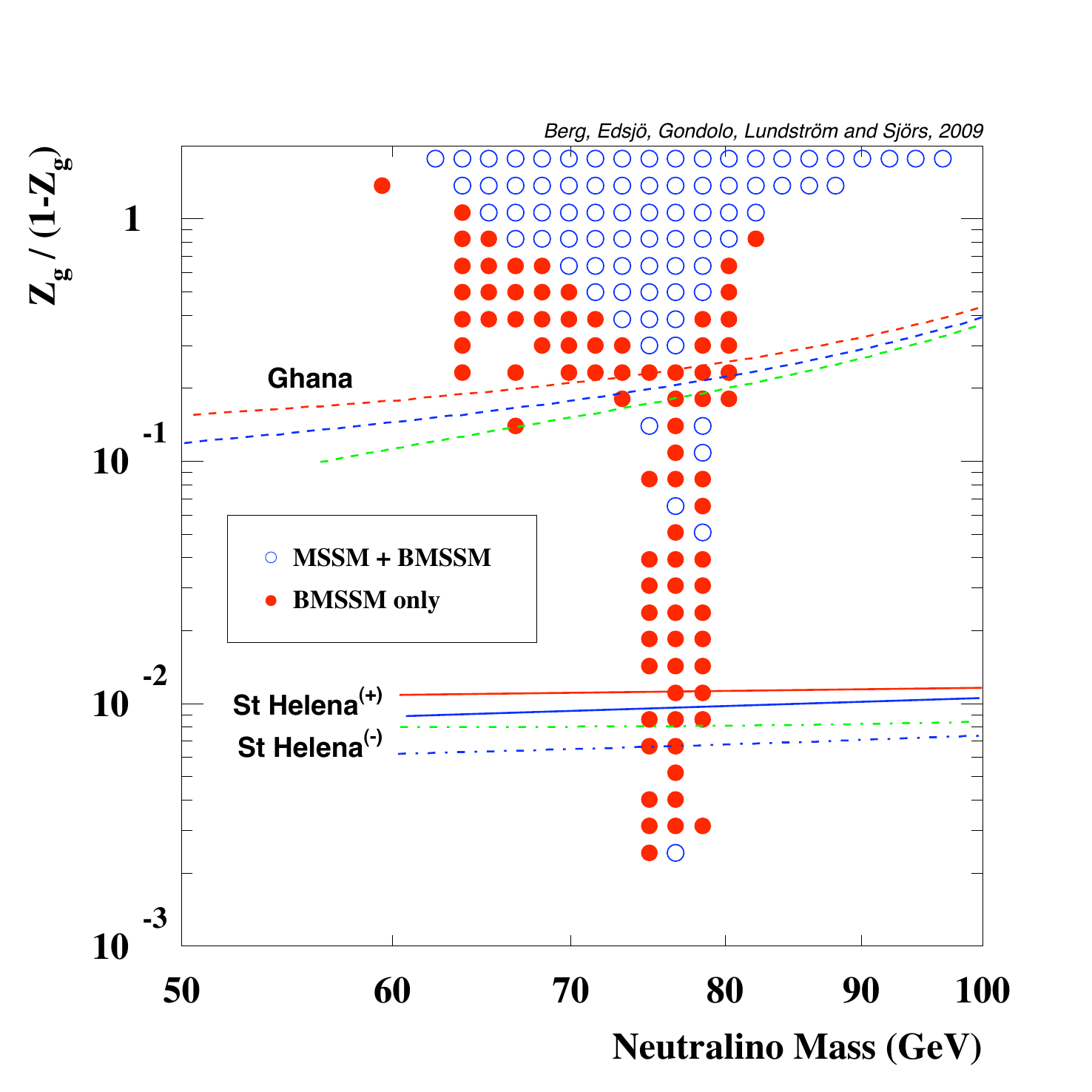}
\caption{Zoom-in of the bottom-left region of figure \ref{fig:africa}.
Shown in the figure are our benchmark scans of sections
\ref{sthelena} (St Helena) and \ref{nigeria} (Ghana). 
In each group of curves, the red curves are positive $\epsilon_1$,
green curves are negative $\epsilon_1$, and
the blue curves are the corresponding MSSM models ($\epsilon_{1,2}=0$). }
\label{fig:zoom}
\end{center}
\end{figure}
\begin{figure}[h]
\includegraphics[width=8cm]{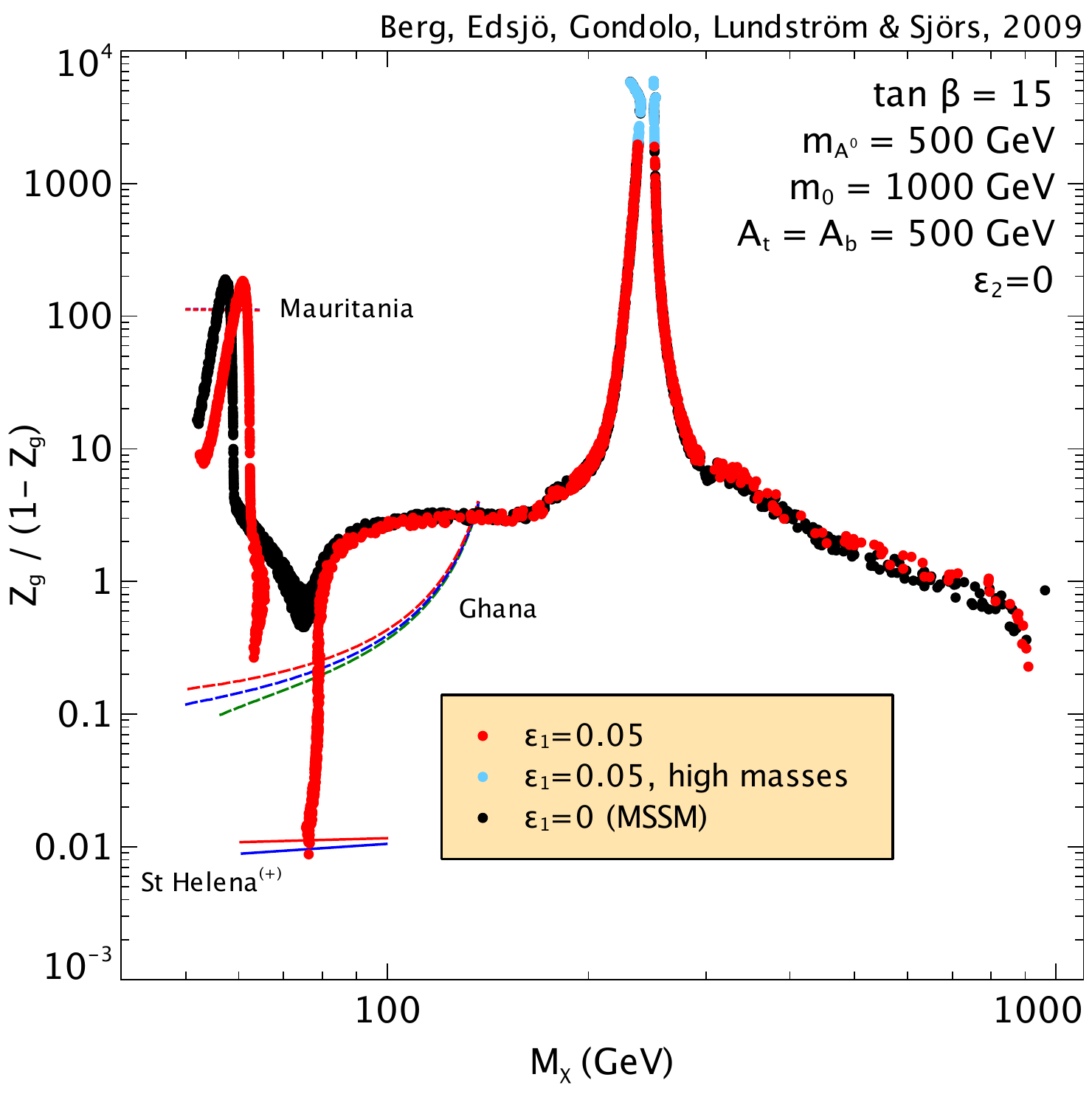}
\includegraphics[width=8cm]{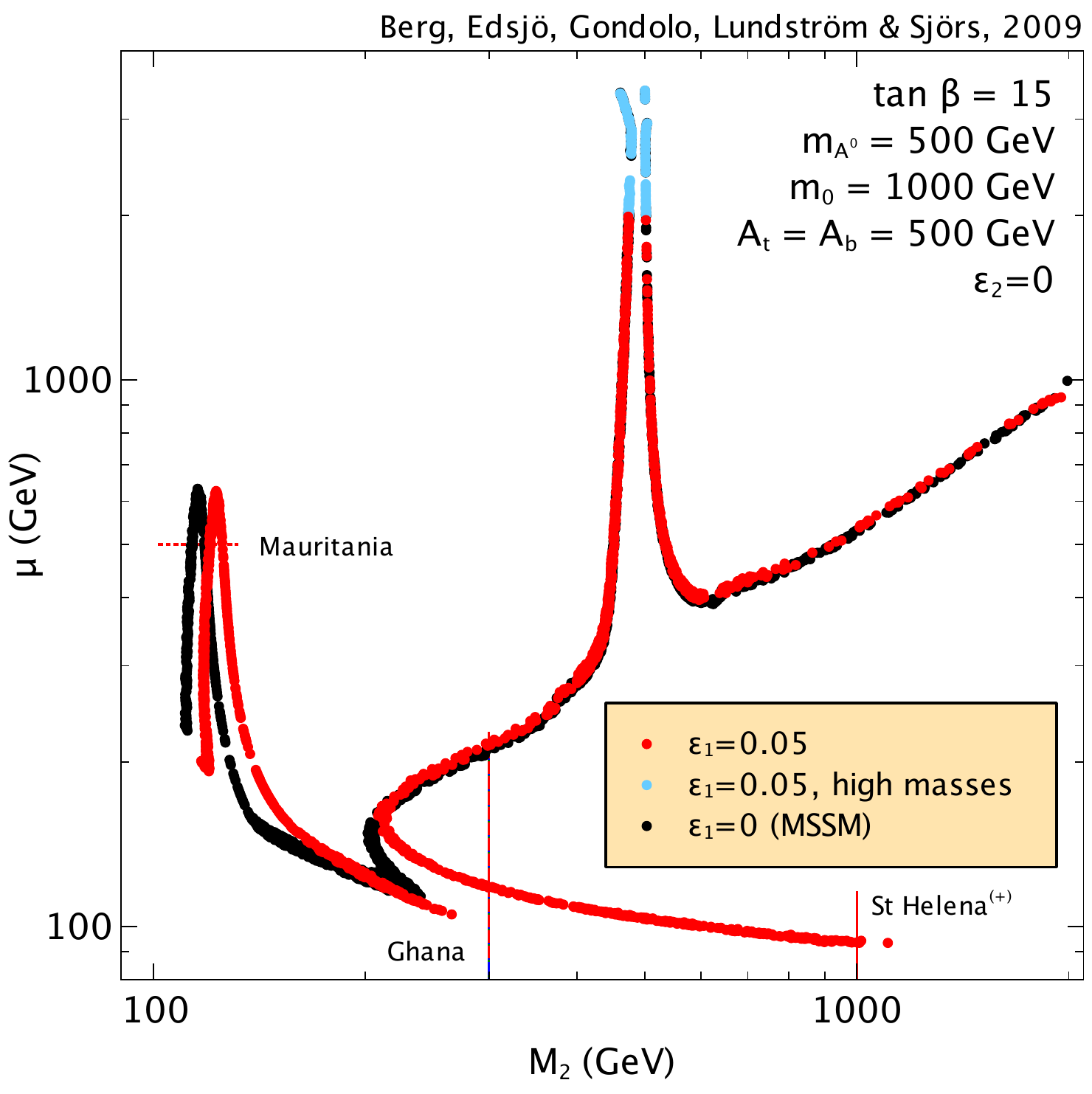}
\caption{Slices through parameter space, in the $(m_{\rm LSP},Z_{\rm g}/(1-Z_{\rm g}))$ plane and  in the   $(M_2,\mu)$  plane. 
The red points are $\epsilon_1=0.05$, and the black points
are $\epsilon_1=0$. Here $\epsilon_2=0$.
The blue points are $\epsilon_1=0.05$, but with $|\mu|>2$ TeV.
Shown in the figure are 
the locations of our benchmark scans of sections
\ref{sthelena} (St Helena${}^{(+)}$),  \ref{saharah} (Mauritania)
and \ref{nigeria} (Ghana).
In each group of benchmark scans, the red curves are positive $\epsilon_1$,
green curves are negative $\epsilon_1$, and
the black curves are the corresponding MSSM models ($\epsilon_{1,2}=0$).
}
\label{fig:slice1}
\vspace{-4mm}
\end{figure}
The blue circles and red dots in the figure
both represent regions in the $(m_{\rm LSP},
Z_{\rm g}/(1-Z_{\rm g}))$ plane where we find models consistent with all imposed
accelerator constraints
and that provide the correct relic abundance of eq.\ (\ref{eq:oh2}). The red dots 
correspond to regions where we only find models with
$\epsilon_1\neq 0$ and/or $\epsilon_2\neq 0$, i.e. regions which the ordinary MSSM
is unable to reach. Since the set of all MSSM models in figure \ref{fig:africa}
bears a vague resemblance to a  map of Africa,
we label various interesting regions by African countries
(and one island) in the appropriate locations.

The only feature of the BMSSM 
that is immediately obvious in figure \ref{fig:africa} is the existence of completely new models with
light ($m_{\rm LSP} \sim 75$ GeV) higgsino-like neutralinos;
the red dots around the regions marked ``St Helena" and ``Ghana". 
We show a plot zoomed in on this region in fig.\ \ref{fig:zoom},
and  elaborate on this  in
sections \ref{sthelena} and \ref{nigeria}, respectively.  

The BMSSM can also by construction naturally accommodate light  top squarks without 
tension with the mass bound on the lightest Higgs bosons,
unlike the ordinary MSSM (see section \ref{sec:effhiggs} above
for more details on this so-called ``little hierarchy problem"). This
property is investigated in the dark matter context for the ``Niger" region  in section \ref{sec:saharat}.
We note  that these light-stop BMSSM models,
and many of the other ones we study below, are not visible in
the commonly used ``projected" parameter plot in fig.\ \ref{fig:africa}, i.e.\
when projected onto the $(m_{\rm LSP},
Z_{\rm g}/(1-Z_{\rm g}))$ plane they are covered by MSSM models.
Thus, the fact that fig.\ \ref{fig:africa} is dominated by blue
circles should not be taken to mean that 
new BMSSM physics occurs {\it only} where the red dots are located.
This will be discussed further  in the following sections.

In the remaining sections
\ref{sthelena} through \ref{nigeria}
 we investigate these more subtle aspects of BMSSM models
through a few selected benchmark scans.

\subsection{A slice through parameter space}
\label{sec:slice}

Figure \ref{fig:africa} is useful for hints
for experimental searches
by showing what the genuinely new BMSSM models in the $(m_{\rm LSP},
Z_{\rm g}/(1-Z_{\rm g}))$ plane are, but it does not express much of 
the dark matter physics of individual classes of models.
To investigate this in more detail, we consider a specific slice
of  the full BMSSM parameter space.  For this slice, we
fix all parameters but $M_2$ and $\mu$
according to $\tan \beta=15$, $m_{A^0}=500$ GeV, $m_0 =1000$ GeV,
$A_t=A_b=500$ GeV, $\epsilon_2=0$. 

The result is shown in
fig.\ \ref{fig:slice1}, where the black and red points correspond to models with $\epsilon_1=0$ and
$\epsilon_1=0.05$, respectively, that are consistent with accelerator constraints and provide a 
dark matter relic density  in agreement with the WMAP $2\sigma$ bounds in eq.\ (\ref{eq:oh2}).
Note that fig.\ \ref{fig:africa} is the projection of many different slices similar to
the left panel of fig.\ \ref{fig:slice1} on top
of each other. (The position of the particular slice in
  fig.\ \ref{fig:slice1}  in the full scan of fig.\ \ref{fig:africa} can 
  be understood from the labelled benchmark scans.) This should be interpreted as follows.
  Although models in the red BMSSM regions of fig.\ \ref{fig:slice1} that
  are not on top of some  corresponding MSSM models in black may seem to 
  provide physically
  ``new" models, this is not always so, since we might be able to reach
  physically similar points by varying the  seven parameters in the MSSM-7 (i.e.\ leaving this slice) instead of varying
   the BMSSM parameters $\epsilon_{1,2}$.

But sometimes, there is something that can only be achieved in the MSSM-7 by varying
 {\it a few specific} parameters. For example, near the
  slice through parameter space shown in fig.\ \ref{fig:slice1}, the position and width of  the ${A^0}$ resonance
in fig.\ \ref{fig:slice1} is fixed for fixed $m_{A^0}$ and $\tan \beta$
(see section \ref{sec:bmssmhiggs}). 
In the MSSM slice (black),
this means the tree-level Higgs mass is then also fixed,
and the $h^0$ resonance on the left (in the ``Mauritania" region)
can only move by adjusting the loop corrections through $m_0$ 
and $A_t$. In the BMSSM,
the position of the $h^0$ resonance can be adjusted by adjusting $\epsilon_{1,2}$,
as can clearly be seen in the figure (the motion of the peak to the right when we turn on $\epsilon_1$). 
So, we do expect physically ``new" models in the BMSSM case,
even though they appear on top of each other in 
the projected figure \ref{fig:africa}. 
We investigate the motion of the $h^0$ peak
 in the Mauritania benchmark scan in section \ref{saharah}. 

The differences between the MSSM and the BMSSM 
around the mixed gaugino-higgsino region 
labelled ``Ghana" are  explored in
section \ref{nigeria}.

For convenience we also express the
 parameter space slice of this section in terms of the input parameters $M_2$ and
$\mu$. The result, which is a deformed version of the left panel in fig.\ \ref{fig:slice1}, is shown in 
the right panel of fig.\  \ref{fig:slice1}.

\begin{table}[h]
\begin{center}
\begin{tabular}{|l|c|c|c|c|c|}  \hline 
Parameter & St Helena${}^{(-)}$ & St Helena${}^{(+)}$ & Ghana${}^{(+)}$ & Mauritania & Niger \\  \hline  \hline
$M_2$ & 1000 &  1000 & 300  & 120.5 & 500 \\
$\mu$ & $-$101 & 94  & 118 & 500 & 700 \\
tan $\beta$ & 10 & 15 & 15 & 15 & 3\\
$m_{A^0}$ & 500 & 500 & 500 & 500& 1000  \\
$m_0$ & 1000 & 1000 & 1000 & 1000  & 298 \\
$A_t$ & 2000  & 500 & 500 & 500 & 0 \\ 
$A_b$ & 2000  & 500 & 500 & 500 & 0 \\ 
$\epsilon_1$ &$-$0.05 &  0.05 & 0.05 & 0.05 & 0.05 \\
$\epsilon_2$ &0 & 0 & 0 & 0 & 0.05 \\  \hline
\end{tabular}
\caption{Benchmark models, as referred to in the main text.
All parameters with mass dimension are
given in GeV.}
\label{bench}
\end{center}
\vspace{-7mm}
\end{table}
\begin{table}[h]
\begin{center}
\begin{tabular}{|l|c|c|c|c|c|}  \hline 
 & St Helena${}^{(-)}$ & St Helena${}^{(+)}$ & Ghana${}^{(+)}$ &Mauritania& Niger \\  \hline  \hline
$m_{\rm LSP}$ & 78 &  76 & 79  & 60 & 248 \\
$Z_{\rm g}/(1-Z_{\rm g})$ & $8.1 \cdot 10^{-3}$  & $1.1 \cdot 10^{-2}$ & $2.5 \cdot 10^{-1}$ & 
$1.1 \cdot 10^2$  & $1.2 \cdot 10^2$ \\ 
$m_{\wt \chi^0_2} $ & 115 & 118  & 143 & 116 & 478 \\
$m_{\wt \chi^0_3} $ & 505 & 505  & 165 & 509 & 706 \\
$m_{\wt \chi^0_4} $ & 1006 & 1007  & 325 & 510 & 727 \\
$m_{\wt \chi^{\pm}}$ & 99 & 95 & 108 & 116 & 478 \\
$m_h$ & 117 & 122 & 125 & 124 & 130 \\
$m_{\tilde{t}_1}$ & 825 & 1055 & 971 & 973 & 276 \\
$m_{\tilde{t}_2}$ & 1173 & 971 & 1055 & 1053 & 397 \\
%$m_{C2}$ & 1006 & 1007 & 325 & 514 \\
$\Omega_{\chi} h^2$ & 0.114 & 0.108 & 0.115 & 0.105  & 0.111 \\  \hline
\end{tabular}
\caption{Spectra and relic density for the benchmark models
of table \ref{bench}.  All parameters with mass dimension are
given in GeV.}
\label{benchspec}
\end{center}
\vspace{-7mm}
\end{table}
\begin{table}[h]
\begin{center}
\begin{tabular}{|l|c|c|c|c|c|c|}  \hline 
& StHelena${}^{(-)}$ & St Helena${}^{(+)}$ & Ghana${}^{(-)}$ & Ghana${}^{(+)}$ & Mauritania & Niger  \\  \hline  \hline
\multicolumn{6}{|l|}{BMSSM models } \\ \hline
$M_2$ & 1000 &  1000 & 300  & 300  & 102 \ldots 132 & 485 \ldots 515 \\
$\mu$ & $-$86 \ldots $-121$ & 79 \ldots 116  & 65 \ldots 218  & 88 \ldots 226 & 500 & 700 \\  \hline
\multicolumn{6}{|l|}{Corresponding MSSM models ($\epsilon_{1}=\epsilon_2=0$)} \\ \hline
$M_2$ & 1000 &  1000 & 300 & 300  & 102 \ldots 132 & 485 \ldots 515 \\
$\mu$ & $-$64 \ldots $-104$ & 67 \ldots 106  & 72 \ldots 220 & 72 \ldots 220 & 500  & 700\\  \hline
\end{tabular}
\caption{Benchmark scans, as referred to in the main text. All parameters with mass dimension are
given in GeV.}
\label{benchscan}
\end{center}
\vspace{-4mm}
\end{table}

\subsection{Light higgsino LSP (St Helena)}
\label{sthelena}
To obtain a dark matter relic density in the region
favored by WMAP, the light higgsino region  in the MSSM is problematic for several reasons. 
The first challenge is that whenever $m_{\rm LSP}\gtrsim m_W\approx 80$ GeV, annihilation of LSPs into pairs of W bosons is effective and pushes the relic density far below the value favored by WMAP, so higgsino-like neutralinos can 
typically provide good dark matter candidates
only if they are sufficiently heavy ($\gtrsim 900$ GeV). But what about neutralinos with masses
$m_{\rm LSP}\lesssim  80$ GeV below the W threshold,  where the annihilation cross section should be significantly lower? Here we instead run into two other problems, as we now describe. 

Firstly, the 94 GeV lower limit on the chargino mass requires the mass splitting $\Delta m_{{\pm}}=m_{\tilde{\chi}^{\pm}}-m_{\rm LSP}$ between the lightest chargino and neutralino
(see eq.\ (\ref{massplit}))
to be at
least 14 GeV when
 $m_{\rm LSP}\lesssim  80$ GeV.  However, in the higgsino region, the mass scales of both the lightest (and next to lightest) neutralino and the lightest chargino are set by the same value $\mu$. Typically the particles are nearly degenerate in mass, which means that all such models will be excluded by the chargino mass bound.

Secondly, even if  the chargino bound could be relaxed somewhat
in particular models (for example for $\Delta m_{\pm}\lesssim 3$ GeV \cite{Amsler:2008zzb},
see also section \ref{strategy} for further comments on the chargino bound), chargino coannihilation is very important at such small $\Delta m_{\pm}$ and causes the relic density 
to come out low also when $m_{\rm LSP}<m_W$.

To summarize, it is generally difficult to find light higgsino LSPs
in the MSSM with the relic density favored by WMAP.

This conclusion can be substantially altered in the BMSSM. As was discussed in section \ref{sec:chargino} and shown in fig.\ \ref{fig:massdiff}, the $\epsilon_1$ parameter can introduce a larger mass splitting $\Delta m_{\pm}$, and hence avoid both the chargino mass bound and the coannihilations at the same time. This is the reason for the existence of new models with higgsino-like neutralinos just below the W threshold in figures \ref{fig:africa}-\ref{fig:slice1}.

Note that it is not completely impossible to find valid MSSM models even in this region, as can be seen in figures \ref{fig:africa} and \ref{fig:zoom}. However, these  seem to appear only if parameters are finely tuned. The blue circle at the very bottom of the light higgsino-like LSP region actually only includes a single MSSM model, which is very special in the following sense. 
It has $M_2 > \mu$ but not $M_2 \gg \mu$, to be precise
$M_2 = 255$ GeV, $\mu=-77.2$ GeV,
which would typically produce a mixed higgsino-gaugino LSP
for generic values of the other parameters. 
However, this MSSM model has the specific (small) $\tan \beta$ value of $\tan \beta=1.16$,
and then the diagonalization
of the mass matrix $\cal M_{\wt \chi^0}$ (which
is given by (\ref{neumass_tree}) at tree level in the MSSM, though we 
also include loop corrections) is such that the LSP is higgsino-like despite the relatively low value of $M_2$. 
In the MSSM, mixed or gaugino-like LSPs  have no problem
giving a sufficiently large neutralino-chargino mass splitting, and
this particular higgsino-like LSP inherits this virtue, 
so it clears both the WMAP lower bound and the chargino mass bound.
 The BMSSM, on the other hand, naturally provides higgsino-like models with large neutralino-chargino mass splittings without the need for any tuning of parameters. This property is apparent in fig.\ \ref{fig:zoom}, where we see that the BMSSM models dominate the higgsino-like region even though the 
 MSSM and BMSSM samples contain a similar number of models for the region shown
 in that figure.

In table \ref{bench} we show the parameters of two benchmark models, St Helena${}^{(-)}$ and St Helena${}^{(+)}$, with exactly the BMSSM properties discussed above. As can be seen in table \ref{benchspec},  both of them provide a lightest neutralino mass below the W threshold and relic density in agreement with eq.\ (\ref{eq:oh2}), and without violating any accelerator bounds. The models differ in the signs of $\epsilon_1$ and $\mu$, and also come with different values of $\tan\beta$, $A_t$ and $A_b$. The magnitudes of $\mu$ have been chosen   for the models to give the correct relic density.

 \begin{figure}[h]
\includegraphics[width=8cm]{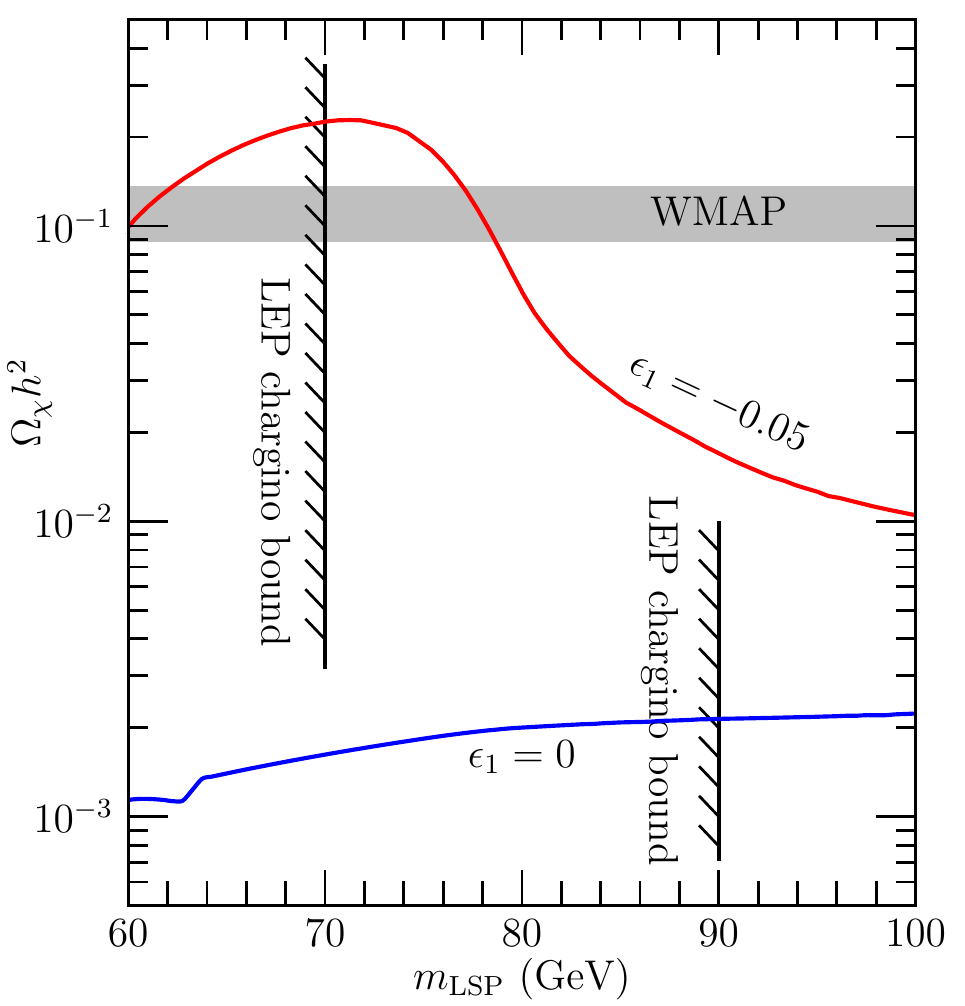}\hspace{-3mm}
\includegraphics[width=8cm]{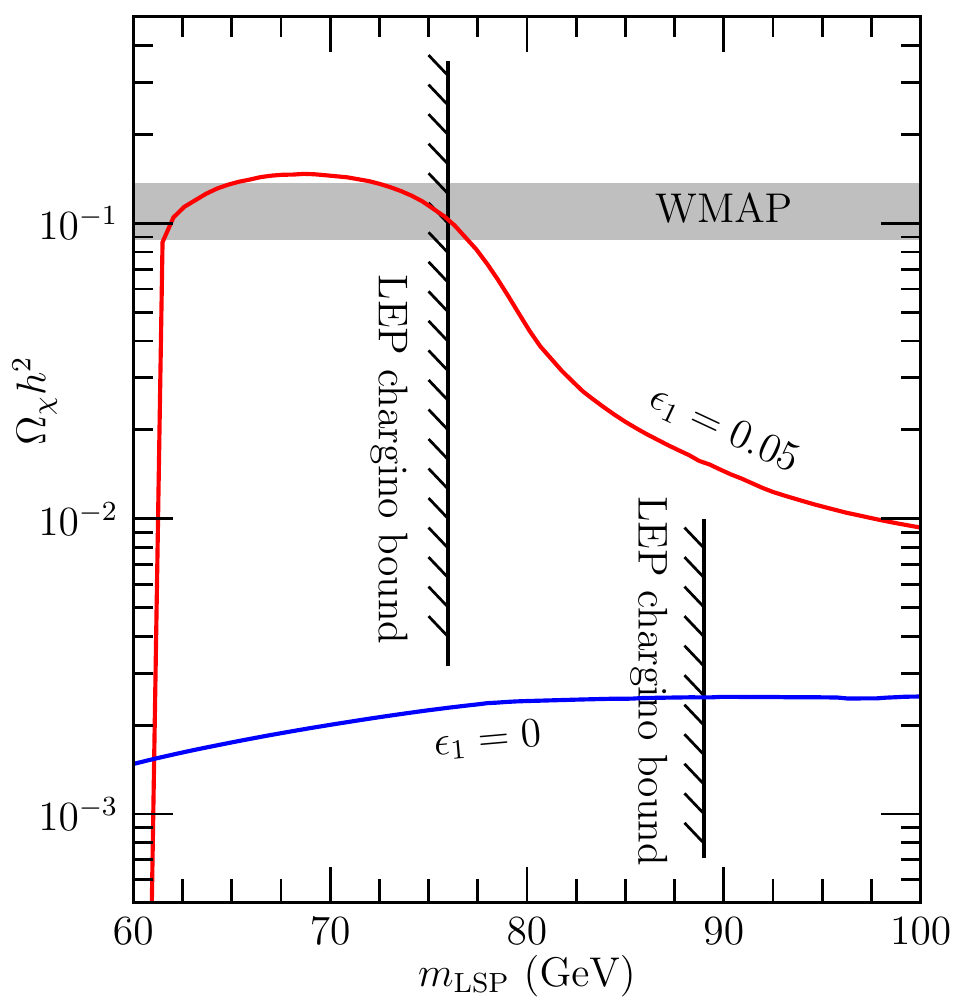}
\caption{Relic density $\Omega_{\chi} h^2$ along the St Helena${}^{(-)}$ and St Helena${}^{(+)}$ benchmark scans, respectively (see table \ref{benchscan}). The BMSSM models ($\epsilon_1\ne 0$) are shown in red, while the blue curves show the corresponding MSSM models ($\epsilon_1=0$).
The WMAP $2\sigma$ band is shown, as well as the LEP chargino bounds expressed
in $m_{\rm LSP}$ for the individual scans.  
For discussion, see the main text.}
\label{fig:sthelena}
\end{figure}
By scanning over values of $\mu$, we produce
 figure \ref{fig:sthelena}, where we plot the relic density $\Omega_{\chi} h^2$ as a function of the neutralino mass $m_{\rm LSP}$ for  the same set of parameters as those of the 
 St Helena${}^{(-)}$ and St\-Helena${}^{(+)}$ benchmark points (except for the $\mu$ value, of course). We also show the results for the corresponding MSSM models, i.e. using the same set of parameters except for setting $\epsilon_1=0$ (and again scanning over $\mu$). The ranges of $\mu$ corresponding to the neutralino mass range 60-100 GeV are shown in Table \ref{benchscan}. The reach of the  St Helena${}^{(-)}$ and St Helena${}^{(+)}$ scans in the $(m_{\rm LSP},Z_{\rm g}/(1-Z_{\rm g}))$ plane is indicated in figs.\ \ref{fig:africa} and \ref{fig:zoom}. The St\-Helena${}^{(+)}$ scans are part of the parameter space slice of 
 fig. \ref{fig:slice1}, and hence appear there.

For each individual scan, the $m_{\tilde{\chi}^{\pm}}>94$ GeV constraint is translated into a lower bound on $m_{\rm LSP}$, as indicated by the jagged vertical lines in fig.\ \ref{fig:sthelena}. As $\epsilon_1$ is switched on, and consequently as $\Delta m_{\pm}$ is increased, we see how the chargino bound is effectively weakened, \ie  moved to the left when expressed in $m_{\rm LSP}$. The shaded horizontal bands display the WMAP favored region of eq.\ (\ref{eq:oh2}). We see that there are models in both the  St Helena${}^{(-)}$ and St Helena${}^{(+)}$ scans that can satisfy the chargino and WMAP bounds simultaneously. The resulting relic densities for the corresponding MSSM models are reduced because of (chargino) coannihilations. We also repeat
(see section \ref{strategy}) that the entire parameter range of these scans also satisfy all our remaining imposed accelerator constraints, including bounds on the lightest Higgs boson mass and b $\rightarrow$ s$\gamma$.

The light higgsino LSP region was also studied by \cite{Cheung:2009qk} 
using analytical estimates for a few coannihilation processes and using a fixed
relative velocity. It is a priori difficult to say to what extent this 
reproduces  a full analysis with 
all relevant processes and actual velocity distributions.
They bring up a few of the points above, but we find it difficult to 
compare their preliminary analysis to ours. In particular,
we cannot reproduce  the relic density as in their main dark matter result, 
fig.\ 4 of  \cite{Cheung:2009qk}. Instead, we find that the models with their $\epsilon\geq 0$ (which corresponds to our $\epsilon_1\leq 0$
to get the same sign as theirs in the chargino mass,
see footnote \ref{footnote:cheung}) are
excluded by bounds on the Higgs mass.

\subsection{Heavy gaugino LSP (Niger)}
\label{sec:saharat}

Within the MSSM, most models have heavy top squarks $\tilde{t}_{1,2}$, due to the well-known need for large loop corrections to the lightest Higgs boson mass 
$m_{h^0}$ in order for a given model to pass current accelerator bounds.
(See section \ref{sec:mssmhiggs} for a discussion of this.)
Roughly speaking, $m_{\tilde{t}}$ needs to be about $\gtrsim 1$ TeV to  
make $h^0$ clear the bound. (See section \ref{computational} for comments on the
meaning of the  LEP Higgs bound in models other than the nonsupersymmetric Standard Model.)

Since the new BMSSM parameters $\epsilon_1$ and $\epsilon_2$ introduce corrections to the Higgs boson masses according to eq.\ (\ref{higgscorr}), these can be used to raise the Higgs boson mass even in models with light top squarks. This
issue has been discussed in the particle phenomenology literature
 many times (e.g.\  \cite{Brignole:2003cm,Casas:2003jx,Dine:2007xi}), to which we refer for details
 beyond those we have given in earlier sections.

\begin{figure}[h]
\begin{center}
\includegraphics[width=9cm]{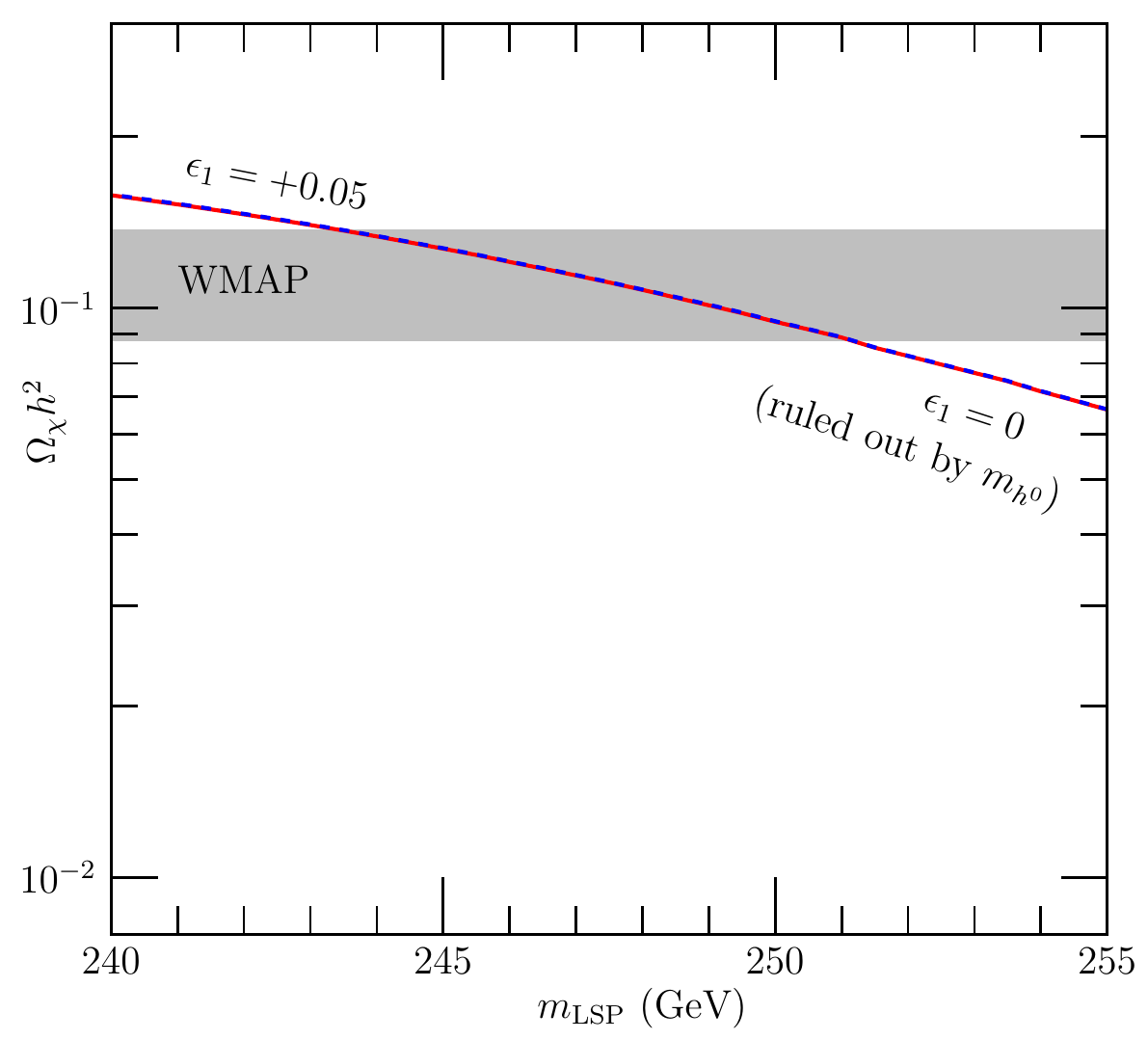}
\caption{Relic density $\Omega_{\chi} h^2$ along the Niger benchmark scans
(see table \ref{benchscan}). The BMSSM models ($\epsilon_{1,2}= +0.05$) are shown in red, while the blue curve shows the corresponding MSSM models ($\epsilon_{1,2}=0$). The curves
are practically on top of each other.
The WMAP $2\sigma$ band is shown, while the LEP chargino bounds appear first at much lower masses than shown here.
The MSSM models are excluded by the LEP Higgs bound
for all values of $m_{\rm LSP}$ in the plot.
For discussion, see the main text.}
\label{fig:saharat}
\end{center}
\end{figure}

An example of a gaugino-like model with light squarks, ``Niger", that passes all accelerator constraints is shown in tables \ref{bench} and \ref{benchspec}. By adjusting the value of $m_0$, we have introduced the right amount of squark coannihilations to find a relic density consistent with eq.\ (\ref{eq:oh2}). Note that the corresponding MSSM model, \ie with
the same parameters except for $\epsilon_{1,2}=0$, is excluded due to its low Higgs boson mass ($m_{h^0}\approx$ 91 GeV). We expect this to be a fairly generic phenomenon
in BMSSM models, since the alleviated
``little hierarchy problem" is one of the distinguishing characteristics of
the BMSSM. 

In figure \ref{fig:saharat} we show, by scanning over $M_2$, the relic density as a function of the neutralino mass, for the same set of parameters as those of the Niger benchmark point (except for $M_2$, that is now scanned over). The shaded horizontal band displays the WMAP-favored region of eq.\ (\ref{eq:oh2}). The figure also shows the resulting relic density for a scan over the corresponding MSSM models; the two curves are almost exactly on top of each other. This is as expected, simply because all $\epsilon_1$ corrections in the neutralino/chargino sector go as
$\epsilon_1/\mu$ (see eq.\ (\ref{expllag})), which is smaller for gaugino-like than
for higgsino-like models. Except for the Higgs mass bound, which is satisfied for all BMSSM points but none of the MSSM points, all models in the scans satisfy the remaining imposed accelerator constraints. The ranges of $M_2$ corresponding to the neutralino mass range 240-255 GeV are shown in Table \ref{benchscan}. The position of the Niger scans in the $(m_{\rm LSP},Z_{
\rm g}/(1-Z_{\rm g}))$ plane is indicated in the
overview plot  in fig.\ \ref{fig:africa} of section \ref{sec:generalscan}.
 
% We expect the Z coupling to be modified by $\epsilon$,
% as also pointed out in \cite{Cheung:2009qk}.
%\hmm{let's see if we get any mileage out of this question before we give up...}

\subsection{Light gaugino LSP (Mauritania)}
\label{saharah}
While light gaugino-like neutralinos in general provide relic densities
that are too high, they could become valid dark matter candidates if the annihilation cross section is increased by virtue of an $s$-channel Higgs boson resonance. In fig.\ \ref{fig:slice1} we can clearly see both the $h^0$ resonance around $m_{\rm LSP}\sim 60$ GeV and the $A^0$ resonance at $m_{\rm LSP}=500$ GeV.  Since we use the physical $m_{A^0}$ as an input parameter in both the MSSM and the BMSSM, the position of the $A^0$-resonance cannot move when we change $\epsilon_1$ and $\epsilon_2$. 
The $h^0$ resonance will move, however, by an amount roughly given by the analytical expression
in  eq.\ (\ref{higgscorr}), and this is clearly manifest in fig.\ \ref{fig:slice1}.  

In tables \ref{bench} and \ref{benchspec} we give an example of a BMSSM model, ``Mauritania", which passes all accelerator constraints and for which the $h^0$ resonance brings down the relic density into the 
region favored by WMAP. (Recall that in this region, $\Omega_{\chi} h^2$ tends to 
come out too high and needs to be brought down, whereas in the light higgsino
region in section \ref{sthelena}, $\Omega_{\chi} h^2$ needed to be increased.)
\begin{figure}
\begin{center}
\includegraphics[width=9cm]{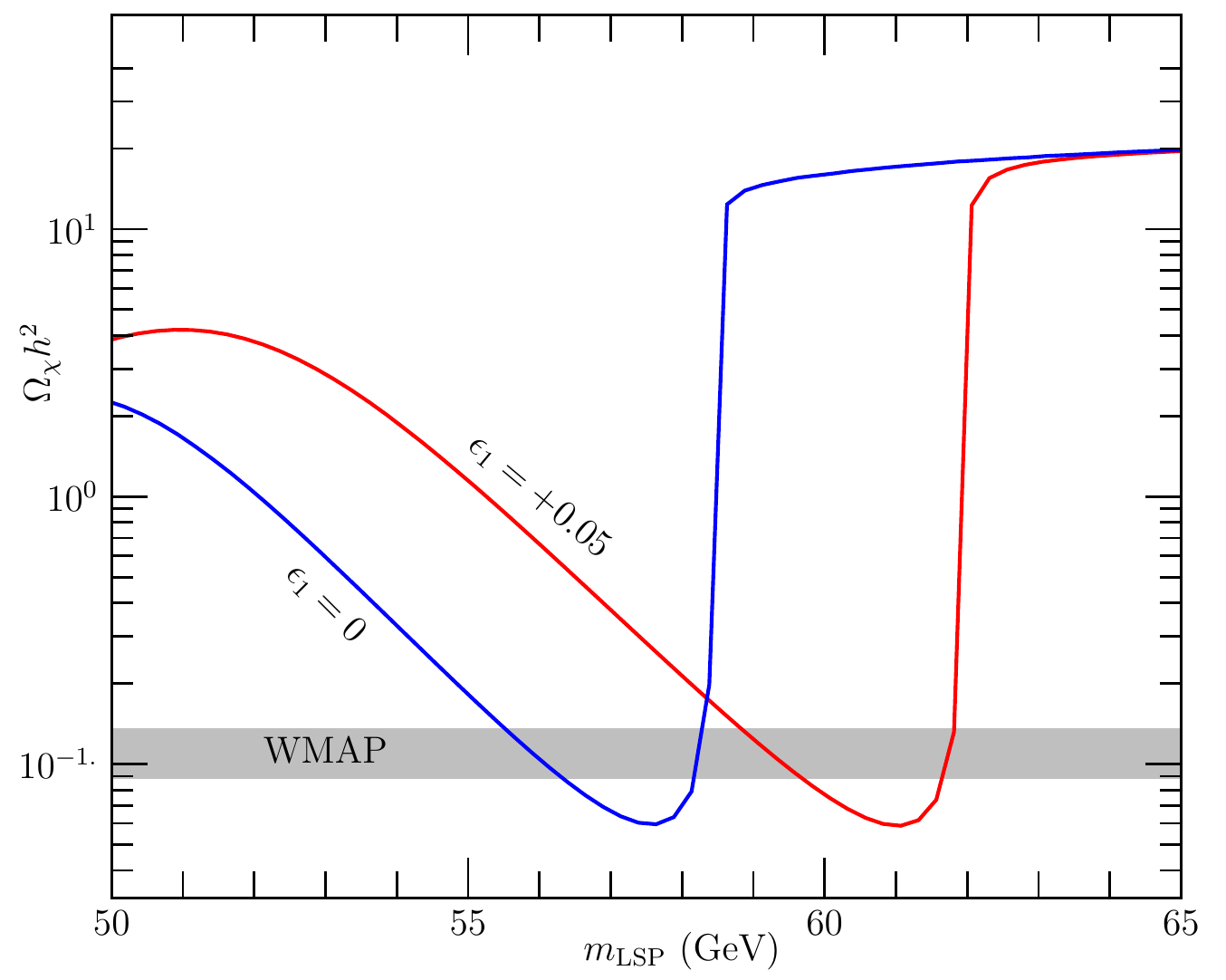}
\caption{
Relic density $\Omega_{\chi} h^2$ along the Mauritania benchmark scans
(see table \ref{benchscan}). The BMSSM models ($\epsilon_{1}=+0.05$) are shown in red, while the blue curve shows the corresponding MSSM models ($\epsilon_{1}=0$).
The WMAP $2\sigma$ band is shown, while the LEP chargino bounds appear first at lower masses than shown here. For discussion, see the main text.}
\label{fig:saharah}
\end{center}
\end{figure}

In figure \ref{fig:saharah} we show the relic density as a function of the neutralino mass 
$m_{\rm LSP}$ for  the same set of parameters as those of the Mauritania benchmark point,
apart from $M_2$, which is scanned over. We also plot the results for the corresponding MSSM models, \ie with $\epsilon_1=0$ instead, which gives a Higgs boson mass of $m_{h^0}\approx 117$ GeV. The ranges of $M_2$ corresponding to the neutralino mass range $m_{\rm LSP}=$ 50-65 GeV are shown in table \ref{benchscan}. All models in the scans pass our imposed accelerator constraints. The shaded horizontal band displays the WMAP bound of eq.\ (\ref{eq:oh2}), as before, and we can see how the  neutralino masses favored by dark matter constraints slide to the right  along with the $\epsilon_1$ correction to the light Higgs boson mass. The asymmetric shape of the curves originates from the Boltzmann tail of the velocity distribution in the early Universe. The reach of the Mauritania scans in the $(m_{\rm LSP}, Z_{\rm g}/(1-Z_{\rm g}))$ plane is indicated in figures \ref{fig:africa} and \ref{fig:slice1}.

\subsection{Mixed higgsino-gaugino LSP (Ghana)}
\label{nigeria} 

As can be seen from the separation of the red and black curves
in figure \ref{fig:slice1}, the effects that give rise to the new higgsino-like models of section \ref{sthelena} are present already at quite moderate values of $Z_{\rm g}/(1-Z_{\rm g})$, i.e. for fairly mixed neutralinos.  In tables \ref{bench} and \ref{benchspec} we give an example of a 
benchmark model ``Ghana$^{(+)}$" that provides correct relic density and passes all accelerator constraints, in a part of the light mixed neutralino region that the MSSM models of figure \ref{fig:slice1} are not capable of reaching ---  as long as we stay in the slice through parameter space
shown in figure  \ref{fig:slice1}. 
See 
section \ref{sec:outlook} for some further comments on what this means. 

\begin{figure}[h]
\includegraphics[width=8cm]{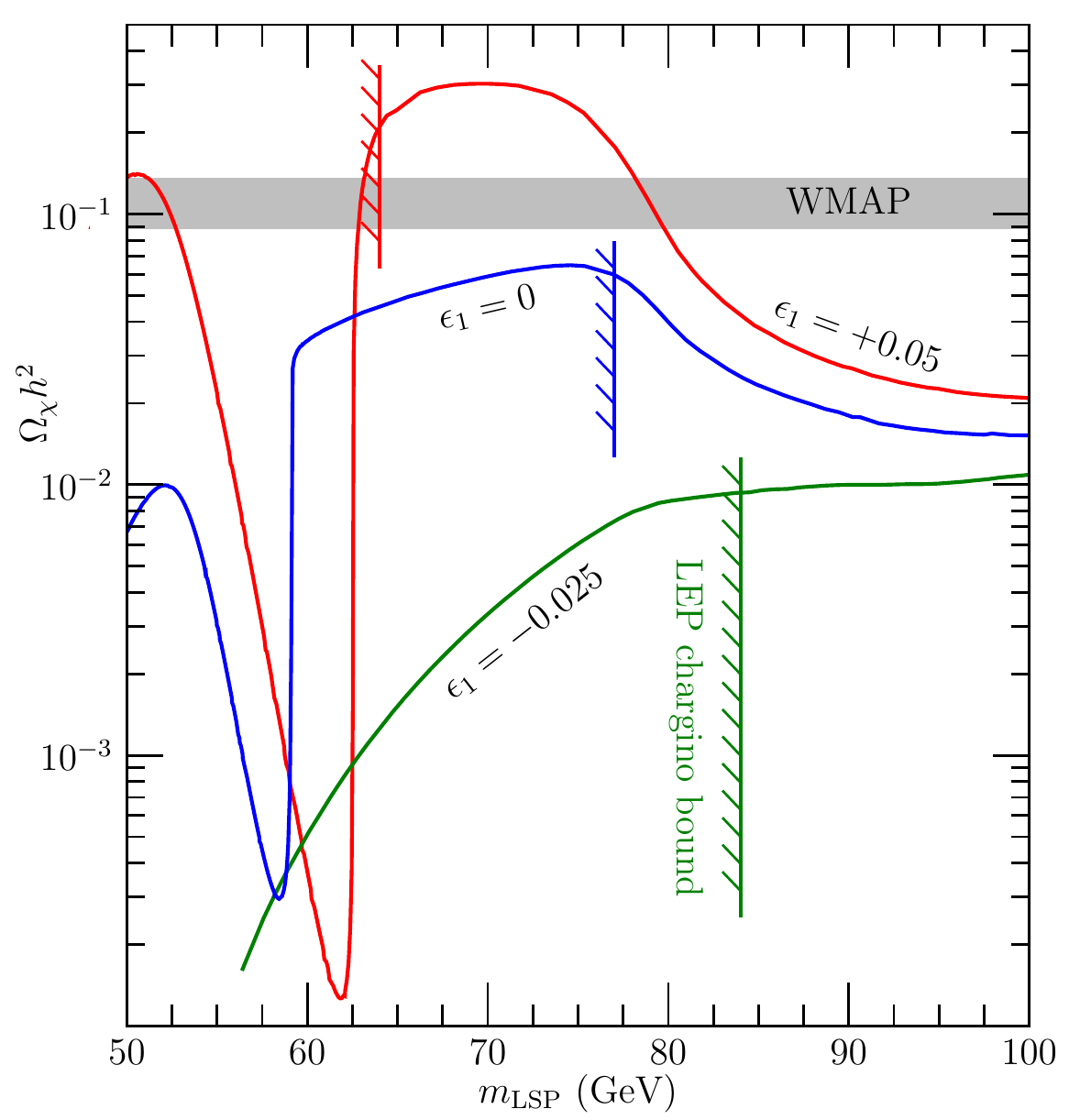}\hspace{-1mm}
\includegraphics[width=7.8cm]{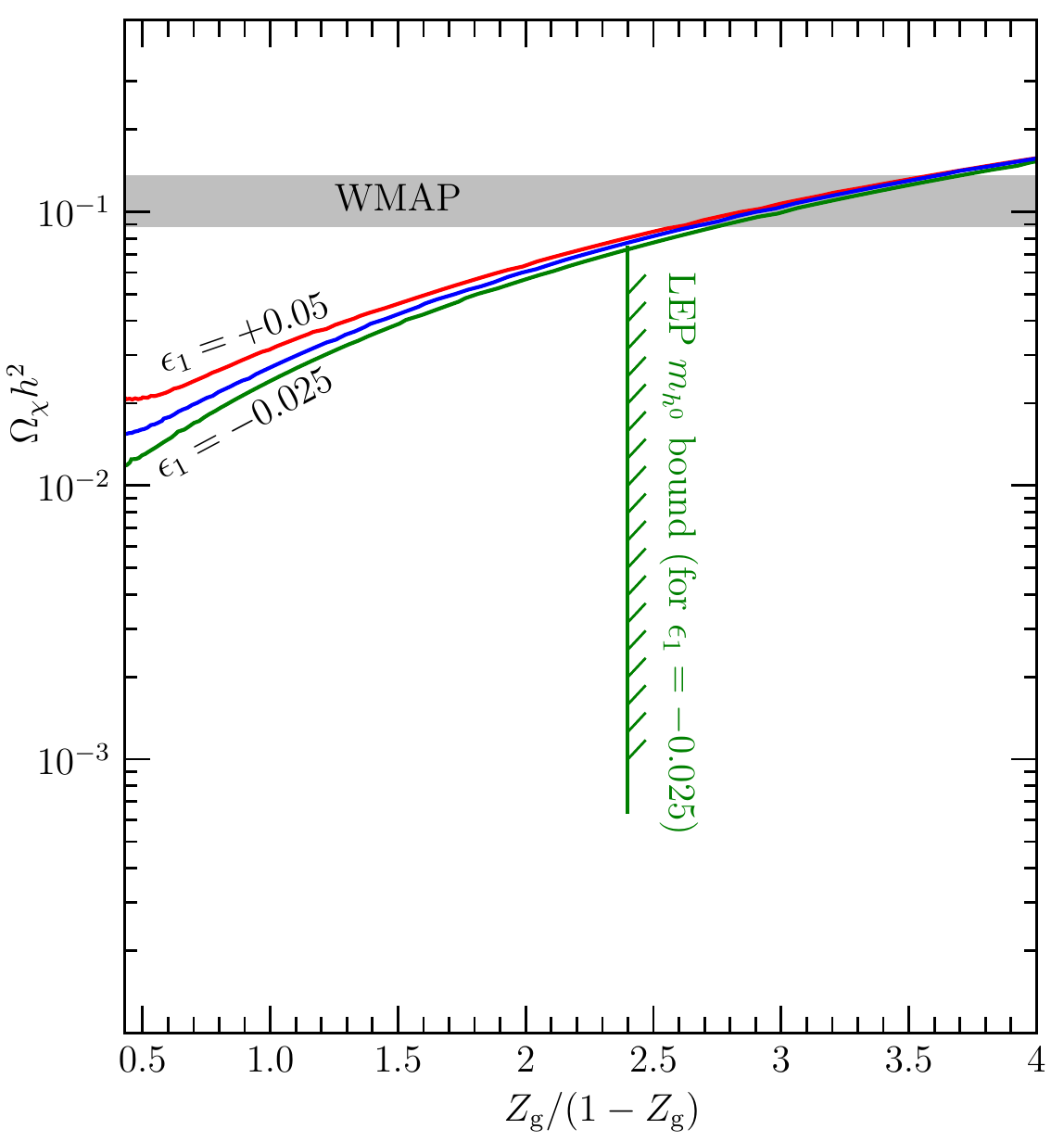}
\begin{center}
\vspace{-4mm}
\caption{Relic density $\Omega_{\chi} h^2$ along the
Ghana  benchmark scans
(see table \ref{benchscan}). 
The BMSSM models ($\epsilon_1\neq 0$) are shown in red and green while the blue
curve shows the corresponding MSSM models ($\epsilon_{1}=0$). 
The WMAP $2\sigma$ band is shown, as well as the LEP chargino bounds
expressed in $m_{\rm LSP}$ for the individual scans. 
The right panel begins at $Z_{\rm g}=0.49$, which
in this slice (see fig.\ \ref{fig:africa}
or \ref{fig:slice1}) corresponds to $m_{\rm LSP}=100$ GeV, where the left panel ends.
The Ghana${}^{(-)}$ benchmark scan (in green) is also constrained by
the LEP Higgs mass bound, as indicated in the figure. 
For discussion, see the main text.}
\label{fig:nigeria}
\end{center}
 \vspace{-0.7cm}
\end{figure}

We investigate the mixed region in more detail by scanning over $\mu$, for parameter values 
that are otherwise the same as in the Ghana$^{(+)}$ benchmark model. In figure \ref{fig:nigeria}, the relic density is shown first as a function of $m_{\rm LSP}$ (left panel), and then as a function of  $Z_{\rm g}/(1-Z_{\rm g})$ (right panel) in intervals such that the graphs transition
smoothly from the left into the right panel. We also plot the relic density for the corresponding MSSM models (i.e. with $\epsilon_1=0$), as well as for another set of BMSSM models, the Ghana${}^{(-)}$ scan, in which we instead set $\epsilon_1=-0.025$. The ranges of $\mu$ used in fig. \ref{fig:nigeria} are stated in table \ref{benchscan}. Except for the chargino bound and (only for high values of $Z_{\rm g}/(1-Z_{\rm g})$ in the Ghana${}^{(-)}$ scan) the Higgs bound indicated in the figures, all models satisfy our imposed accelerator constraints. The ranges of these scans are indicated in figures \ref{fig:africa} and \ref{fig:slice1}.

The $m_{\wt \chi^\pm}>94$ GeV constraint  is translated into a lower bound on $m_{\rm LSP}$ for each individual scan, as is indicated
by a jagged line across each separate curve in the left panel of figure \ref{fig:nigeria}. 
(Recall from section \ref{sthelena} that as
$\epsilon_1$ is increased, $\Delta m_{{\pm}}=m_{\tilde{\chi}^{\pm}}-m_{\rm LSP}$ is increased,
so the bound expressed in $m_{\rm LSP}$ moves to the left.)
Also shown as a shaded horizontal band is the WMAP-favored region of eq. (\ref{eq:oh2}). We see that there are models in the Ghana${}^{(+)}$ scan that can provide correct relic density while at the same time avoiding the chargino bound. As pointed out in section \ref{sthelena}, this is thanks to the drop in annihilation rate right below the W pair production threshold, in combination with larger neutralino-chargino mass splitting that makes the coannihilations less important
in the BMSSM model than 
in the corresponding MSSM model with $\epsilon_{1,2}=0$ (the blue curve). Also note how the presence of the $m_{h^0}$ resonance makes the relic density drop sharply for even lower ($\sim 60$ GeV) neutralino masses. In the Ghana$^{(+)}$ scan, however, the chargino bound  just about rules these models out, but the Higgs resonance is clearly visible in fig. \ref{fig:slice1} for only slightly higher gaugino fractions. Neither of the other two scans shown provide any dark matter candidates near the W pair production threshold. The reason that the Ghana$^{(-)}$ scan (in green) ends abruptly  for $m_{\rm LSP}\lesssim 65$ GeV is that the chargino becomes the LSP.

The right panel of figure $\ref{fig:nigeria}$ shows how the relic density increases, and finally passes through the WMAP-favored values, when we go to higher gaugino fractions $Z_{\rm g}$. For the Ghana${}^{(-)}$ scan (in green) only, the Higgs mass drops below the LEP bound before the WMAP-favored region is reached, as is shown by the jagged vertical line.

\section{Summary and outlook}
\label{sec:outlook}
In this paper, we considered effective field theory corrections to the MSSM,
and incorporated them in existing packages for accurately calculating the relic density
of dark matter. 
We found that the corrections can make a difference for dark matter
in certain regions of parameter space, and we performed a first scan of parameter
space of these models.

An important question that
we addressed (but did not fully answer)  in sections
\ref{sec:slice}, \ref{sthelena} and \ref{nigeria} 
 is whether
the new models generated by turning on the BMSSM corrections $\epsilon_{1,2}$
can effectively be recreated by modifying the {\it other} pre-existing parameters of the ordinary MSSM,
like $m_0$ or $\mu$.  
To be concrete, take
the BMSSM models in the benchmark scans in the slice through parameter
space in figure\ \ref{fig:slice1}. Let us call the models in the red BMSSM curve that are not on top
of some parts of the black MSSM curve ``new BMSSM models". 
Do ordinary MSSM models that are in some sense physically similar
to the new BMSSM models exist anywhere in some {\it other} slice of parameter space
 than that fixed in figure \ref{fig:slice1}?
If they do, then whether they
should be considered equally interesting as the BMSSM models that
satisfied the constraints ``naturally" in this slice depends on the details.
For example, as argued in section \ref{sec:slice}, there could  be only a restricted
subset of parameters that could be varied to reach those MSSM models,
and those parameters could change e.g.\ accelerator physics of these models
in characteristic ways. Or, it could require finetuning of parameters
so awkward (see e.g.\ section \ref{sthelena}) that the price of reaching these
models in the MSSM is not worth paying.  
This will of course depend on one's 
idea of naturalness. We leave this question here,
and hope that our detailed remarks in previous sections give some useful guidance for further studies
of the importance of naturalness for MSSM vs.\ BMSSM dark matter. 

There are many directions one could take this further. First, we imposed many restrictions on the models, some of which are well motivated, some of which could be relaxed.
The most obvious direction is perhaps to perform
large dedicated scans to study certain characteristics of these models more systematically.
It is generally challenging to give precise boundaries in parameter
space of the ordinary MSSM where the model is ruled out by experiment (and thus can be ruled in
by the BMSSM),
since one can often tune MSSM parameters to very special points
to evade a given bound. 
Thus, to really understand what BMSSM models are truly ``new" and
do not exist {\it anywhere} in MSSM parameter space (especially
in models more general than the MSSM-7),
a Markov Chain Monte Carlo (MCMC) analysis might be more enlightening
than the random grid scans we have performed here. 
The light Higgsino region would
be an example that would be worth studying further in this respect. 

Next, one can of course generalize
our rather ``minimal" BMSSM implementation further.
For example, it might be interesting to study
 CP violating processes,
allowing for imaginary parts of the $\epsilon_{1,2}$, or
physics directly related to the $1/M^2$ operators.   
One comment about this last point:
as discussed in section \ref{bounds},
it is interesting that the $1/M^2$ operators
in the K\"ahler potential are rather stringently
restricted by the precision electroweak observables $S$ and $T$ 
\cite{Blum:2008ym}. In the effective theory,
the coefficients $\xi_i$ of those operators \cite{Dine:2007xi}
are independent of the coefficients $\epsilon_{1,2}$ we have considered,
but in an underlying theory there may be relations between $\epsilon_i$ and $\xi_i$
(see \cite{Dine:2007xi} for an example). As a highly optimistic scenario, consider
a future experiment
where dark matter has been detected that is well described
by the neutralino in  the light Higgsino region of our plot. Let us interpret that  as evidence for new physics beyond the MSSM. By the logic in this paragraph and section \ref{bounds}, 
 constraints from electroweak precision observables 
could be used to further constrain the underlying theory in this situation.

Also, one could study the interpretations of specific experimental/observational data in the light of these corrections; some recent studies of accelerator physics
that we think could be 
relevant to combine with further work on our dark matter calculations
include
\cite{Batra:2004vc,Randall:2007as,Batra:2008rc,Cassel:2009ps,Mason:2009iq}.
Also the work
on BMSSM baryogenesis \cite{Blum:2008ym}  would be interesting to 
study further in the light of our calculations.

 Finally,
especially in BMSSM models with more new parameters, it would be useful to 
try to obtain additional restrictions on those classes of models to keep the analysis practical. A related question is if it is worth pushing the size of the BMSSM corrections further than we have. Suggestions in this direction include allowing
for the creation of new vacua, but imposing the BDH criterion \cite{Blum:2009na},
and thereby trying to make sure the model is stable at least on timescales
of the age of the universe.
\begin{appendix}
\section{Notation and conventions}
\label{appconv}

We denote the two Higgs doublets with hypercharge $Y=+1/2$ and $Y=-1/2$ by $H_u$ and $H_d$ respectively. In the literature \cite{Edsjo:1997hp,Haber:1984rc} one also often  finds the notation $H_2$ and $H_1$, $i.e.$, $H_u=H_2$ and $H_d=H_1$. In $SU(2)$ components the doublets are written
\be
 H_u = \begin{pmatrix} H_u^+ \\ H_u^0 \end{pmatrix}, \quad H_d = \begin{pmatrix} H_d^0 \\ H_d^- \end{pmatrix}.
\ee
All suppressed $SU(2)$ doublet indices are contracted in the main text. Doublets transforming in conjugate representations are contracted with the Kronecker delta $\delta_{ab}$, for example we have
\be
 H_u^\dagger H_u = \delta_{ab} (H_u^{a})^* H_u^b = |H_u^+|^2 + |H_u^0|^2
\ee
while doublets transforming in the same representation are contracted with the epsilon symbol $\epsilon_{ab}$, with $\epsilon_{12} = -1$, for example we have
\be
 H_u H_d = \epsilon_{ab} H_u^a H_d^b = H_u^0 H_d^0 - H_u^+ H_d^-.
\ee
It should be clear which contraction is intended where. When multiple contractions appear we use brackets to denote which fields to contract, for example we have
\be
 (H_u H_d)(H_u^\dagger H_u) = \epsilon_{ab} \delta_{cd} H_u^a H_d^b (H_u^c)^* H_u^d \; . 
\ee
Our sign of $\mu$ agrees with the sign in {\tt DarkSUSY}
and is opposite to that of DST \cite{Dine:2007xi}. The parameters $\epsilon_1$ and $\epsilon_2$
that we introduced in (\ref{V1}) and (\ref{V2}) have the same sign as in DST;
a positive $\epsilon_1$ correction {\it increases} the lightest Higgs mass $m_{h^0}$,
as can be seen in figure (\ref{fig:higgs}).

\subsection{Superpotential interactions}
The dynamical field content of a left-chiral super field $\Phi$ is a complex scalar $\phi$ and a left-handed two-spinor $\psi$. Given a superpotential $W=W(\Phi_i)$ for a family of left-chiral superfields $\Phi_i$, we calculate the fermionic interactions using
\be  \label{fermionint}
 \lag_{(\text{fermion})^2} = -\frac{1}{2} \left( \left. \sum_{i,j}\frac{\p^2 W}{\p \Phi_i \p \Phi_j} \right|_{\Phi=\phi} \psi_i \psi_j + \hc \right)
\ee
and the F-term scalar potential using
\be
V_F = \sum_{i} \left| \frac{\p W}{\p\Phi_i} \right|^2_{\Phi=\phi} \; . 
\ee
These formulas hold for any $W$,  not just renormalizable ones.
\section{Higher-dimensional operators}
\label{eft}
There are two obvious questions about 
effective field theories with higher-dimensional operators:
how we compute with them, and where they come from. 
In this short appendix, we try to elucidate this as briefly and as simply as possible.
Some of this is textbook material \cite[Ch. 12.3]{Weinberg:1995mt},
some is available in review articles like \cite{Georgi:1994qn,Pich:1998xt}.

Consider the following ``microscopic'' toy-model Lagrangian
for two neutral scalars $H_u^0$ and $H_d^0$, and three neutral fermions $\tilde{S}$, $\tilde{H}_u^0$ and $\tilde{H}_d^0$:
\be  \label{L}
\lag_{\rm microscopic} &=& \lag_{\rm free}  +  \lag_{{\rm free}, S}   
 - \lambda_S \tilde{S} H_u^0  \tilde{H}_d^0 - \lambda_S \tilde{S} \tilde{H}_u^0  H_d^0   + \mbox{h.c.}
\ee
where 
$ \lag_{\rm free}$ are the usual kinetic and
mass terms for $H_u^0$, $H_d^0$, $\tilde{H}_u^0$ and $\tilde{H}_d^0$, and
$ \lag_{{\rm free},S}$ are the kinetic and mass terms for the fermion $\tilde{S}$,
and
we introduced the Yukawa coupling $\lambda_S$, that we set to be equal
for the two terms.
We want to calculate an effective Lagrangian $\lag_{\rm eff}$
that incorporates the lowest-order interactions
between $H_u^0$, $H_d^0$, $\tilde{H}_u^0$ and $\tilde{H}_d^0$ due to 
the exchange of $\tilde{S}$, but no longer contains $\tilde{S}$ explicitly. 
In other words,
we want to approximately integrate out $\tilde{S}$. (We write ``approximately''
to distinguish this from performing the functional integral
over $\tilde{S}$, which would not be an approximation.)

To this end, recall how the four-fermion coupling in Fermi theory
is  a low-energy approximation
of  four-fermion interactions in the electroweak theory, valid
for external momenta much smaller than the mass of the exchanged gauge boson. 
Analogously, the  theory (\ref{L}) gives rise to the following operator in 
the effective Lagrangian at tree level:
\be
\hspace{0mm}
 \parbox{42mm}{\includegraphics{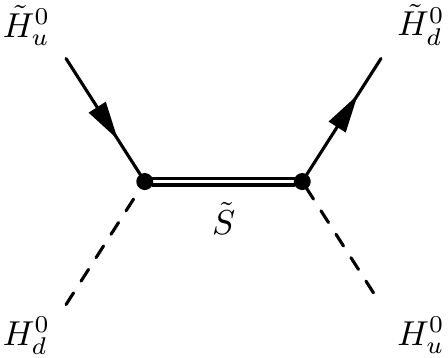}} 
 \hspace{0mm} \stackrel{p^2 \ll M_{\tilde{S}}^2 }{\longrightarrow}\quad 
\hspace{-4mm}\parbox{42mm}{\includegraphics{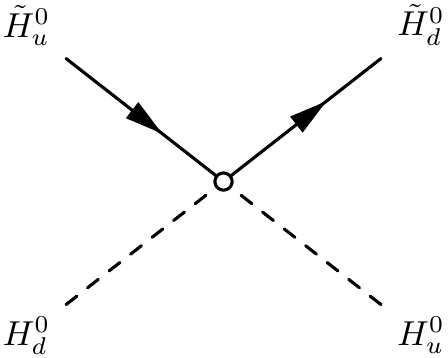}}  
\hspace{2mm}
 \sim {\lambda_S^2 \over M_{\tilde{S}}} H_u^0 H_d^0 \tilde{H}_u^0 \tilde{H}_d^0
\nonumber
\ee
On the other hand,  as in the body of the paper (see eq.\ (\ref{Weff}))
we can write a {\it general} effective dimension-5 operator (the little circle in the second diagram above) of this form with a dimensionless coupling $\lambda$:
\be  \label{L5}
 \lag_{\rm dim-5} = {\lambda  \over M} H_u H_d \tilde{H}_u \tilde{H}_d
\ee
The $1/M$ suppression for some mass scale $M$ is just by dimensional analysis. 
Here, the energy scale $M$ is supposed to be the ``scale of new physics'', the energy
at which our effective theory starts to deviate from the microscopic theory.
For the microscopic theory above, clearly
this will be $M \sim M_{\tilde{S}}$. (Whether $M$ is exactly this scale $M_{\tilde{S}}$ or somewhere
close to this scale is a matter of definition --- this is equivalent to stating when $p^2 \ll M^2$
fails to be satisfied  to
a prescribed accuracy.). We now {\it match}
the effective field theory coupling $\lambda$ to the microscopic coupling $\lambda_S$
at a prescribed renormalization group
(RG) scale $\Lambda$, say $\Lambda=M_{\tilde{S}}$:
\be  \label{matching}
{\lambda }(\Lambda)= {\lambda_S^2 }(\Lambda) \quad \mbox{at $\Lambda=M_{\tilde{S}}$.}
\ee
The couplings may run differently below and above this scale, since the
field content is effectively different, so the $\beta$ functions will be different.
We see that it is useful to think of $\lambda(\Lambda)$ and $\lambda_S(\Lambda)$ as
two different quantities, that are related by $\lambda=\lambda_S^2$ only at the matching point in 
(\ref{matching}). 

In our examples 
in the paper,
we
express the coupling in terms of  $\epsilon_1 := -\lambda \mu^* /M$ (see eq.\ (\ref{epsilon1})), so 
equivalently
\be  \label{L5_2}
 \lag_{\rm dim-5} = -{\epsilon_1  \over \mu^*} H_u H_d \tilde{H}_u \tilde{H}_d
\ee

To summarize, the effective theory valid below the energy $E=M$ is
\be  \label{Leff}
\lag_{\rm eff} =
\underbrace{{\lag}_{\rm free}}_{\mbox{from (\ref{L})} }
+ \underbrace{{\lag}_{\rm dim-5}}_{ \mbox{from  (\ref{L5})}}
\ee
where $\tilde{S}$ does not appear at all. We distinguish this 
from the microscopic theory (\ref{L}), where $\tilde{S}$ appears explicitly. 
If we use $\lag_{\rm eff}$ to compute the cross section 
$\sigma(\tilde{H}_u H_d \rightarrow \tilde{H_u} H_d)$, 
we have the low energy expansion
\be\label{crossexp}
\sigma(\tilde{H}_u H_d \rightarrow \tilde{H_u} H_d) = 
\sigma_0 + \left(E \over M\right) \sigma_1 + \left(E \over M\right)^{\!2} \sigma_2 + \ldots
\ee
where $\sigma_0$ is the cross section for $\epsilon_1=0$,
$\sigma_1$ is the leading correction, and 
each term in the low-energy expansion itself has a loop expansion ($\sigma_0 = \sigma_{0,0} + \lambda \sigma_{0,1} + \ldots$).
We find that for $E \ll M$, the cross section (\ref{crossexp}) with just the linear term
is not much different from the complete result in the underlying theory 
$\lag_{\rm microscopic}$.  

At loop level, as long as we use a mass-independent scheme such as $\overline{\rm MS}$, the expansion in $(E/M)$ will remain consistent. This
does  introduce apparent divergences,
but they cancel when we match the theory to the microscopic Lagrangian (\ref{L})
at the RG scale $\Lambda = M$,
by a relation similar to (\ref{matching}) (see e.g.\ \cite{Pich:1998xt}). In the same vein, 
apparent UV divergences due to hard supersymmetry
breaking operators  like (\ref{V2}) are no more harmful
than soft supersymmetry breaking terms when matching to the microscopic theory,
where supersymmetry breaking is usually  spontaneous
(one clear example of this in the MSSM is \cite{Martin:1999hc}). 

So to answer the questions posed at
 the beginning of this appendix, we calculate with the higher-dimension operators
as with any other Lagrangian, as long as we remember that the approximation is only valid
at energies much smaller than  the scale of new physics $M$, which can be set to $M_{\tilde{S}}$ in this example. 
It is also important that in the effective field theory,
we can work without direct reference to any parameters in $\lag_{\rm microscopic}$,
and view $\epsilon_1$ in $\lag_{\rm eff}$ as a model parameter. 

Now for the origin
of the operators, and the generality of the effective theory, versus that of the microscopic theory.
We have seen that for a {\it given} microscopic theory, we can express the low-energy parameters
 in terms of microscopic quantities,
as in (\ref{matching}). It is important, however, that (\ref{L}) is not the {\it only} theory that can give rise to new operators that can be approximated by the form (\ref{Leff}). For example, if we instead integrate out a field with $SU(2)$ charge,
we can obtain a similar expression for 
$\lag_{\rm eff}$. The effective theory captures aspects of several possible underlying theories,
not just eq.\ (\ref{L}). In this sense it is more general than any particular microscopic theory. 

The effective theory is also less general than any particular  underlying theory,
in the following sense. 
The particular Lagrangian in (\ref{L}) is actually part of the Next-to-Minimal Supersymmetric Standard Model, or NMSSM.
In this theory, $\tilde{S}$ (sometimes called the ``singlino'') can be the LSP in some part of parameter space.
That is clearly not possible in the effective theory (\ref{Leff}),
where the singlino does not appear at all. In other words,
we only consider indirect effects of any new physics on the interactions between existing
fields in our theory, we do not consider any new fields beyond the field content of the usual MSSM.  
In this sense, the effective theory does not capture all aspects of a given microscopic theory.

One final comment on the range of validity of the effective theory.
If we require  the microscopic theory to be perturbative, e.g.\ $\lambda < 1$, then 
if we view $\epsilon_1$ as fixed at say $0.01$, 
we find a constraint
\be \label{condition}
{|\mu| \over M}  = { \epsilon_1 \over \lambda} = {0.01 \over \lambda} > 0.01
\ee
so for example for $M \sim 5$ TeV, we have $\mu > 50$ GeV for perturbativity. 
Similar comments hold for $\epsilon_2$ and $m_{\rm SUSY}$.
For practical purposes, we will
allow $M$ to vary slightly,
 such that (\ref{condition}) is always satisfied
for our (reasonably small) range of the parameters $\mu$ and $m_{\rm SUSY}$.
Thus, we do not bring
this up in the discussion in section \ref{bounds}. 
 One can also
contemplate viewing the effective operators as not arising from a particular perturbative microscopic theory at all, as for the chiral Lagrangian in QCD, but we would like to keep
perturbativity as an option.

\section{Feynman rules}
\label{feynmanrules}

In this appendix we present the Feynman rules generated by the 
Lagrangian in eq.\ (\ref{expllag}). These include interactions between charginos, neutralinos and Higgs bosons as well as Higgs boson self-interactions. Concerning the Higgs boson self-interactions, we present only three-point couplings and no four-point couplings, as the latter contribute negligibly to relic density calculations.

When a coupling contains two fermions, our convention
(following \cite{Edsjo:1997bg}) is that the first one always appears with a bar in the Lagrangian while the second one always appears without the bar, i.e., the first fermion is outgoing and
 the second fermion is ingoing. The scalar bosons all appear in the couplings exactly as they appear in the Lagrangian. For example, to the following Lagrangian term
\be
\frac{1}{2} g^L_{H^+ H^- \widetilde\chi^0_i \widetilde\chi^0_j} H^{+} H^{-} \overline{\widetilde\chi^0_i} P_L \widetilde\chi^0_j
\ee
we assign the following coupling:
\be
g^L_{H^+ H^- \widetilde\chi^0_i \widetilde\chi^0_j} \; .
\ee
Here we assumed that the coupling is symmetrized in $i,j$.

All Feynman rules contain an $i$ that we do not write explicitly, i.e.\ all couplings below should be multiplied by a factor of $i$. The {\tt DarkSUSY} convention is to have
the vertices divided by a factor of $i$, so that  the vertices below are coded into {\tt DarkSUSY}
exactly as written.

{\allowdisplaybreaks

\subsection{$H$-$H$-$H$ vertices}
The vertex is
\vertex{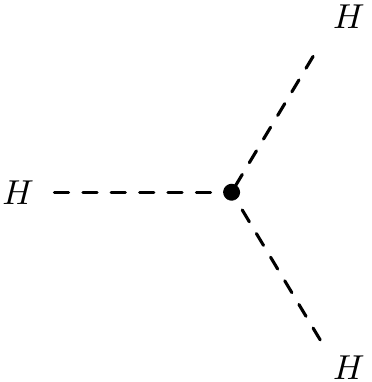}{g_{HHH}}
\begin{align} % requires amsmath; align* for no eq. number
	g_{H^0_1 H^0_1 H^0_1}&= {-\epsilon_1} v \left( 12 \sqrt{2} \sin ( \beta + \alpha ) - 6 \sqrt{2} \cos 2\alpha \sin ( \beta - \alpha ) \right)  \nonumber   \\
                             &+ \epsilon_2 v \left( - 3 \sqrt{2} \cos ( \beta - \alpha ) + 3 \sqrt{2} \cos 2\alpha \cos ( \beta + \alpha ) \right) \\
	g_{H^0_1 H^0_1 H^0_2}&= {-\epsilon_1} v \left( 6 \sqrt{2} \cos 2\alpha \cos ( \beta - \alpha ) \right)  \nonumber \\
                             &+ \epsilon_2 v \left( \sqrt{2} \sin ( \beta - \alpha ) - 3 \sqrt{2} \cos 2\alpha \sin ( \beta + \alpha ) \right) \\
	g_{H^0_1 H^0_2 H^0_2}&= {-\epsilon_1} v \left( 6 \sqrt{2} \cos 2\alpha \sin ( \beta - \alpha ) \right) \nonumber \\
                             &+ \epsilon_2 v \left( \sqrt{2} \cos ( \beta - \alpha ) - 3 \sqrt{2} \cos 2\alpha \cos ( \beta + \alpha ) \right) \\
	g_{H^0_2 H^0_2 H^0_2}&= {-\epsilon_1} v \left( 12 \sqrt{2} \cos ( \beta + \alpha ) - 6 \sqrt{2} \cos 2\alpha \cos ( \beta - \alpha ) \right) \nonumber \\
                             &+ \epsilon_2 v \left( - 3 \sqrt{2} \sin ( \beta - \alpha ) + 3 \sqrt{2} \cos 2\alpha \sin ( \beta + \alpha ) \right) \\
	g_{H^0_1 H^0_3 H^0_3}&= {-\epsilon_1} v \left( - 2 \sqrt{2} \cos 2\beta \sin ( \beta - \alpha ) \right) \nonumber \\
                             &+ \epsilon_2 v \left( 3 \sqrt{2} \cos ( \beta - \alpha ) - \sqrt{2} \cos 2\beta \cos ( \beta + \alpha ) \right) \\
	g_{H^0_1 H^0_3 G^0}&= {-\epsilon_1} v \left( 2 \sqrt{2} \cos 2\beta \cos ( \beta - \alpha ) \right) \nonumber \\
                             &+ \epsilon_2 v \left( \sqrt{2} \sin ( \beta - \alpha ) - \sqrt{2} \cos 2\beta \sin ( \beta + \alpha ) \right) \\
	g_{H^0_1 G^0 G^0}&= {-\epsilon_1} v \left( 4 \sqrt{2} \sin ( \beta + \alpha ) + 2 \sqrt{2} \cos 2\beta \sin ( \beta - \alpha ) \right)\nonumber   \\
                             &+ \epsilon_2 v \left( - \sqrt{2} \cos ( \beta - \alpha ) + \sqrt{2} \cos 2\beta \cos ( \beta + \alpha ) \right) \\
	g_{H^0_2 H^0_3 H^0_3}&= {-\epsilon_1} v \left( 2 \sqrt{2} \cos 2\beta \cos ( \beta - \alpha ) \right) \nonumber  \\
                             &+ \epsilon_2 v \left( 3 \sqrt{2} \sin ( \beta - \alpha ) + \sqrt{2} \cos 2\beta \sin ( \beta + \alpha ) \right) \\
	g_{H^0_2 H^0_3 G^0}&= {-\epsilon_1} 
	   v \left( 2 \sqrt{2} \cos 2\beta \sin ( \beta - \alpha ) \right) \nonumber \\
                             &+ \epsilon_2 v \left( - \sqrt{2} \cos ( \beta - \alpha ) - \sqrt{2} \cos 2\beta \cos ( \beta + \alpha ) \right) \\
	g_{H^0_2 G^0 G^0}&= {-\epsilon_1} v \left( 4 \sqrt{2} \cos ( \beta + \alpha ) - 2 \sqrt{2} \cos 2\beta \cos ( \beta - \alpha ) \right) \nonumber  \\
                             &+ \epsilon_2 v \left( - \sqrt{2} \sin ( \beta - \alpha ) - \sqrt{2} \cos 2\beta \cos ( \beta + \alpha ) \right) \\
	g_{H^0_1 H^+ H^-}&= {-\epsilon_1} 
	v \left( - 2 \sqrt{2} \cos 2\beta \sin ( \beta - \alpha ) \right) \nonumber  \\
                             &+ \epsilon_2 v \left( \sqrt{2} \cos ( \beta - \alpha ) - \sqrt{2} \cos 2\beta \cos ( \beta + \alpha ) \right) \\
	g_{H^0_1 H^+ G^-}&= {-\epsilon_1} v \left( 2 \sqrt{2} \cos 2\beta \cos ( \beta - \alpha ) \right) \nonumber  \\	
                             &+ \epsilon_2 v \left( - \sqrt{2} \cos 2\beta \sin ( \beta + \alpha ) \right) \\
	g_{H^0_1 G^+ H^-}&= {-\epsilon_1} v \left( 2 \sqrt{2} \cos 2\beta \cos ( \beta - \alpha ) \right) \nonumber  \\
                             &+ \epsilon_2 v \left( - \sqrt{2} \cos 2\beta \sin ( \beta + \alpha ) \right) \\
	g_{H^0_1 G^+ G^-}&= {-\epsilon_1} v \left( 4 \sqrt{2} \sin ( \beta + \alpha ) + 2 \sqrt{2} \cos 2\beta \sin ( \beta - \alpha ) \right) \nonumber  \\
                             &+ \epsilon_2 v \left( - \sqrt{2} \cos ( \beta - \alpha ) + \sqrt{2} \cos 2\beta \cos ( \beta + \alpha ) \right) \\
	g_{H^0_2 H^+ H^-}&= {-\epsilon_1} v \left( 2 \sqrt{2} \cos 2\beta \cos ( \beta - \alpha ) \right)\nonumber   \\
                             &+ \epsilon_2 v \left( \sqrt{2} \sin ( \beta - \alpha ) + \sqrt{2} \cos 2\beta \sin ( \beta + \alpha ) \right) \\
	g_{H^0_2 H^+ G^-}&= {-\epsilon_1} v \left( 2 \sqrt{2} \cos 2\beta \sin ( \beta - \alpha ) \right)\nonumber  \\
                             &+ \epsilon_2 v \left( - \sqrt{2} \cos 2\beta \cos ( \beta + \alpha ) \right) \\
	g_{H^0_2 G^+ H^-}&= {-\epsilon_1} v \left( 2 \sqrt{2} \cos 2\beta \sin ( \beta - \alpha ) \right)\nonumber  \\
                             &+ \epsilon_2 v \left( - \sqrt{2} \cos 2\beta \cos ( \beta + \alpha ) \right) \\
	g_{H^0_2 G^+ G^-}&= {-\epsilon_1} v \left( 4 \sqrt{2} \cos ( \beta + \alpha ) - 2 \sqrt{2} \cos 2\beta \cos ( \beta - \alpha ) \right) \nonumber  \\
                             &+ \epsilon_2 v \left( - \sqrt{2} \sin ( \beta - \alpha ) - \sqrt{2} \cos 2\beta \sin ( \beta + \alpha ) \right) \\
	g_{H^0_3 H^+ G^-}&= \epsilon_2 v \left( -i \sqrt{2} \right) \\
	g_{H^0_3 G^+ H^-}&= \epsilon_2 v \left( i \sqrt{2} \right)
\end{align}

\subsection{$H$-$\wt\chi$-$\wt\chi$ vertices}

\subsubsection{$H^0$-$\wt\chi^0_i$-$\wt\chi^0_j$ vertices}
The vertex is
\vertex{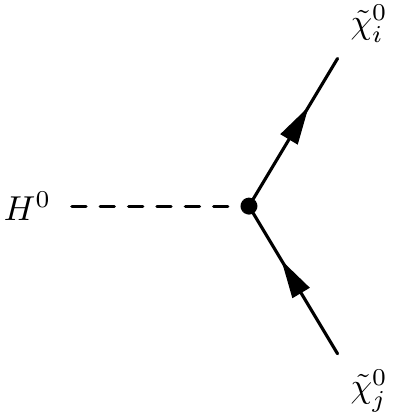}{g^L_{H^0 \tilde\chi^0_i \tilde\chi^0_j} P_L + g^R_{H^0 \tilde\chi^0_i \tilde\chi^0_j} P_R}
\begin{align}
 g^L_{H^0_1 \wt\chi^0_i \wt\chi^0_j} & = 2 \left( -\frac{\epsilon_1}{\mu^*} \right) \Big(- v\sqrt{2} \cos\beta \cos\alpha N^*_{i4} N^*_{j4} - v \sqrt{2} \sin\beta \sin\alpha N^*_{i3} N^*_{j3} \nonumber \\
& - 2\sqrt{2} v \sin(\alpha+\beta) \frac{1}{2}(N^*_{i4} N^*_{j3} + N^*_{j4} N^*_{i3})\Big) \\
 g^R_{H^0_1 \wt\chi^0_i \wt\chi^0_j} & = \left( g^L_{H^0_1 \wt\chi^0_i \wt\chi^0_j} \right)^* \\
 g^L_{H^0_2 \wt\chi^0_i \wt\chi^0_j} & = 2 \left( -\frac{\epsilon_1}{\mu^*} \right) \Big( v\sqrt{2} \cos\beta \sin\alpha N^*_{i4} N^*_{j4} - v \sqrt{2} \sin\beta \cos\alpha N^*_{i3} N^*_{j3} \nonumber \\
& - 2\sqrt{2} v \cos(\alpha+\beta) \frac{1}{2}(N^*_{i4} N^*_{j3} + N^*_{j4} N^*_{i3}) \Big) \\
 g^R_{H^0_2 \wt\chi^0_i \wt\chi^0_j} & = \left( g^L_{H^0_2 \wt\chi^0_i \wt\chi^0_j} \right)^* \\
 g^L_{H^0_3 \wt\chi^0_i \wt\chi^0_j} & = 2 \left( -\frac{\epsilon_1}{\mu^*} \right) \Big( -i v\frac{1}{\sqrt{2}} \sin{2\beta} N^*_{i4} N^*_{j4} - i v \frac{1}{\sqrt{2}} \sin{2\beta} N^*_{i3} N^*_{j3} \nonumber \\
& - i 2 \sqrt{2} v \frac{1}{2}(N^*_{i4} N^*_{j3} + N^*_{j4} N^*_{i3}) \Big) \\
 g^R_{H^0_3 \wt\chi^0_i \wt\chi^0_j} & = \left( g^L_{H^0_3 \wt\chi^0_i \wt\chi^0_j} \right)^* \\
 g^L_{G^0 \wt\chi^0_i \wt\chi^0_j} & = 2 \left( -\frac{\epsilon_1}{\mu^*} \right) \left( iv\sqrt{2} \cos^2\beta N^*_{i4} N^*_{j4} - i v \sqrt{2} \sin^2\beta N^*_{i3} N^*_{j3} \right) \nonumber \\
 g^R_{G^0 \wt\chi^0_i \wt\chi^0_j} & = \left( g^L_{H^0_1 \wt\chi^0_i \wt\chi^0_j} \right)^*
\end{align}

\subsubsection{$H^-$-$\wt\chi^0_j$-$\wt\chi^+_c$ vertices}
The vertex is
\vertex{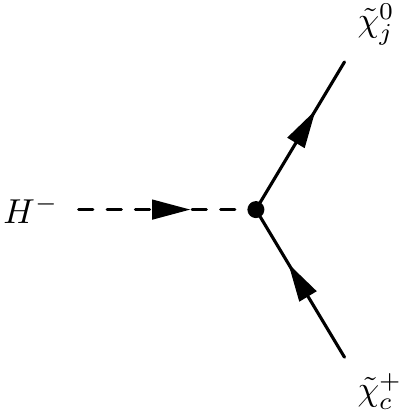}{g^L_{H^- \tilde\chi^0_j \tilde\chi^+_c} P_L + g^R_{H^0 \tilde\chi^0_j \tilde\chi^+_c} P_R}
\begin{align}
 g^L_{G^- \wt\chi^0_j \wt\chi^+_c} & = \left( -\frac{\epsilon_1}{\mu^*} \right)\left( - 2 v \cos^2\beta N^*_{j4} V^*_{c2} - v \sin{2\beta} N^*_{j3} V^*_{c2} \right) \\
 g^R_{G^- \wt\chi^0_j \wt\chi^+_c} & = \left( -\frac{\epsilon_1^*}{\mu} \right)\left( 2 v \sin^2\beta N_{j3} U_{c2} + v \sin{2\beta} N_{j4} U_{c2} \right) \\
 g^L_{H^- \wt\chi^0_j \wt\chi^+_c} & = \left( -\frac{\epsilon_1}{\mu^*} \right)\left( v \sin{2\beta} N^*_{j4} V^*_{c2} + 2 v \sin^2{\beta} N^*_{j3} V^*_{c2} \right) \\
 g^R_{H^- \wt\chi^0_j \wt\chi^+_c} & = \left( -\frac{\epsilon_1^*}{\mu} \right)\left( v \sin{2\beta} N_{j3} U_{c2} + 2 v \cos^2\beta N_{j4} U_{c2} \right)
\end{align}
Changing the directions of the arrows gives
\vertex{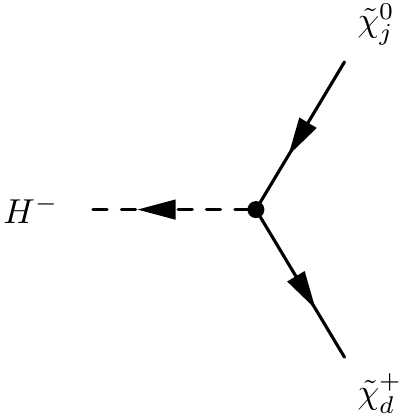}{\left(g^R_{H^- \tilde\chi^0_j \tilde\chi^+_c}\right)^* P_L + \left(g^L_{H^0 \tilde\chi^0_j \tilde\chi^+_c}\right)^* P_R}

\subsubsection{$H^0$-$\wt\chi^+_c$-$\wt\chi^+_d$ vertices}
The vertex is
\vertex{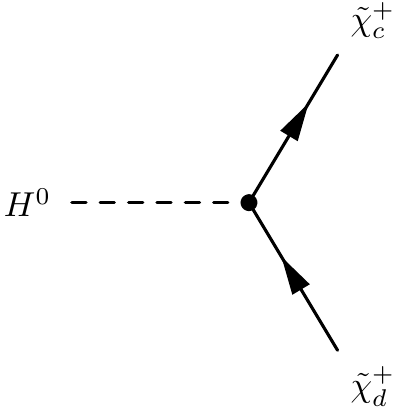}{g^L_{H^0 \tilde\chi^+_c \tilde\chi^+_d} P_L + g^R_{H^0 \tilde\chi^+_c \tilde\chi^+_d} P_R}
\begin{align}
 g^L_{H^0_1 \wt\chi^+_c \wt\chi^+_d} & = \left( -\frac{\epsilon_1}{\mu^*} \right) \sqrt{2} v \sin(\alpha+\beta) U^*_{c2} V^*_{d2} \\
 g^R_{H^0_1 \wt\chi^+_c \wt\chi^+_d} & = \left( -\frac{\epsilon_1^*}{\mu} \right) \sqrt{2} v \sin(\alpha+\beta) U_{d2} V_{c2} \\
 g^L_{H^0_2 \wt\chi^+_c \wt\chi^+_d} & = \left( -\frac{\epsilon_1}{\mu^*} \right) \sqrt{2} v \cos(\alpha+\beta) U^*_{c2} V^*_{d2} \\
 g^R_{H^0_2 \wt\chi^+_c \wt\chi^+_d} & = \left( -\frac{\epsilon_1^*}{\mu} \right) \sqrt{2} v \cos(\alpha+\beta) U_{d2} V_{c2} \\
 g^L_{H^0_3 \wt\chi^+_c \wt\chi^+_d} & = \left( -\frac{\epsilon_1}{\mu^*} \right) i\sqrt{2} v U^*_{c2} V^*_{d2} \\
 g^R_{H^0_3 \wt\chi^+_c \wt\chi^+_d} & = \left( -\frac{\epsilon_1^*}{\mu} \right) (-i\sqrt{2} v) U_{d2} V_{c2}
\end{align}

\subsection{$H$-$H$-$\wt\chi$-$\wt\chi$ vertices}
\subsubsection{$H^0$-$H^0$-$\wt\chi^0_i$-$\wt\chi^0_j$ vertices}
The vertex is
\vertex{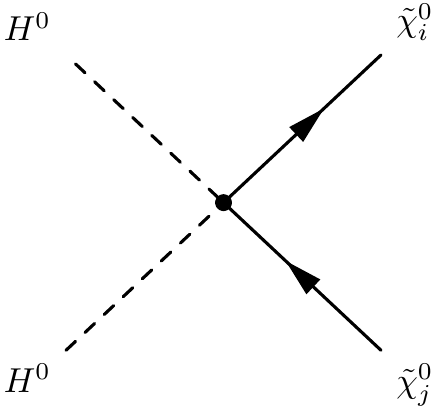}{g^L_{H^0 H^0 \tilde\chi^0_i \tilde\chi^0_j} P_L + g^R_{H^0 H^0 \tilde\chi^0_i \tilde\chi^0_j} P_R}
\begin{align}
 g^L_{H^0_1 H^0_1 \wt\chi^0_i \wt\chi^0_j} & = 4 \left( -\frac{\epsilon_1}{\mu^*} \right) \Big( -\frac{1}{2} \cos^2\alpha N^*_{i4} N^*_{j4} - \frac{1}{2} \sin^2\alpha N^*_{i3} N^*_{j3} \nonumber \\ 
& - \sin{2\alpha} \frac{1}{2}(N^*_{i4} N^*_{j3} + N^*_{j4} N^*_{i3}) \Big) \\
 g^R_{H^0_1 H^0_1 \wt\chi^0_i \wt\chi^0_j} & = \left(g^L_{H^0_1 H^0_1 \wt\chi^0_i \wt\chi^0_j}\right)^* \\
 g^L_{H^0_2 H^0_2 \wt\chi^0_i \wt\chi^0_j} & = 4 \left( -\frac{\epsilon_1}{\mu^*} \right) \Big( -\frac{1}{2} \sin^2\alpha N^*_{i4} N^*_{j4} - \frac{1}{2} \cos^2\alpha N^*_{i3} N^*_{j3} \nonumber \\
& + \sin{2\alpha} \frac{1}{2}(N^*_{i4} N^*_{j3} + N^*_{j4} N^*_{i3}) \Big) \\
 g^R_{H^0_2 H^0_2 \wt\chi^0_i \wt\chi^0_j} & = \left(g^L_{H^0_2 H^0_2 \wt\chi^0_i \wt\chi^0_j}\right)^* \\
 g^L_{H^0_3 H^0_3 \wt\chi^0_i \wt\chi^0_j} & = 4 \left( -\frac{\epsilon_1}{\mu^*} \right) \Big( \frac{1}{2} \sin^2\beta N^*_{i4} N^*_{j4} + \frac{1}{2} \cos^2\beta N^*_{i3} N^*_{j3} \nonumber \\
& + \sin{2\beta} \frac{1}{2}(N^*_{i4} N^*_{j3} + N^*_{j4} N^*_{i3}) \Big) \\
 g^R_{H^0_3 H^0_3 \wt\chi^0_i \wt\chi^0_j} & = \left(g^L_{H^0_3 H^0_3 \wt\chi^0_i \wt\chi^0_j}\right)^* \\
 g^L_{G^0 G^0 \wt\chi^0_i \wt\chi^0_j} & = 4 \left( -\frac{\epsilon_1}{\mu^*} \right) \Big( \frac{1}{2} \cos^2\beta N^*_{i4} N^*_{j4} + \frac{1}{2} \sin^2\beta N^*_{i3} N^*_{j3} \nonumber \\
& - \sin{2\beta} \frac{1}{2}(N^*_{i4} N^*_{j3} + N^*_{j4} N^*_{i3})\Big) \\
 g^R_{G^0 G^0 \wt\chi^0_i \wt\chi^0_j} & = \left(g^L_{G^0 G^0 \wt\chi^0_i \wt\chi^0_j}\right)^* \\
 g^L_{H^0_1 H^0_2 \wt\chi^0_i \wt\chi^0_j} & = 2 \left( -\frac{\epsilon_1}{\mu^*} \right) \Big( \frac{1}{2} \sin{2\alpha} N^*_{i4} N^*_{j4} - \frac{1}{2} \sin{2\alpha} N^*_{i3} N^*_{j3} \nonumber \\
& - 2\cos{2\alpha} \frac{1}{2}(N^*_{i4} N^*_{j3} + N^*_{j4} N^*_{i3})\Big) \\
 g^R_{H^0_1 H^0_2 \wt\chi^0_i \wt\chi^0_j} & = \left(g^L_{H^0_1 H^0_2 \wt\chi^0_i \wt\chi^0_j}\right)^* \\
 g^L_{H^0_1 H^0_3 \wt\chi^0_i \wt\chi^0_j} & = 2 \left( -\frac{\epsilon_1}{\mu^*} \right) \Big( -i \cos\alpha \sin\beta N^*_{i4} N^*_{j4} - i \sin\alpha\cos\beta N^*_{i3} N^*_{j3} \nonumber \\
& - 2i \cos(\alpha-\beta) \frac{1}{2}(N^*_{i4} N^*_{j3} + N^*_{j4} N^*_{i3}) \Big) \\
 g^R_{H^0_1 H^0_3 \wt\chi^0_i \wt\chi^0_j} & = \left(g^L_{H^0_1 H^0_3 \wt\chi^0_i \wt\chi^0_j}\right)^* \\
 g^L_{H^0_1 G^0 \wt\chi^0_i \wt\chi^0_j} & = 2 \left( -\frac{\epsilon_1}{\mu^*} \right) \Big( i \cos\alpha \cos\beta N^*_{i4} N^*_{j4} - i \sin\alpha\sin\beta N^*_{i3} N^*_{j3}  \nonumber \\
& + 2i \sin(\alpha-\beta) \frac{1}{2}(N^*_{i4} N^*_{j3} + N^*_{j4} N^*_{i3})\Big) \nonumber \\
 g^R_{H^0_1 G^0 \wt\chi^0_i \wt\chi^0_j} & = \left(g^L_{H^0_1 G^0 \wt\chi^0_i \wt\chi^0_j}\right)^* \\
 g^L_{H^0_2 H^0_3 \wt\chi^0_i \wt\chi^0_j} & = 2 \left( -\frac{\epsilon_1}{\mu^*} \right) \Big( i \sin\alpha \sin\beta N^*_{i4} N^*_{j4} - i \cos\alpha\cos\beta N^*_{i3} N^*_{j3} \nonumber \\
& + 2i \sin(\alpha-\beta) \frac{1}{2}(N^*_{i4} N^*_{j3} + N^*_{j4} N^*_{i3}) \Big) \\
 g^R_{H^0_2 H^0_3 \wt\chi^0_i \wt\chi^0_j} & = \left(g^L_{H^0_2 H^0_3 \wt\chi^0_i \wt\chi^0_j}\right)^*  \\
 g^L_{H^0_2 G^0 \wt\chi^0_i \wt\chi^0_j} & = 2 \left( -\frac{\epsilon_1}{\mu^*} \right) \Big( -i \sin\alpha \cos\beta N^*_{i4} N^*_{j4} - i \cos\alpha\sin\beta N^*_{i3} N^*_{j3} \nonumber \\
& + 2i \cos(\alpha-\beta) \frac{1}{2}(N^*_{i4} N^*_{j3} + N^*_{j4} N^*_{i3}) \Big) \\
 g^R_{H^0_2 G^0 \wt\chi^0_i \wt\chi^0_j} & = \left(g^L_{H^0_2 G^0 \wt\chi^0_i \wt\chi^0_j}\right)^* \\
 g^L_{H^0_3 G^0 \wt\chi^0_i \wt\chi^0_j} & = 2 \left( -\frac{\epsilon_1}{\mu^*} \right) \Big( -\frac{1}{2} \sin{2\beta} N^*_{i4} N^*_{j4} + \frac{1}{2}\sin{2\beta} N^*_{i3} N^*_{j3} \nonumber \\
& - 2 \cos{2\beta} \frac{1}{2}(N^*_{i4} N^*_{j3} + N^*_{j4} N^*_{i3}) \Big) \\
 g^R_{H^0_3 G^0 \wt\chi^0_i \wt\chi^0_j} & = \left(g^L_{H^0_3 G^0 \wt\chi^0_i \wt\chi^0_j}\right)^*
\end{align}

\subsubsection{$H^+$-$H^-$-$\wt\chi^0_i$-$\wt\chi^0_j$ vertices}
The vertex is
\vertex{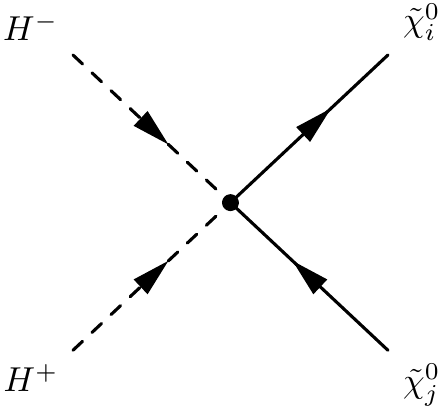}{g^L_{H^+ H^- \tilde\chi^0_i \tilde\chi^0_j} P_L + g^R_{H^+ H^- \tilde\chi^0_i \tilde\chi^0_j} P_R}
\begin{align}
 g^L_{G^+ G^- \wt\chi^0_i \wt\chi^0_j} & = 2 \left( -\frac{\epsilon_1}{\mu^*} \right) \left( - \sin{2\beta} \frac{1}{2}(N^*_{i4} N^*_{j3} + N^*_{j4} N^*_{i3})\right) \\
 g^R_{G^+ G^- \wt\chi^0_i \wt\chi^0_j} & = \left(g^L_{G^+ G^- \wt\chi^0_i \wt\chi^0_j}\right)^* \\
 g^L_{H^+ H^- \wt\chi^0_i \wt\chi^0_j} & = 2 \left( -\frac{\epsilon_1}{\mu^*} \right) \left( + \sin{2\beta} \frac{1}{2}(N^*_{i4} N^*_{j3} + N^*_{j4} N^*_{i3})\right) \\
 g^R_{H^+ H^- \wt\chi^0_i \wt\chi^0_j} & = \left(g^L_{H^+ H^- \wt\chi^0_i \wt\chi^0_j}\right)^* \\
 g^L_{G^- H^+ \wt\chi^0_i \wt\chi^0_j} & = 2 \left( -\frac{\epsilon_1}{\mu^*} \right) \left( - 2 \cos^2\beta \frac{1}{2}(N^*_{i4} N^*_{j3} + N^*_{j4} N^*_{i3}) \right) \\
 g^R_{G^- H^+ \wt\chi^0_i \wt\chi^0_j} & = 2 \left( -\frac{\epsilon_1^*}{\mu} \right) \left( + 2 \sin^2\beta \frac{1}{2}(N_{i4} N_{j3} + N_{j4} N_{i3}) \right) \\
 g^L_{G^+ H^- \wt\chi^0_i \wt\chi^0_j} & = 2 \left( -\frac{\epsilon_1}{\mu^*} \right)  \left( + 2 \sin^2\beta \frac{1}{2}(N^*_{i4} N^*_{j3} + N^*_{j4} N^*_{i3}) \right) \\
 g^R_{G^+ H^- \wt\chi^0_i \wt\chi^0_j} & = 2 \left( -\frac{\epsilon_1^*}{\mu} \right) \left( - 2 \cos^2\beta \frac{1}{2}(N_{i4} N_{j3} + N_{j4} N_{i3}) \right)
\end{align}

\subsubsection{$H^0$-$H^0$-$\wt\chi^+_c$-$\wt\chi^+_d$ vertices}
The vertex is
\vertex{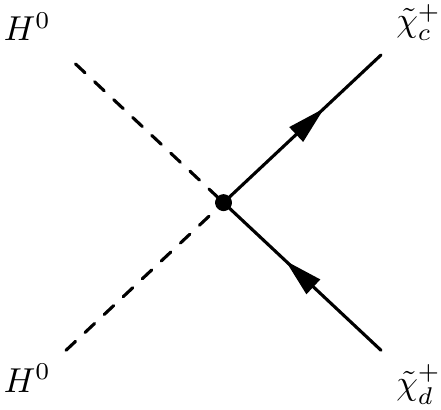}{g^L_{H^0 H^0 \tilde\chi^+_c \tilde\chi^+_d} P_L + g^R_{H^0 H^0 \tilde\chi^+_c \tilde\chi^+_d} P_R}
\begin{align}
 g^L_{H^0_1 H^0_1 \wt\chi^+_c \wt\chi^+_d} & = 2 \left( -\frac{\epsilon_1}{\mu^*} \right) \left( \frac{1}{2} \sin{2\alpha} V^*_{d2} U^*_{c2} \right) \\
 g^R_{H^0_1 H^0_1 \wt\chi^+_c \wt\chi^+_d} & = 2 \left( -\frac{\epsilon_1^*}{\mu} \right) \left( \frac{1}{2} \sin{2\alpha} V_{c2} U_{d2} \right) \\
 g^L_{H^0_2 H^0_2 \wt\chi^+_c \wt\chi^+_d} & = 2 \left( -\frac{\epsilon_1}{\mu^*} \right) \left( - \frac{1}{2} \sin{2\alpha} V^*_{d2} U^*_{c2} \right) \\
 g^R_{H^0_2 H^0_2 \wt\chi^+_c \wt\chi^+_d} & = 2 \left( -\frac{\epsilon_1^*}{\mu} \right) \left( - \frac{1}{2} \sin{2\alpha} V_{c2} U_{d2} \right) \\
 g^L_{H^0_3 H^0_3 \wt\chi^+_c \wt\chi^+_d} & = 2 \left( -\frac{\epsilon_1}{\mu^*} \right) \left( -\frac{1}{2} \sin{2\beta} V^*_{d2} U^*_{c2} \right) \\
 g^R_{H^0_3 H^0_3 \wt\chi^+_c \wt\chi^+_d} & = 2 \left( -\frac{\epsilon_1^*}{\mu} \right) \left( -\frac{1}{2} \sin{2\beta} V_{c2} U_{d2} \right) \\
 g^L_{G^0 G^0 \wt\chi^+_c \wt\chi^+_d} & = 2 \left( -\frac{\epsilon_1}{\mu^*} \right) \left( \frac{1}{2} \sin{2\beta} V^*_{d2} U^*_{c2} \right) \\
 g^R_{G^0 G^0 \wt\chi^+_c \wt\chi^+_d} & = 2 \left( -\frac{\epsilon_1^*}{\mu} \right) \left( \frac{1}{2} \sin{2\beta} V_{c2} U_{d2} \right) \\
 g^L_{H^0_1 H^0_2 \wt\chi^+_c \wt\chi^+_d} & = \left( -\frac{\epsilon_1}{\mu^*} \right) \left( \cos{2\alpha} V^*_{d2} U^*_{c2} \right) \\
 g^R_{H^0_1 H^0_2 \wt\chi^+_c \wt\chi^+_d} & = \left( -\frac{\epsilon_1^*}{\mu} \right) \left( \cos{2\alpha} V_{c2} U_{d2} \right) \\
 g^L_{H^0_1 H^0_3 \wt\chi^+_c \wt\chi^+_d} & = \left( -\frac{\epsilon_1}{\mu^*} \right) \left( i \cos(\beta-\alpha) V^*_{d2} U^*_{c2} \right) \\
 g^R_{H^0_1 H^0_3 \wt\chi^+_c \wt\chi^+_d} & = \left( -\frac{\epsilon_1^*}{\mu} \right) \left( -i \cos(\beta-\alpha) V_{c2} U_{d2} \right)  \\
 g^L_{H^0_1 G^0 \wt\chi^+_c \wt\chi^+_d} & = \left( -\frac{\epsilon_1}{\mu^*} \right) \left( i \sin(\beta-\alpha) V^*_{d2} U^*_{c2} \right) \\
 g^R_{H^0_1 G^0 \wt\chi^+_c \wt\chi^+_d} & = \left( -\frac{\epsilon_1^*}{\mu} \right) \left( -i \sin(\beta-\alpha) V_{c2} U_{d2} \right) \\
 g^L_{H^0_2 H^0_3 \wt\chi^+_c \wt\chi^+_d} & = \left( -\frac{\epsilon_1}{\mu^*} \right) \left( i \sin(\beta-\alpha) V^*_{d2} U^*_{c2} \right) \\
 g^R_{H^0_2 H^0_3 \wt\chi^+_c \wt\chi^+_d} & = \left( -\frac{\epsilon_1^*}{\mu} \right) \left( - i \sin(\beta-\alpha) V_{c2} U_{d2} \right) \\
 g^L_{H^0_2 G^0 \wt\chi^+_c \wt\chi^+_d} & = \left( -\frac{\epsilon_1}{\mu^*} \right) \left( -i \cos(\beta-\alpha) V^*_{d2} U^*_{c2} \right) \\
 g^R_{H^0_2 G^0 \wt\chi^+_c \wt\chi^+_d} & = \left( -\frac{\epsilon_1^*}{\mu} \right) \left( + i \cos(\beta-\alpha) V_{c2} U_{d2} \right) \\
 g^L_{H^0_3 G^0 \wt\chi^+_c \wt\chi^+_d} & = \left( -\frac{\epsilon_1}{\mu^*} \right) \left(\cos{2\beta} V^*_{d2} U^*_{c2} \right) \\
 g^R_{H^0_3 G^0 \wt\chi^+_c \wt\chi^+_d} & = \left( -\frac{\epsilon_1^*}{\mu} \right) \left( \cos{2\beta} V_{c2} U_{d2} \right)
\end{align}

\subsubsection{$H^+$-$H^-$-$\wt\chi^+_c$-$\wt\chi^+_d$ vertices}
The vertex is
\vertex{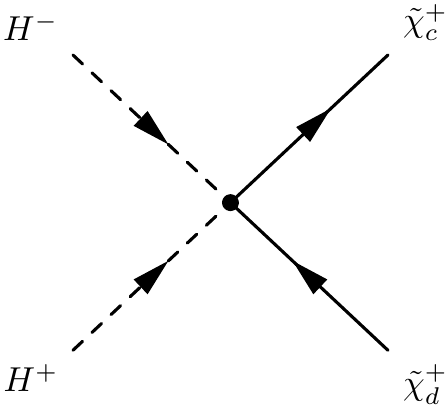}{g^L_{H^+ H^- \tilde\chi^+_c \tilde\chi^+_d} P_L + g^R_{H^+ H^- \tilde\chi^+_c \tilde\chi^+_d} P_R}
\begin{align}
 g^L_{H^+ H^- \wt\chi^+_c \wt\chi^+_d} & = \left( -\frac{\epsilon_1}{\mu^*} \right) \left( -2 \sin{2\beta} V^*_{d2} U^*_{c2} \right) \\
 g^R_{H^+ H^- \wt\chi^+_c \wt\chi^+_d} & = \left( -\frac{\epsilon_1^*}{\mu} \right) \left( -2 \sin{2\beta} V_{c2} U_{d2} \right) \\
 g^L_{G^+ G^- \wt\chi^+_c \wt\chi^+_d} & = \left( -\frac{\epsilon_1}{\mu^*} \right) \left( 2 \sin{2\beta} V^*_{d2} U^*_{c2} \right) \\
 g^R_{G^+ G^- \wt\chi^+_c \wt\chi^+_d} & = \left( -\frac{\epsilon_1^*}{\mu} \right) \left( 2 \sin{2\beta} V_{c2} U_{d2} \right) \\
 g^L_{G^- H^+ \wt\chi^+_c \wt\chi^+_d} & = \left( -\frac{\epsilon_1}{\mu^*} \right) \left( 4 \cos^2{\beta} V^*_{d2} U^*_{c2} \right) \\
 g^R_{G^- H^+ \wt\chi^+_c \wt\chi^+_d} & = \left( -\frac{\epsilon_1^*}{\mu} \right) \left( -4 \sin^2{\beta} V_{c2} U_{d2} \right) \\
 g^L_{G^+ H^- \wt\chi^+_c \wt\chi^+_d} & = \left( -\frac{\epsilon_1}{\mu^*} \right) \left( -4 \sin^2{\beta} V^*_{c2} U^*_{d2} \right) \\
 g^R_{G^+ H^- \wt\chi^+_c \wt\chi^+_d} & = \left( -\frac{\epsilon_1^*}{\mu} \right) \left( 4 \cos^2{\beta} V_{d2} U_{c2} \right)
\end{align}

\subsubsection{$H^-$-$H^-$-$(\wt\chi^+_c)^c$-$\wt\chi^+_d$ vertices}
The vertex is
\vertex{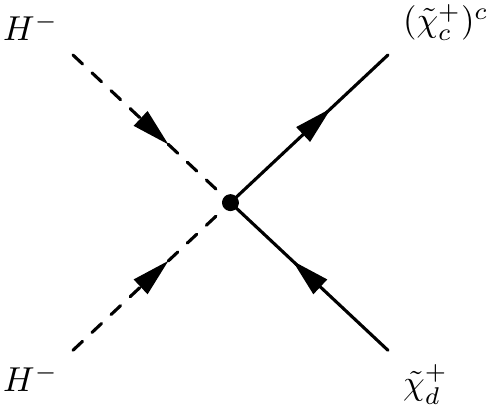}{g^L_{H^- H^- (\tilde\chi^+_c)^c \tilde\chi^+_d} P_L + g^R_{H^- H^- (\tilde\chi^+_c)^c \tilde\chi^+_d} P_R}
\begin{align}
 g^L_{H^- H^- (\wt\chi^+_c)^c \wt\chi^+_d} & = 4 \left( -\frac{\epsilon_1}{\mu^*} \right) \left( -\sin^2{\beta} V^*_{c2} V^*_{d2} \right) \\
 g^R_{H^- H^- (\wt\chi^+_c)^c \wt\chi^+_d} & = 4 \left( -\frac{\epsilon_1^*}{\mu} \right) \left( -\cos^2{\beta} U_{c2} U_{d2} \right) \\
 g^L_{G^- G^- (\wt\chi^+_c)^c \wt\chi^+_d} & = 4 \left( -\frac{\epsilon_1}{\mu^*} \right) \left( - \cos^2{\beta} V^*_{c2} V^*_{d2} \right) \\
 g^R_{G^- G^- (\wt\chi^+_c)^c \wt\chi^+_d} & = 4 \left( -\frac{\epsilon_1^*}{\mu} \right) \left( -\sin^2{\beta} U_{c2} U_{d2} \right) \\
 g^L_{G^- H^- (\wt\chi^+_c)^c \wt\chi^+_d} & = 2 \left( -\frac{\epsilon_1}{\mu^*} \right) \left( \sin{2\beta} V^*_{c2} V^*_{d2} \right) \\
 g^R_{G^- H^- (\wt\chi^+_c)^c \wt\chi^+_d} & = 2 \left( -\frac{\epsilon_1^*}{\mu} \right) \left( -\sin{2\beta} U_{c2} U_{d2} \right)
\end{align}
Changing the directions of the arrows gives
\vertex{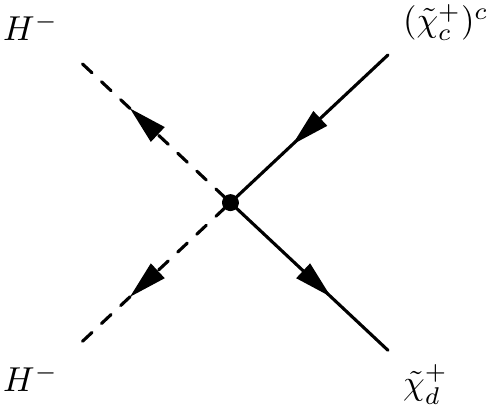} {\left(g^R_{H^- H^- (\tilde\chi^+_c)^c \tilde\chi^+_d}\right)^* P_L + \left(g^L_{H^- H^- (\tilde\chi^+_c)^c \tilde\chi^+_d}\right)^* P_R}

\subsubsection{$H^0$-$H^-$-$\wt\chi^0_j$-$\wt\chi^+_c$ vertices}
The vertex is
\vertex{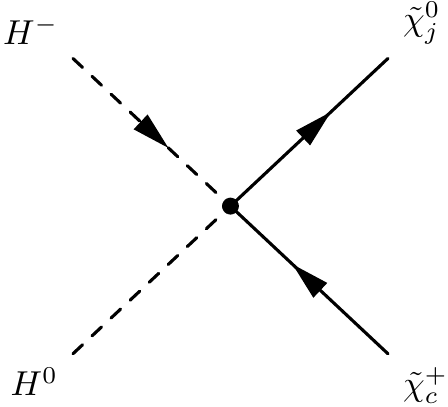}{g^L_{H^0 H^- \tilde\chi^0_j \tilde\chi^+_c} P_L + g^R_{H^0 H^- \tilde\chi^0_j \tilde\chi^+_c} P_R}
\begin{align}
 g^L_{H^0_1 G^- \wt\chi^0_j \wt\chi^+_c} & = \left( -\frac{\epsilon_1}{\mu^*} \right) \left( -\sqrt{2} \cos\alpha \cos\beta N^*_{j4} V^*_{c2} - \sqrt{2} \cos\beta \sin\alpha N^*_{j3} V^*_{c2} \right) \\
 g^R_{H^0_1 G^- \wt\chi^0_j \wt\chi^+_c} & = \left( -\frac{\epsilon_1^*}{\mu} \right) \left( \sqrt{2} \sin\alpha \sin\beta N_{j3} U_{c2} + \sqrt{2} \cos\alpha \sin\beta N_{j4} U_{c2} \right) \\
 g^L_{H^0_1 H^- \wt\chi^0_j \wt\chi^+_c} & = \left( -\frac{\epsilon_1}{\mu^*} \right) \left( \sqrt{2} \cos\alpha \sin\beta N^*_{j4} V^*_{c2} + \sqrt{2} \sin\alpha \sin\beta N^*_{j3} V^*_{c2} \right) \\
 g^R_{H^0_1 H^- \wt\chi^0_j \wt\chi^+_c} & = \left( -\frac{\epsilon_1^*}{\mu} \right) \left( \sqrt{2} \sin\alpha \cos\beta N_{j3} U_{c2} + \sqrt{2} \cos\alpha \cos\beta N_{j4} U_{c2} \right) \\
 g^L_{H^0_2 G^- \wt\chi^0_j \wt\chi^+_c} & = \left( -\frac{\epsilon_1}{\mu^*} \right) \left( \sqrt{2} \sin\alpha \cos\beta N^*_{j4} V^*_{c2} - \sqrt{2} \cos\alpha \cos\beta N^*_{j3} V^*_{c2} \right) \\
 g^R_{H^0_2 G^- \wt\chi^0_j \wt\chi^+_c} & = \left( -\frac{\epsilon_1^*}{\mu} \right) \left( \sqrt{2} \cos\alpha \sin\beta N_{j3} U_{c2} - \sqrt{2} \sin\alpha \sin\beta N_{j4} U_{c2} \right) \\
 g^L_{H^0_2 H^- \wt\chi^0_j \wt\chi^+_c} & = \left( -\frac{\epsilon_1}{\mu^*} \right) \left( -\sqrt{2} \sin\alpha \sin\beta N^*_{j4} V^*_{c2} + \sqrt{2} \cos\alpha \sin\beta N^*_{j3} V^*_{c2} \right) \\
 g^R_{H^0_2 H^- \wt\chi^0_j \wt\chi^+_c} & = \left( -\frac{\epsilon_1^*}{\mu} \right) \left( \sqrt{2} \cos\alpha \cos\beta N_{j3} U_{c2} - \sqrt{2} \sin\alpha \cos\beta N_{j4} U_{c2} \right) \\
 g^L_{H^0_3 G^- \wt\chi^0_j \wt\chi^+_c} & = \left( -\frac{\epsilon_1}{\mu^*} \right) \left( -i\frac{1}{\sqrt{2}} \sin{2\beta} N^*_{j4} V^*_{c2} - i \sqrt{2} \cos^2\beta N^*_{j3} V^*_{c2} \right) \\
 g^R_{H^0_3 G^- \wt\chi^0_j \wt\chi^+_c} & = \left( -\frac{\epsilon_1^*}{\mu} \right) \left( -i\frac{1}{\sqrt{2}} \sin{2\beta} N_{j3} U_{c2} -i \sqrt{2} \sin^2\beta N_{j4} U_{c2} \right) \\
 g^L_{H^0_3 H^- \wt\chi^0_j \wt\chi^+_c} & = \left( -\frac{\epsilon_1}{\mu^*} \right) \left( i\sqrt{2} \sin^2\beta N^*_{j4} V^*_{c2} + i\frac{1}{\sqrt{2}} \sin{2\beta} N^*_{j3} V^*_{c2} \right) \\
 g^R_{H^0_3 H^- \wt\chi^0_j \wt\chi^+_c} & = \left( -\frac{\epsilon_1^*}{\mu} \right) \left( -i \sqrt{2} \cos^2\beta N_{j3} U_{c2} - i\frac{1}{\sqrt{2}} \sin{2\beta} N_{j4} U_{c2} \right) \\
 g^L_{G^0 G^- \wt\chi^0_j \wt\chi^+_c} & = \left( -\frac{\epsilon_1}{\mu^*} \right) \left( +i\sqrt{2} \cos^2\beta N^*_{j4} V^*_{c2} - i \frac{1}{\sqrt{2}} \sin{2\beta} N^*_{j3} V^*_{c2} \right) \\
 g^R_{G^0 G^- \wt\chi^0_j \wt\chi^+_c} & = \left( -\frac{\epsilon_1^*}{\mu} \right) \left( -i\sqrt{2} \sin^2\beta N_{j3} U_{c2} +i \frac{1}{\sqrt{2}} \sin{2\beta} N_{j4} U_{c2} \right) \\
 g^L_{G^0 H^- \wt\chi^0_j \wt\chi^+_c} & = \left( -\frac{\epsilon_1}{\mu^*} \right) \left( -i\frac{1}{\sqrt{2}} \sin{2\beta} N^*_{j4} V^*_{c2} + i\sqrt{2} \sin^2\beta N^*_{j3} V^*_{c2} \right) \\
 g^R_{G^0 H^- \wt\chi^0_j \wt\chi^+_c} & = \left( -\frac{\epsilon_1^*}{\mu} \right) \left( -i \frac{1}{\sqrt{2}} \sin{2\beta} N_{j3} U_{c2} + i\sqrt{2} \cos^2\beta N_{j4} U_{c2} \right)
\end{align}

Changing the directions of the arrows gives
\vertex{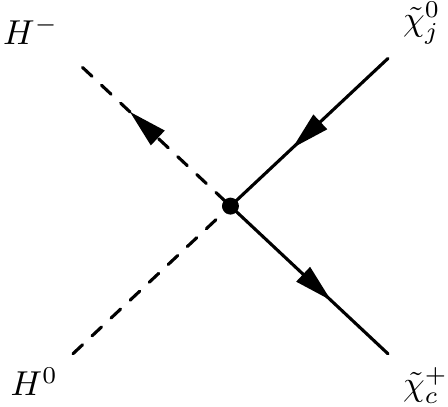}{\left(g^R_{H^0 H^- \tilde\chi^0_j \tilde\chi^+_c}\right)^* P_L + \left(g^R_{H^0 H^- \tilde\chi^0_j \tilde\chi^+_c}\right)^* P_R \; .}

}
\section{Acknowledgments} 
\label{acknow}
We thank 
Ulf Danielsson,
Emilian Dudas, Gordon Kane, Maxim Perelstein, Aaron Pierce, 
and Scott Thomas for useful discussions. 
We also thank Thomas Hahn and Sven Heinemeyer
for extensive help with {\tt FeynHiggs}. 
MB and JE thank the Swedish Research
Council (VR) for support. MB also thanks 
the Swedish Foundation for International Cooperation in Research
and Higher Education
(STINT) for partial support. 
PG was partially supported by NSF award PHY-0456825. PG thanks Stockholm
University for support during the completion of this work.
\end{appendix}
\bibliographystyle{kasper}
\bibliography{bmssm}

\end{document}